\documentclass{article}

\usepackage{arxiv}

\usepackage[utf8]{inputenc} 
\usepackage[T1]{fontenc}    
\usepackage{hyperref}       
\usepackage{url}            
\usepackage{array}
\newcolumntype{R}[1]{>{\raggedleft\let\newline\\\arraybackslash\hspace{0pt}}m{#1}}
\usepackage{booktabs}       
\usepackage{multirow}
\usepackage{amsfonts}       
\usepackage{nicefrac}       
\usepackage{microtype}      
\usepackage{lipsum}
\usepackage{amsmath}
\usepackage{amssymb}
\numberwithin{equation}{section}
\usepackage[colorinlistoftodos]{todonotes}
\usepackage{color}
\usepackage{makecell}
\usepackage[normalem]{ulem}
\usepackage[nameinlink,capitalise]{cleveref}
\usepackage{graphicx}
\usepackage{caption}
\usepackage{subcaption}
\usepackage{soul}
\usepackage{enumitem}%
\usepackage{calc}
\usepackage{tasks}
\usepackage[section]{placeins}

\newcommand{\mean}[1]{\langle#1\rangle}
\newcommand{\E}{\mathbf{E}}
\newcommand{\LN}{\mathcal{LN}}
\newcommand{\dist}{\langle \mathrm{dist} \rangle}
\newcommand{\fit}{1+\LN(\ln(2),0.25)}

\title{Inferring urban social networks from publicly available data}

\author{
Stefano Guarino\textsuperscript{*},
Enrico Mastrostefano,
Massimo Bernaschi, 
Alessandro Celestini,\\
\textbf{Marco Cianfriglia,
Davide Torre,
Lena Zastrow}\\
Istituto per le Applicazioni del Calcolo ``Mauro Picone''\\Consiglio Nazionale delle Ricerche\\
\texttt{*s.guarino@iac.cnr.it}
}

\begin{document}
\maketitle

\begin{abstract}
The emergence of social networks and the definition of suitable generative models for synthetic yet realistic social graphs are widely studied problems in the literature.
By not being tied to any real data, random graph models cannot capture all the subtleties of real networks and are inadequate for many practical contexts -- including areas of research, such as computational epidemiology, which are recently high on the agenda.
At the same time, the so-called \emph{contact} networks describe interactions, rather than relationships, and are strongly dependent on the application and on the size and quality of the sample data used to infer them.
To fill the gap between these two approaches, we present a data-driven model for urban social networks, implemented and released as open source software.
Given a territory of interest, and only based on widely available aggregated demographic and social-mixing data, we construct an age-stratified and geo-referenced synthetic population whose individuals are connected by ``strong ties'' of two types: intra-household (\emph{e.g.}, kinship) or friendship.
While household links are entirely data-driven, we propose a parametric probabilistic model for friendship, based on the assumption that distances and age differences play a role, and that not all individuals are equally sociable.
The demographic and geographic factors governing the structure of the obtained network, under different configurations, are thoroughly studied through extensive simulations focused on three Italian cities of different size.

\keywords{Urban social network \and Graph  model \and Data-driven \and Open source \and Simulator.}

\end{abstract}


\section{Introduction and Background}
\label{sec:intro}
Defining accurate models for real-world social networks is instrumental in several research fields, \emph{e.g.}, in sociology~\cite{book405595}, epidemiology~\cite{Cauchemez-2011} or marketing~\cite{Cynthi-2004}.
In combination with computer simulations these models may represent a valuable tool to understand social phenomena, along with classic analytical studies.
Dynamic processes, such as the spread of a disease or a rumour, can be represented upon suitable networks that encode the patterns of connection and interaction among the individuals of a population.
Moreover, the comparison of synthetic networks produced by different generative models helps to infer how each factor contributes to the emergence of experimentally measured properties of real networks~\cite{alizadeh2017generating}.

In this paper, we present a novel computational model for urban social networks, that combines a data-driven framework with a set of adjustable parameters.
A fully operational open source implementation of the model is available under the GPL v3 at \href{https://gitlab.com/cranic-group/usn}{gitlab.com/cranic-group/usn}.
The software allows to generate a synthetic social network of ``strong ties''~\cite{krackhardt2003strength} among geo-referenced and age-stratified individuals.
The graph encodes information on the urban social fabric and, as such, it increases the plausibility of dynamic (\emph{e.g.}, transmission) processes that may be influenced by preferences and actions of agents and groups of related agents.
On the one hand, our social graph may be used to simulate the fact that friends and relatives may go out together, organize public or private meetings, and are, in general, more likely to interact.
This can be achieved by using proximity, measured on the network, as a driving factor in the simulation of contacts and events.
On the other hand, the fact that the social graph is embedded into the urban landscape makes it possible to consider both geographical constraints and social ties to comprehend how specific places of aggregation foster interactions.
Finally, this network can be used as a standalone tool to characterize urban social relations patterns and to understand how these patterns are influenced by the geography and the demography of a given territory.

The edges of our spatial network describe stable interpersonal relationships of two types: intra-household (\emph{e.g.}, kinship) and friendship.
Once the organization of the population into households has been inferred from the available data, the household network can be defined quite naturally as a set of \emph{cliques} (\emph{i.e.}, complete subgraphs), one for each household.
The design of an accurate model for friendships is not equally straightforward.
Most simulation-based social studies claim a lack of reliable data and therefore model social networks by using well-known random graph models, such as Erdos-Renyi (ER), Small-world, or Scale-free graphs~\cite{amblard2015models}.
However, survey-based interaction networks, notoriously biased towards strong relationships~\cite{eagle2009inferring}, do not seem to follow any well-known simple random model~\cite{barrett2009generation}.

Inspired by both empirical studies~\cite{palla2007quantifying, herrera2015anatomy, scellato2011socio, onnela2007analysis} and previous modeling attempts~\cite{caldarelli2002scale} of real world social networks, we designed our friendship network model on top of the fundamental assumption that the probability $\Pr[u,v]$ of two agents $u$ and $v$ being friends is ultimately governed by three key factors:

\begin{itemize}
    \item their age group, whose role in the edge creation process can be described by means of a $n\times n$ social mixing matrix $S = \{s_{i,j}\}$, where $s_{i,j}$ measures the frequency of relationships between individuals of age groups $i$ and $j$;
    \item their geographical distance, whose impact can be controlled through a suitable non-increasing function $D(u,v)$ of the distance $d(u,v)$ between $u$ and $v$;
    \item the sociability of the two individuals (\emph{i.e.}, the propensity of each of them to have friends), measured in our model by a \emph{social-fitness score} $f_u$ associated to each agent $u$;
\end{itemize}
While $f_u$ and $D$ might be adjusted to the type and strength of the social ties that one aims at reproducing, the coefficients $s_{i,j}$ should instead be derived from social survey data, when possible.
Even in the absence of information about social \emph{relationships}, publicly available data regarding patterns of physical \emph{contacts} and \emph{interactions} should be regarded as a valuable source to estimate the ratios of intra- and inter-age group connections.
In the proposed model, $D$ and $f_u$ thus only impact on the selection of edges that exist between any two groups $i$ and $j$, whereas the overall social mixing structure defined by the data-driven $S$ is preserved, at least on average.
For the scope of the present paper, we estimate the coefficients $s_{i,j}$ 
based on data from Polymod~\cite{Mossong-2008}, extracted through the recently released SOCRATES~\cite{willem2020socrates} Data Tool\footnote{\url{https://lwillem.shinyapps.io/socrates_rshiny/}.}.
The tool allows to easily specify parameters such as age breaks, gender, day of the week, duration or location of the contacts, and it produces a social contact matrix drawing from the best public survey datasets for the country of interest.
Any suitable mixing matrix can, however, be fed to the simulator through a dedicated configuration file.

Besides the available real-world data, the resulting network depends on a number of design choices.
Based on the related literature, we identified a set of minimum requirements to guide the definition of suitable range of values for the parameters of our model, which can be summarized as follows:
\begin{description}[style=sameline, leftmargin=\widthof{Tie}]
\item[Tie strength]
Our model aims at representing strong ties as the union of two layers: the layer of households and the layer of friendship.
The former is entirely data-driven, and so is its average degree $\nu$.
Friendship ties, on the other hand, are generated using a probabilistic model and, to retain control over the density of this layer, the average number of friends $\mu$ is an input parameter.  There are at least two reasons to only consider fairly small values of $\mu$: 
(i) friendship links should represent social ties comparable to kinship and, as such, rare \cite{krackhardt2003strength}, with acquaintances modeled by short, but $>1$, distances on the graph;
(ii) a small value of $\mu$ ensures that our network is quite sparse, a necessary feature in most practical applications.

\item[Minimum connectivity]
Recent work highlighted that spatial social networks are generally well connected, with a single giant component covering the great majority of the graph, at least for urban areas~\cite{onnela2007analysis, herrera2015anatomy, liben2005geographic, scellato2011socio}.
To represent social relationships in a city in a faithful way, we therefore argue that, while small enough to mimic strong ties, $\mu$ must also be large enough so that the combination of household and friendship edges guarantees connectivity.

\item[Heterogeneity]
The parameters $f_u$ and $D$ are meant to break the homogeneity of the network, in such a way to mimic well known features of real-world spatial social and contact networks.
In particular, previous works agree on a heavy- but not fat-tailed degree distribution~\cite{lambiotte2008geographical, herrera2015anatomy, liben2005geographic, illenberger2013role} and on an inverse-power-law dependence on the distance with exponents in the range $[0.5,2]$~\cite{lambiotte2008geographical,latane1995distance,onnela2011geographic}.
Special attention will be therefore paid to the combination of $f_u$ following a (shifted) Lognormal distribution with limited skewness and $D= d(u,v)^{-\beta}$ for suitable $\beta$.

\end{description}

Our network model is described in details in Section~\ref{sec:methods}.
The model is experimentally evaluated by looking at a set of empirical regularities often observed in related real-world social networks.
In Section~\ref{sec:parameters} we address the impact of single parameters and provide evidence in support to the main design choices -- including using all the available data.
In Section~\ref{sec:USG} we instead focus on a set of selected configurations and we provide insights into the structural properties of the resulting social graph.
Finally, a summary of the results, a few guidelines for the users of our simulator and suitable directions for future research activities are discussed in Section~\ref{sec:discussion}.
For an overview of the model, that includes the analysis of an epidemic use case, we refer the reader to~\cite{guarino2021model}.



\subsection{Related Work}
\label{sec:related}

\subsubsection{Synthetic Population}

The approaches used in the literature for the definition of a synthetic population can be broadly classified in two major groups: \emph{Synthetic Reconstruction} (SR) and \emph{Combinatorial Optimization} (CO).
According to the \emph{SR}~\cite{SRBeckman} approach, the population is generated by giving to each agent some relevant socio-demographic variables drawn from suitable joint-distributions deduced by putting together aggregate data covering the whole population with  disaggregated data from a sample (usually gathered from surveys).
In the \emph{CO}~\cite{COVoas} approach, instead, the area of interest is divided into mutually exclusive zones for which a set of marginal distributions is available, then a sample of real individuals from the target population is directly used to generate the whole population through replication/resampling methods.
A comparison between SR and CO can be found in~\cite{SRCOanalysis}.

A peculiar shortcoming of the SR approach in its basic implementation~\cite{SRBeckman} is that it is not possible to satisfy joint distributions of attributes at either household level or individual level simultaneously.
A way to overcome this problem is the reconstruction algorithm proposed by Guo and Bhat~\cite{Guo2007}.
Ye \emph{et al.}~\cite{Ye2009} proposed to extend the usual Iterative Proportional Fitting method (usually employed in the SR approach) with the Iterative Proportional Updating (IPU) method where the algorithm adjusts and reallocates weights among households until both household and individual-level attributes match. 
Unfortunately both CO and SR approaches require the collection of several kind of data which can be difficult to obtain.
To avoid this problem, a few sample-free models were proposed.
The main objective of these methods is to achieve a synthesis of the population starting from the most disaggregated level that is actually available.
Barthelemy and Toint~\cite{Barthelemy2013} extended the SR approach by generating a synthetic population based on households' and individuals' attributes for 589 municipalities of Belgium.
For a comparison between sample-free and sample-based methods, see Lenormand and Deffuant~\cite{Lenormand_2013}.

Barrett \emph{et al.}~\cite{barrett2009generation} proposed a ``first principle'' method for synthesizing populations on both urban and national scale. 
Their method is capable of reconstructing the contact network of the United State by simulating individuals in the population, their household structure, demographics attributes and a 24-hour activity sequence. 
To do so, they relied on a large data set, including census data, activity location, traffic data and roads network as well as several thousand responses to an activity, time-use survey, etc.
The definition of our synthetic population, described in Section~\ref{sec:population} and analyzed in Appendix~\ref{app:population}, follows a partially similar approach to~\cite{barrett2009generation}, but it requires significantly less data.

\subsubsection{Social and Contact Network Models}

A number of random graph models, proposed across decades of research, have been widely used in computational social sciences~\cite{amblard2015models}.
The lack of available data about real-world social ties fostered the choice of rather simple models (\emph{e.g.}, Erdös-Rényi~\cite{ERgraph}, Barabasi-Albert~\cite{barabasi2000graph} and Watts-Strogatz~\cite{smallworld}) which rely on just a few, easy to explain, assumptions. 
While these models capture in a simple and elegant way some essential features of different kinds of complex networks, it is well know that, in practice, they have significant limitations.
Models designed to mimic the scale-free degree distribution emerging in many real networks, for instance, may fail to yield the expected clustering structure~\cite{cointet2007realistic}. 
Exponential random graphs have been shown to overcome some of these limits \cite{robins2007recent, daraganova2012networks}.

Generally speaking, a possible adjustment to those models consists in introducing a \emph{homophily} principle~\cite{holzhauer2013considering}, according to the widely acknowledged insight that individuals tend to socialize with their peers \cite{mcpherson2001birds, festinger1950social}.
Among other aspects, such as education or economy, age emerged as a critical element in the formation of social ties~\cite{marsden1988homogeneity, palla2007quantifying, thelwall2009homophily, HUANG2013969}, possibly thanks to the availability of age-related data at different spatial scales~\cite{worldpop}. 
As far as we know, however, there are no quantitative studies that report the relative frequency of social links (\emph{i.e.}, relationships) by age, and previous network models that incorporate real data on social mixing by age are designed to generate synthetic physical interactions.
Another widely studied type of homophily is spatial proximity, which gives rise to the so-called
\emph{spatial networks}.
Most authors considered variations of well known random network models obtained by embedding the vertices in a metric space.
The imposed spatial constraints influence the topological properties of the network~\cite{barthelemy2011spatial, wong2006spatial} and the imposed penalty on ``long'' edges causes the spatial distribution of the vertices to impact on clusters, path lengths, degree distribution, and more~\cite{bullock2010spatial, alizadeh2017generating}.
Again, there is no much research about data-driven spatial social network models, in which the location of the individuals can be retrieved from real data.

A somehow symmetrical problem consists in inferring social or contact networks from real data.
For \emph{virtual} populations, \emph{e.g.}, online social networks, relationships and actually occurred interactions may be directly retrieved~\cite{scellato2011socio, liben2005geographic, bailey2020social, Guarino2020}, at least to some extent.
The dependability of this information is however debatable, because the virtual population may convey significantly different information with respect to self-reported friendships~\cite{Mastrandrea-2015}.
When the focus is, instead, on the physical interactions among the individuals of a population, a synthetic version of the population may be created on the basis of census and/or survey data~\cite{mistry2020inferring}.
The final goal is the definition of a set of \emph{human mixing patterns} (\emph{i.e.}, the frequency of contacts among people of different ages and/or in different places) which allow to reproduce the dynamics of the network, for instance to model the diffusion of a disease in real populations \cite{read2014china, AJELLI20171, klepac2020contacts, mistry2020inferring}. 
This approach usually requires to focus on a set of primary social settings (\emph{e.g.}, households, schools and workplaces) and to collect sample data (\emph{e.g.}, surveys, questionnaires, diaries, mobility, etc.)~\cite{Mossong-2008,Prem-2017,AJELLI20171}, possibly integrated with mobile/traffic/wearable sensors data~\cite{eagle2009inferring, aleta2020modelling, Mastrandrea-2015, cattuto2010dynamics} or online tools \cite{klepac2020contacts}.
These data may be used to directly extract setting-specific \emph{contact matrices}~\cite{Mossong-2008,barrett2009generation, Prem-2017,AJELLI20171}, or to recreate \emph{realistic} instances of such social settings and synthetic agendas used to feed an agent-based simulator by which agent-to-agent interactions are reproduced~\cite{eubank2004modelling, delvalle2007episims, barrett2009generation, aleta2020modelling, iozzi2010little}.
It is worth noting that these studies target physical contacts rather than relationships.
While they provide valuable information about social mixing patterns, they do not aim to define a network of strong social ties.
Moreover, in a recent call to action several experts expressed criticisms about most already existing tools and raised the attention towards the need for accurate yet flexible and replicable approaches~\cite{squazzoni2020}.

\subsubsection{Properties of Empirical Spatial Networks}

Spatial social networks have been found to form a large connected component on both mobile phone networks \cite{onnela2007analysis, herrera2015anatomy, liben2005geographic} and online social networks \cite{scellato2011socio}.
The average degree is typically small, albeit dependent on the type of network considered: it ranges from $\approx5$ in mobile phone and online social networks~\cite{onnela2007analysis, scellato2011socio, liben2005geographic}, to $\approx10$ (when only strong ties are considered) or $\approx20$ when the network is inferred from social surveys \cite{kowald2013distance, van2009size, axhausen2007size}.
Despite the limited average degree, most spatial social networks show the typical small-world effect, with a short average path length close to the well-known ``6 degrees of separation'' rule~\cite{scellato2011socio, herrera2015anatomy}.
Another typical feature of social networks is the high clustering coefficient (\emph{i.e.}, number of triangles) compared to random ER-like networks.
Indeed, a clustering value of $\approx0.2$ has been found in mobile phone~\cite{scellato2011socio, herrera2015anatomy} online social~\cite{liben2005geographic} and survey-based~\cite{illenberger2013role} networks.
A clear peculiarity of geographical social networks is instead the absence of very large hubs, which are widely observed across multiple fields, including web graphs~\cite{bernaschi2017exploring}, the Internet~\cite{newman2010networks} and online social networks~\cite{Guarino2020}. 
The maximum degree in geographical social networks is apparently limited to a few hundreds, as emerged from both mobile phone networks \cite{onnela2007analysis, lambiotte2008geographical, herrera2015anatomy} and the LiveJournal Blog with geo-localization \cite{liben2005geographic}, and in line with sociological studies~\cite{granovetter1983strength, duck1991friends, krackhardt2003strength}.
In \cite{onnela2007analysis, lambiotte2008geographical} the authors found a power-law distribution with $\approx8$ and $\approx5$ exponent, respectively, whereas \cite{herrera2015anatomy} and \cite{liben2005geographic} only highlighted the occurrence of a long-tailed distribution.
In \cite{illenberger2013role} the degree distribution is well approximated by a lognormal with $\sigma=0.9$ and $\mu=2.6$

A widely studied problem is characterizing the role of spatial proximity and population density in the edge creation process.
The dependence of friendship on distance has been found empirically as an inverse power-law whose exponent $\beta$ usually lies in the range $[0.5, 2]$~\cite{kowald2013distance, axhausen2007size, lambiotte2008geographical, scellato2011socio, herrera2015anatomy, onnela2011geographic, liben2005geographic, cho2011friendship, illenberger2013role, walsh2011spatial, buchel2020cities}.
The measured edge length distribution is however strongly dependent on the data source and on the size of the considered territory, \emph{e.g.}, it may be almost flat up to 10 km~\cite{scellato2011socio} or present a spike for lengths $<5$ km~\cite{buchel2020cities}.
Mobile phone networks, in particular, may cover hundreds or even thousands of kilometers and generally provide geographic locations, and thus distances, with a very variable granularity -- from 1 km in densely populated urban areas to 10 km in rural regions~\cite{herrera2015anatomy}.
Using the network reconstructed from a survey, in \cite{illenberger2013role} the edge length distribution shows two distinct regimes: $\beta = 0.6$ for short range contacts ($<20km$) and $\beta = 1.8$ for long range contacts.
In general, a value $\beta<1$ is not rare~\cite{liben2005geographic, cho2011friendship} and possibly associated with urban networks, as is the case for the value $\beta=0.44$ obtained for the city of Dublin in~\cite{walsh2011spatial}.
Furthermore, many studies agree on the existence of a significant correlation between geographical proximity and community structure~\cite{herrera2015anatomy, walsh2011spatial, buchel2020cities, bailey2020social, palla2007quantifying}.
A recent analysis of the urban area of New York based on the geo-localization of Facebook accounts shows that most of the connections occur between adjacent zip codes and that the underlying transportation infrastructure has a visible impact~\cite{bailey2020social}.
Studying a phone call network, however, in \cite{onnela2011geographic} the authors found a threshold effect, with a regular increase in the geographical extension of the communities only observable for small clusters with up to 30 members.
This was confirmed by a later work, that detected highly clustered connected components spanning across very large areas of the city~\cite{herrera2015anatomy}.
Population density is also often considered a factor in the formation of social bonds.
In his seminal work~\cite{granovetter1983strength}, Granovetter argued that personal networks in the cities -- with respect to hinterland and peripheries -- are characterized by low clustering values and by a higher number of weak ties.
In \cite{buchel2020cities} an extensive empirical phone call network has been analyzed finding that, while the density of individuals has a negligible impact on the size of each person's network, it induces a higher number of close-range contacts.

Despite the previous findings, there is wide evidence that the geographical factors alone cannot explain the structure of spatial social networks~\cite{scellato2011socio}.
Other forms of social proximity were estimated to be responsible of $\approx 1/3$ of friendships in the LiveJournal Blog network~\cite{liben2005geographic}.
Studying a phone call network, the authors of~\cite{herrera2015anatomy} found that the number of intra-tower connections observed was higher than a pure geographical model would generate, and that the small-world nature of the network was mostly related to other forms of social cohesion, rather than geographical proximity.



\section{Materials and methods}
\label{sec:methods}

Our urban social network is obtained as the combination of a household network with a friendship network.
The construction of these networks requires to first map census data to reconstruct the population and then model interpersonal connections.
Formally, the network is represented by an unweighted undirected graph $G=(V,E)$, where $V$ is the vertex set of size $N=|V|$ and $E$ is the edge set.
In particular, we have $E=E_H\sqcup E_F$, where $E_H$ is the set of \emph{household edges}, $E_F$ is the set of \emph{friendship edges} and $\sqcup$ denotes the disjoint union.
In the following, we explain how $V$, $E_H$ and $E_F$ are defined in our model.
We will often use the expression \emph{household graph} to denote the subgraph $G_H=(V,E_H)$ and the expression \emph{friendship graph} to denote the subgraph $G_F=(V,E_F)$.
The notation and the parameters used throughout this paper are summarized in Table~\ref{tab:parameters}.

\begin{table}[htbp]
    \centering
    \caption{Overview of the main parameters used in this paper.}
    \label{tab:parameters}
    \small
    \begin{tabular}{p{0.13\textwidth}  p{0.34\textwidth} p{0.45\textwidth}}
         \toprule
         \textbf{Notation} & \textbf{Description} & \textbf{Definition}\\
         \midrule
         $G_H=(V,E_H)$ & A layer of edges that represent ties between members of the same household & Data-driven based on ISTAT data (see Section~\ref{sec:population} \\
         \midrule
         $G_F=(V,E_F)$ & A layer of edges that represent ties between people of different households & Generated with a probabilistic model (see Section~\ref{sec:friendship}) \\
         \midrule
         $G=(V,E)$ & The urban social graph whose edges represent generic ``strong'' social ties  & Obtained flattening the $G_H$ and $G_F$ layers onto a single-layer graph \\
         \toprule
         $N=|V|$      & The number of nodes in the graph, equal to the population size & Data-driven, based on the WorldPop database; see Table~\ref{tab:cities} for the population of all cities \\
         \midrule
         $\nu$      & The average degree of the $G_H$ layer & Data-driven, $\approx2$ for all cities (see Table~\ref{tab:cities}) \\
         \midrule
         $\mu$      & The average degree of the $G_F$ layer & An input parameter, set to $1$, $5$ or $10$ in the experiments (see Sections~\ref{sec:parameters} and~\ref{sec:USG})\\
         \midrule
         $K=\nu+\mu$        & The average degree of the graph $G$ & See above \\
         \midrule
         $g_u$      & The age label of node $u$, taking value in $\{0,\ldots,n-1\}$, which determines a partition of the population based on age & Drawn from a data-driven age distribution; the list of age-breaks is parametric; in the experiments we use ISTAT data with age-breaks $(0,18,35,65)$ \\
         \midrule
         $V_i$ (and $|V_i|$)    & The set of nodes having age label $i$ (and the number of such nodes) & Deduced from the age labels assigned to the nodes of the graph\\
         \midrule
         $m_{i,j}$   & The number of pairs of vertices $(u,v)$, with $u\neq v$, such that $u\in V_i$ and $v\in V_j$ & Computed based on $|V_i|$ and $|V_j|$ (see (\ref{eq:m_ij}))\\
         \midrule
         $t_u$      & The tile label of node $u$, taking value in $\{1,\ldots,T\}$, which determines a partition of the population based on the place of residence & Drawn from a data-driven population density; the tessellation is parametric, the tile side is $l$, the grid is composed of $T=T_{\mathrm{lat}}\cdot T_{\mathrm{lon}}$ tiles; in the experiments we used WorldPop data, $l=1$ and grids as reported in Table~\ref{tab:cities}
         \\
         \midrule
         $d(u,v)$   & The approximated euclidean distance between $u$ and $v$ & Computed as $\max\left\{\frac{l}{2},d^*(t_u,t_v)\right\}$, where $d^*(t_u,t_v)$ is the euclidean distance between the center of the two tiles\\
         \midrule
         $D(u,v)$   & A non-increasing function of the distance $d(u,v)$ & An input parameter; in the experiments, we use $D(u,v)=d(u,v)^{-\beta}$ with $\beta\in\{0.5,2\}$\\
         \midrule
         $f_u$      & The real-valued fitness score of node $u$, that quantifies its sociability & Drawn from a probability density function $\phi$ specified as an input parameter; in the experiments, we consider a shifted Lognormal $1+\LN(\ln(2),0.25)$ or a constant distribution\\
         \midrule
         $S=\{s_{i,j}\}$  & The real-valued social mixing matrix whose element $s_{i,j}$ measures the frequency of social ties between age groups $i$ and $j$ & Data driven; in the experiments, we compute $S$ based on contact data from~\cite{Mossong-2008} extracted using~\cite{willem2020socrates}; we also consider a constant $S$ \\
         \bottomrule
    \end{tabular}
\end{table}

\subsection{Territory, Population and Vertex Set}
\label{sec:population}

We define the territory of interest in terms of an \emph{origin} and of a regular lattice of $T=T_{\mathrm{lat}}\cdot T_{\mathrm{lon}}$ square tiles, \emph{i.e.}, by establishing the South-West corner of our bounding box, the side $l$ of each tile and the number of tiles $T_{\mathrm{lat}}$ and $T_{\mathrm{lon}}$ along the latitudinal and longitudinal axes, respectively.
For the scope of this paper, $l=1$ km for all cities.
We then extract a geo-referenced population for the whole area of interest from the WorldPop Project\footnote{\url{https://www.worldpop.org/}}, which provides data of the world population for $100m\times100m$ square cells.
We resort to the \verb|overpass| API of the well known OpenStreetMap database\footnote{\url{https://www.openstreetmap.org/}.} to find the minimal grid that contains the city's boundary; we then select only the tiles of the grid whose center lies inside it (see Appendix~\ref{app:population}).
By mapping the WorldPop data to our tiles, we obtain a very precise estimate of the real population living in each tile.

Albeit WorldPop -- as well as other data sources -- directly provides population densities for different age groups, we believe that all demographic parameters should be acquired from the same data source in order to guarantee the intrinsic consistency of the model.
Any desired age-stratification can thus be easily specified in the simulator's configuration file in the form of a set of age groups and their respective frequencies.
For the scope of this paper, we decided to rely on the Italian Institute of Statistics (ISTAT)\footnote{ISTAT data used in this paper are available at \url{https://www.demo.istat.it/pop2020}.}, which provides age distribution at the provincial level.
The United Nations Statistics Division (UNSD) makes available similar data for many other countries\footnote{\url{https://unstats.un.org/unsd/demographic-social/census/censusdates/.}}.

Each vertex $u\in V$ of our graph, which represents an individual of the population, is thus characterized by three attributes:
\begin{description}[style=sameline, leftmargin=\widthof{Tie}]

\item[Tile] The tile label $t_u\in\{0,\ldots,T-1\}$ is set equal to the unique index of the tile where $u$ resides.

\item[Age] The age label $g_u\in\{0,\ldots,n-1\}$ for vertex $u$ is drawn independently at random from the given age-stratification. 
These labels determine a partition of the vertex set $V$ into $n$ disjoint subsets $V_0,\ldots,V_{n-1}$.

\item[Fitness] Inspired by previous work, that modelled degree heterogeneity by means of a vertex-intrinsic fitness~\cite{caldarelli2002scale}, we extract a sociability fitness attribute $f_u>0$ for each $u$.
The probability of a friendship edge between $u$ and $v$ is then set proportional to $f_u$ and $f_v$ (see Section~\ref{sec:friendship}).

\end{description}

Approximating the position of each vertex with its tile is instrumental in simplifying the computation of the household structure (see Section~\ref{sec:households}) and of pairwise distances.
To this end, we use the approximation $d(u,v)=\max\left\{\frac{l}{2},d^*(t_u,t_v)\right\}$, where $t_u$ is the tile of $u$ and $d^*(t_u,t_v)$ denotes the distance between the center of $t_u$ and the center of $t_v$.

Our model does not put restrictions upon the choice of $f_u$; yet, the distribution of $f_u$ shall be chosen considering its impact on the degree distribution of the friendship graph.
Possible choices include a Lognormal, a Pareto, a uniform and a constant distribution -- all of which are already supported by our simulator.

\subsection{Household edges}
\label{sec:households}

We group individuals into households according to a heuristics that uses the distribution of ``household roles'' per age and the distribution of the number of children per family, under the assumption that: $(i)$ all members of a household live in the same tile; $(ii)$ children are younger than their parents; $(iii)$ partners have, on average, a similar age.
The algorithm is based on the concept of \emph{household role}, represented as a pair of the form (household-type, role), whose possible values, for Italy, are reported in Table~\ref{tab:households}.
For instance, $r_u=$(single-parent, parent) means that $u$ is a parent in a household of type single-parent, where $r_u[0]$=single-parent is the household-type and $r_u[1]$=parent is the role.
We make use of two conditional distributions: $\Pr[r\mid g]$ is the probability that an individual has role $r$ given that she/he belongs to age group $g$; $\Pr[k\mid h]$ is the probability that a household of type $h$ has $k$ members.
This information is made available, as aggregate national data, by the ISTAT for Italy and, \emph{e.g.}, by the UNSD for many other countries.

\begin{table}[htbp]
\centering
\caption{Household types and roles deducible from ISTAT data.}
\label{tab:households}
\begin{tabular}{l p{0.1\linewidth} p{0.18\linewidth} p{0.1\linewidth} p{0.18\linewidth} p{0.1\linewidth}}
\toprule
\multirow{1}{*}{\textbf{Household type}}& singles & single-parent & couples & two-parents & various\\
\midrule
\multirow{2}{*}{\textbf{Role}} & single & parent & peer & parent & various \\
& & child & & child & \\
\bottomrule
\end{tabular}
\end{table}%

At a high level, the heuristics works as follows:
\begin{itemize}
    \item Based on the distribution $\Pr[r\mid g_u]$, extract a role $r$ for each vertex $u$.
    \item For all $u$ such that $r_u[0]\in\{\text{single-parent},\text{two-parents}\}$:
    \begin{itemize}
        \item if $r_u[0]=$two-parents, select a random partner $v$ for $u$ such that $t_v=t_u$, $g_v\in[g_u-1,g_u+1]$ and $r_v[0]=r_u[0]$;
        \item based on the distribution $\Pr[k\mid r_u[0]]$, extract the total number of members $k_u$ for the household of $u$ (and, in case, of $v$) and compute their total number of children $c_u$.
    \end{itemize}
    \item For $i=1,\ldots,\max_{u}c_u$:
    \begin{itemize}
        \item for all $u$ such that $c_u\geq i$, select a random $w$ such that $t_w=t_u$, $g_w<g_u$, $r_w[0]=r_u[0]$ and $r_w[1]$=child, and assign $w$ to the household of $u$.
    \end{itemize}
    \item For all $u$ such that $r_u[0]=$couples, select a random partner $v$ for $u$ such that $t_v=t_u$, $g_v\in[g_u-1,g_u+1]$ and $r_v[0]=r_u[0]$;
    \item based on the distribution $\Pr[k\mid \text{various}]$, randomly compose the households of type various.
\end{itemize}
In our simulations, the number of individuals not assigned to any household by the heuristics is negligible, and the empirical distributions of household types and members per type almost perfectly match the expected ones (see Appendix~\ref{app:population} for details).
The household edges $E_H$ are finally obtained as the union of all the \emph{cliques} that connect all members of the same household.
The average degree of the household edges, denoted $\nu$, is thus entirely data-driven.

\subsection{Friendship Edges}
\label{sec:friendship}

As said, the age labels define a partition of the vertex set $V$ into $n$ disjoint subsets $V_1,\ldots,V_n$.
For each possible pair of groups $(V_i,V_j)$, the number of pairs of vertices $(u,v)$ such that $u\in V_i$ and $v\in V_j$ is
\begin{equation}\label{eq:m_ij} 
m_{i,j} = \begin{cases} |V_i|\cdot |V_j|  &\text{if } i\neq j \\ \frac{|V_i|\cdot (|V_i|-1)}{2} &\text{if } i=j\end{cases} \quad \text{with } \sum_{i\leq j} m_{i,j} = \frac{N\cdot (N-1)}{2}
\end{equation}
Each pair $(V_i,V_j)$ is also characterized by a mutual preference coefficient $s_{i,j}\in [0,1]$.
For the scope of the present paper, we estimate the coefficients $s_{i,j}$ relying on the recent SOCRATES project~\cite{willem2020socrates} that makes it possible to easily extract aggregated social contact matrices from public available and well-established datasets.
In Appendix~\ref{app:social_contact_data}, we describe how we constructed the matrix $S=\{s_{i,j}\}$ and we show that any constant $S$ corresponds to the condition of \emph{age-homogeneous mixing} -- \emph{i.e.}, to edge occurrence being independent of the age.
To each pair $(u,v)\in V\times V$, we finally associate a distance $d(u,v)$ computed based on the tile labels $t_u$ and $t_v$ as explained in Section~\ref{sec:population}.

The probability $\Pr[(u,v)\in E_F]$ that the edge $(u,v)$ exists in the friendship graph $G_F$, simply denoted $\Pr[u,v]$, is defined as follows:
\begin{equation}\label{eq:edge_prob}
    \Pr[u, v] = \frac{\mu\cdot N}{2} \cdot \frac{m_{g_u,g_v} \cdot s_{g_u,g_v}}{\sum_{i\leq j} \left(m_{i,j} \cdot s_{i,j}\right)}\cdot \frac{D(u, v) \cdot f_u \cdot f_v}{\sum_{u'\in V_{g_u}, v'\in V_{g_v}} \left(D(u', v') \cdot f_{u'} \cdot f_{v'}\right)} 
\end{equation}
where the parameter $\mu$ is the average number of friends, \emph{i.e.}, the average degree of the graph $G_F$.

A few features of our friendship graph deserve additional remarks:

\paragraph{Mixing and attraction}
Let us define the \emph{mixing} between age groups $i$ and $j$ as $M(i,j)=m_{i,j}\cdot s_{i,j}$ and the \emph{total mixing} of the given population as $M=\sum_{i\leq j} M(i,j)$.
Similarly, let us define the \emph{attraction} between $u$ and $v$ as $a(u,v) = D(u, v) \cdot f_u \cdot f_v$ and the \emph{total attraction} between groups $i$ and $j$ as
$A(i,j) = \sum_{u\in V_i, v\in V_j}a(u,v)$.
It is trivial to rewrite (\ref{eq:edge_prob}) as 
\begin{equation}
\Pr[u, v] = \frac{\mu\cdot N}{2} \cdot \frac{M(g_u,g_v)}{M} \cdot \frac{a(u,v)}{A(g_u,g_v)} \approx \frac{\mu\cdot N}{2\cdot M\cdot \mean{D}\cdot \mean{f}^2} \cdot s_{g_u,g_v}\cdot a(u,v) 
\label{eq:edge_prob_short}
\end{equation}
where $\mean{D}$ and $\mean{f}$ denote the mean value of the fitness score $f_u$ and of the distance function $D(u,v)$, respectively.
The approximation in (\ref{eq:edge_prob_short}) is obtained under the assumption that $A(i,j)\approx m_{i,j}\cdot \mean{a(u,v)} = m_{i,j} \cdot \mean{D}\cdot \mean{f}^2$ for all $i,j$, which is accurate if all age groups are large enough.

\paragraph{Edge density and social mixing}
To each pair of age groups $(i,j)$ with $i\leq j$ we can associate a subgraph as follows: if $i=j$, $G_F(i,i)=(V_i,E_{F,i,i})$ is the subgraph of $G_F$ induced by the vertices in $V_i$; if $i\neq j$, $G_F(i,j)=(V_i\sqcup V_j,E_{F,i,j})$ is the \emph{bipartite} subgraph of $G_F$ induced by the vertices in the disjoint union $V_i\sqcup V_j$ and where only edges $(u,v)$ such that $u\in V_i$ and $v\in V_j$ are admissible.
It is easy to verify that $G_F$ is the disjoint union of these subgraphs, that is, $E_F=\sqcup_{i\leq j} E_{F,i,j}$.
The coefficient $m_{i,j}$ equals the maximum possible number of edges in each subgraph $G_F(i,j)$, whereas the mixing $M(i,j)$ is the expected number of such edges in a network where the edge probability is uniquely determined by the cohesion $s_{i,j}$. 
The expected value of $E_{F,i,j}$ in our networks is instead 
\begin{equation}
\E[|E_{F,i,j}|] = \sum_{u\in V_i, v\in V_j} \Pr[u,v] =  \frac{\mu\cdot N}{2} \cdot \frac{M(i,j)}{M} \cdot \frac{\sum_{u\in V_i, v\in V_j} a(u, v)}{A(i,j)} = \frac{\mu\cdot N}{2} \cdot \frac{M(i,j)}{M}
\label{eq:condition_groups}
\end{equation}
for all $i\leq j$. 
The expected number of edges in the entire friendship graph is thus
\begin{equation}
\E[|E_F|] = \sum_{i\leq j} \E[|E_{F,i,j}|] = \frac{\mu\cdot N}{2} \cdot \frac{\sum_{i\leq j} M(i,j)}{M} = \frac{\mu\cdot N}{2}
\label{eq:average_degree_total}
\end{equation}
This shows that, thanks to the normalization by $A(g_u,g_v)$, the age-based social mixing is respected, up to a scaling factor.
In particular, the dependence of $\Pr[u,v]$ upon $d(u,v)$ does not affect the (expected) density of the friendship graph, which is entirely controlled by $\mu$.
By tuning $D(u,v)$ we may adjust only the physical length distribution of the edges -- arguably, in order to penalize \emph{long} edges.

\paragraph{Expected degree}
Thanks to the relation $\sum_u \deg_F(u) = 2\cdot|E_F|$, it is easy to verify that the expected friendship degree of a random $u\in V$ is exactly equal to $\mu$.
To guarantee that $\Pr[u,v] \leq 1$ for all $u,v$, the choice of $\mu$ is subject to the condition
\begin{equation}
\mu \leq
\frac{2\cdot M}{N} \cdot \min_{u,v} \left\{ \frac{A(g_u,g_v)}{M(g_u,g_v) \cdot a(u, v)}\right\} 
\approx \frac{2\cdot M \cdot \mean{D}\cdot \mean{f}^2}{N \cdot \max_{u,v}\left\{s_{g_u,g_v} \cdot a(u, v)\right\}}
\label{eq:condition_mu}
\end{equation}
In words, since by increasing $\mu$ we evenly increase the probability of all edges, the choice of $\mu$ is possibly constrained by the existence of pairs $(u,v)$ for which the combined effect of the mixing $M(i,j)$ and the attraction $a(u,v)$ makes the existence of the edge $(u,v)$ significantly more likely than the average.
The expected degree of a vertex $u$ having fitness $f_u$, age $g_u$ and tile $t_u$ can instead be computed as
\begin{align}
\mu_u &= \E[\deg_F(u)\mid f_u,g_u,t_u] = \sum_v \Pr[u,v] = \sum_t \left[ \sum_j \left( \sum_{v\in t, v\in V_j} \Pr[u,v] \right) \right] \nonumber \\
&= \frac{\mu\cdot N \cdot f_u}{2\cdot M} \cdot \sum_t \left[ D(t_u,t) \cdot \sum_j \left( \frac{M(g_u,j)}{A(g_u,j)} \cdot  \sum_{v\in t, v\in V_j} f_v \right) \right]
\label{eq:average_degree_u}
\end{align}
$\mu_u$ is thus proportional to $u$'s social fitness $f_u$ and to $\mu$, as expected.
Equation~(\ref{eq:average_degree_u}) also shows that $\mu_u$ depends upon the tile $t_u$ through a factor given by the average distance of $t_u$ from all other tiles (included $t_u$ itself),
weighted by the (group-weighted) \emph{total sociability} of each tile.

\paragraph{Scale invariance and parameter range}
It is easy to verify that $\Pr[u,v]$ is invariant under multiplication of $s_{i,j}$, $D(u,v)$, $f_u$ and/or $f_v$ by any positive constant.
This means that any two approaches to derive the coefficients $s_{i,j}$ from real data that only differ by a constant factor are equivalent for our model.
Now, for (\ref{eq:edge_prob}) to be meaningful, we need $D(u,v)$ to be upper-bounded by a suitable constant $D_{\max}$.
In our model, this is achieved by imposing that all vertices belonging to the same tile are at distance $d_{\min}=\frac{l}{2}>0$ and that (with a slight abuse of notation) $D_{\max}=D(d_{\min})<+\infty$.
On the other hand, to prevent the occurrence of vertices with expected degree $\approx0$ -- at least, in a sufficiently large network -- $f_u$ must be bounded away from 0.
Summing up, without loss of generality, we may assume that $D(u,v)\in(0,1]$ for all $u,v\in V$, that $f_u$ is drawn from a probability distribution with support $[1,+\infty)$, and that $\sum_{i\leq j}s_{i,j}=1$.
The latter condition suggests to interpret the coefficients $s_{i,j}$ as the probability that a randomly chosen edge of the graph connects groups $i$ and $j$.

\subsubsection{Selected Configurations}
\label{sec:chosen_model}

Although our network model does not impose restrictions upon the choice of $f_u$ and of the function $D$, we believe that the following configuration is of special interest:
\begin{itemize}
    \item $f_u \sim 1+\LN\left(\ln(2),0.25\right)$, where $\LN$ denotes a Lognormal distribution\footnote{Throughout this paper, we use the parameterization $\LN\left(\lambda,\sigma^2\right)$ where $\lambda$ and $\sigma^2$ are the mean and variance of the associated Normal distribution.};
    \item $D(u,v) = d(u,v)^{-\beta}$
     with $\beta\in \left\{0.5,2\right\}$.
\end{itemize}

The choice for the distribution of $f_u$ is motivated by the observation that the degree distribution of our friendship graph will, at least partially, follow the fitness distribution.
Lognormally distributed data occur across different domains~\cite{mitzenmacher2004brief} and recent work suggests that the degree distribution of real-world social networks makes no exception~\cite{liben2005geographic, illenberger2013role}.
In particular, a Lognormal degree distribution would fit the intuition that only a few people have very few social links.
The specific parameters have been chosen to enforce limited skewness and variance, so as to guarantee that most vertices will have a degree close to the average and that the hubs will be limited in both number and size.
It goes without saying that other choices may be preferred, some of which (\emph{e.g.}, a Pareto, a uniform and a constant distribution) are already supported by our simulator.

As to $D$, as extensively discussed in Section~\ref{sec:related}, there is a vast body of research supporting the choice of a power-law decay for the dependence of social ties upon the geographical distance between two individuals.
There is no equal agreement regarding the exponent $\beta$, with empirical findings suggesting a significant variability according to the geographical scale of the analysis and to the type of interaction object of the study.
We decided to compare $\beta=0.5$ and $\beta=2$, which seem to lie at the two ends of the spectrum of the values observed in the literature.
In both cases, as already discussed, since $d(u,v)$ is bounded, so is $D$. 

It is worth to highlight that, for the considered set of models, the upper bound (\ref{eq:condition_mu}) for the choice of $\mu$ is indeed rather loose in most practical cases, as we experimentally verified.


\section{Impact of System Parameters}
\label{sec:parameters}

To sort out the role of the different system parameters in shaping our urban social network we resort to an extensive experimental campaign focused on three Italian cities of different size: Florence (363060 residents), Viterbo (66598) and Sabaudia (21274).
For the sake of readability, the results for Viterbo and Sabaudia are reported in Appendix~\ref{app:viterbo} and Appendix~\ref{app:sabaudia}, respectively.

For each city we selected the administrative boundaries from OpenStreetMap and filtered the population inside the city shape as described in Section~\ref{sec:population}.
For Sabaudia and Viterbo the smallest boundary available is the municipality which encompasses a large rural area with small urban agglomerates around the city.
For Florence we were able to select the actual city boundaries which do not include neighboring areas.
For each city, we extracted the age labels based on ISTAT data aggregated at the provincial level.
We then built the households as described in section \ref{sec:population}.
The obtained territories are depicted in Figure~\ref{fig:city_shape} (and \ref{fig:city_shape_app} in Appendix \ref{app:population}).
An overview of the population and the territory is reported in Table~\ref{tab:cities} for all cities.
For further details see Appendix \ref{app:population}.

\begin{figure}[htbp]
\centering
    \begin{subfigure}[b]{.45\textwidth}
         \centering
         \includegraphics[height=\textwidth]{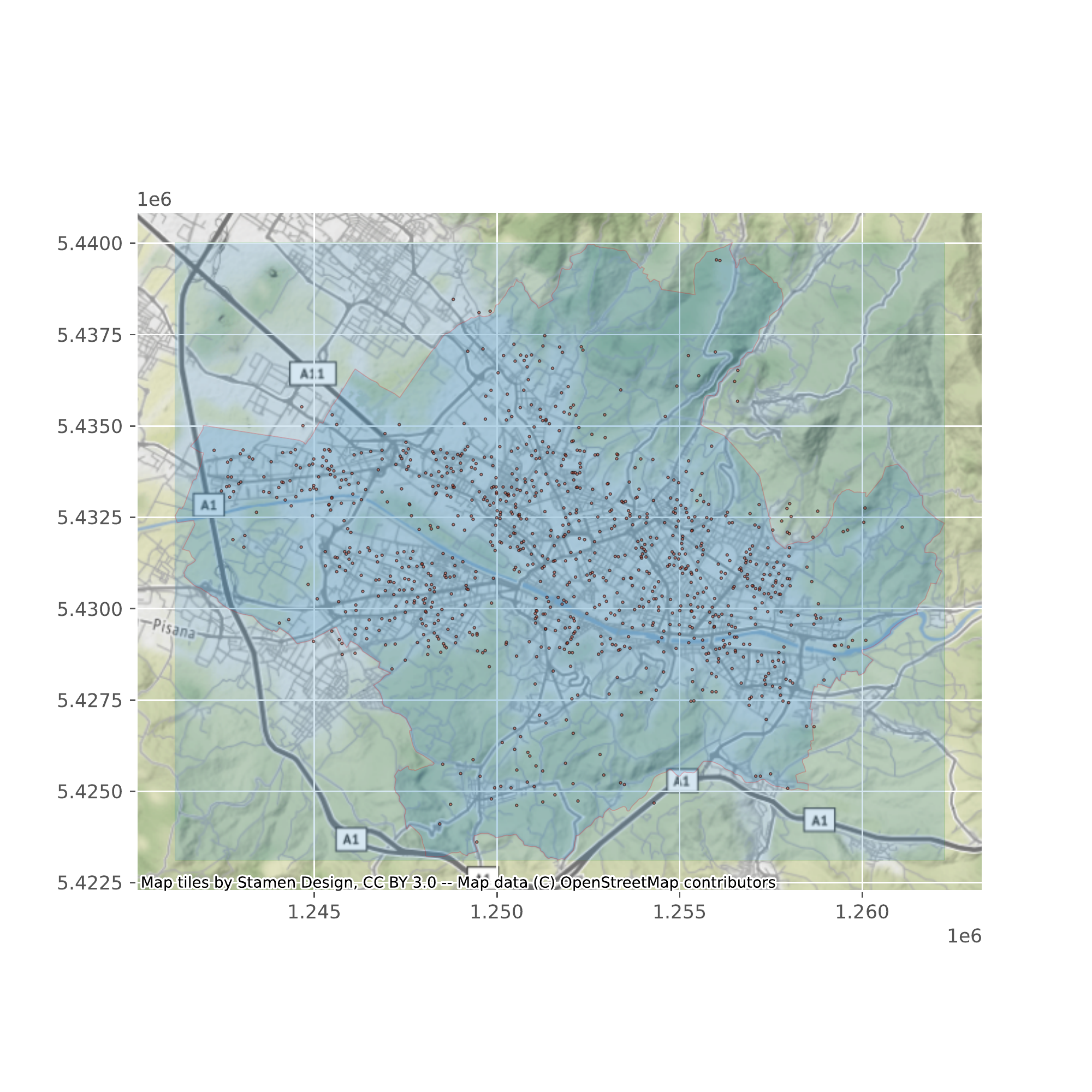}
         \caption{\textbf{Florence}}
         \label{fig:firenze_shape}
    \end{subfigure}
    \hfill
    \begin{subfigure}[b]{.45\textwidth}
         \centering
         \includegraphics[width=\textwidth]{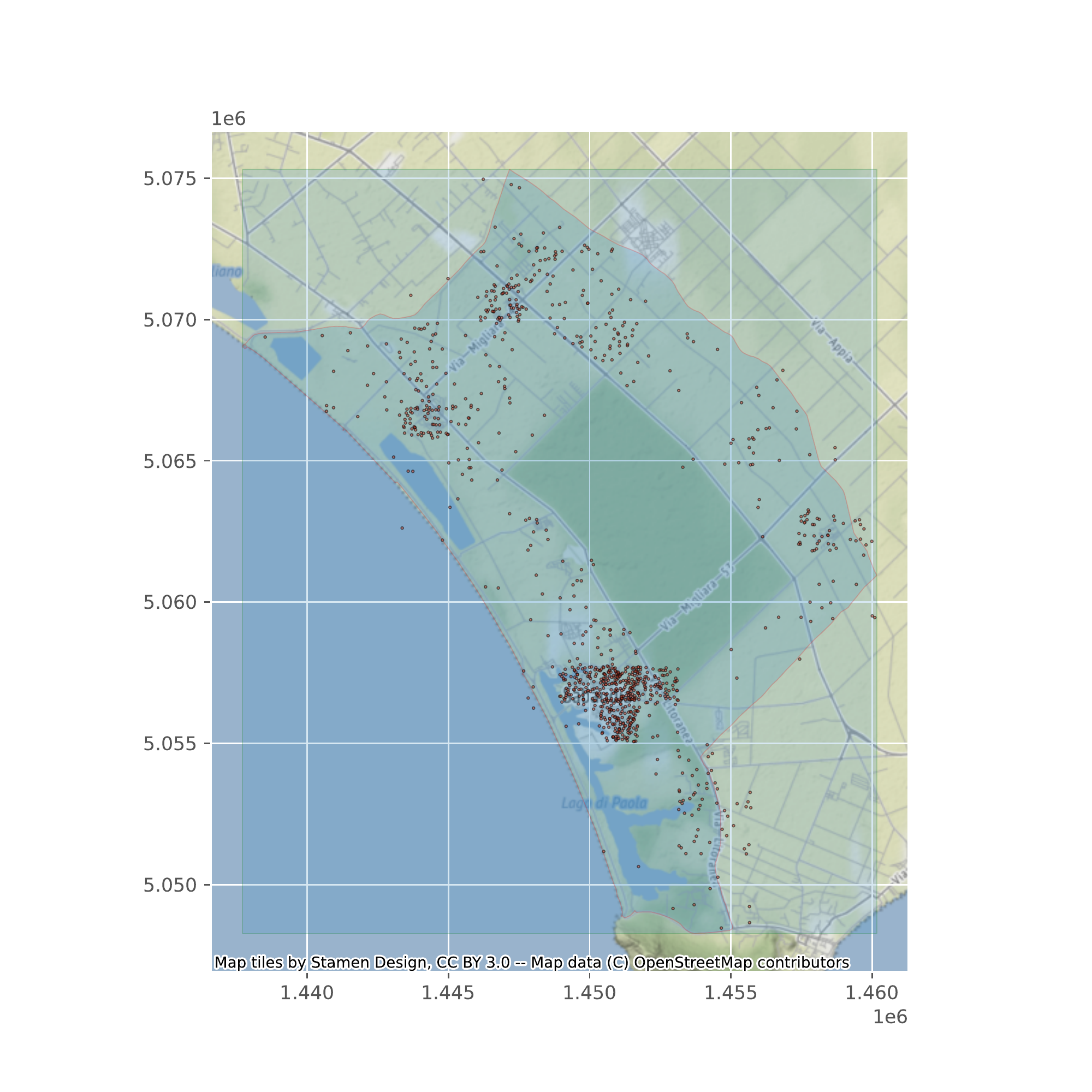}
         \caption{\textbf{Sabaudia}}
         \label{fig:sabaudia_shape}
    \end{subfigure}
    \caption{Random sample of 1000 people on the territory of Florence and Sabaudia (some points may fall outside of the territory if the center of the tile to which they are associated falls inside.}
    \label{fig:city_shape}
\end{figure}

\begin{table}[htbp]
    \centering
    \caption{Reconstructed parameters for the three cities. For the city of Florence we considered only the city shape whereas for Viterbo and Sabaudia all the municipalities, which include rural areas and small villages, have been used. $\nu$ is the average degree of the reconstructed household network $G_H$, $T_{\mathrm{lat}}$ and $T_{\mathrm{lon}}$ are the number of tiles along the two axes of the grid used to cover the city territory, $l$ is the side of each tile.}
    \label{tab:cities}
    \begin{tabular}{llrrrrrrrr}
         \toprule
         City & Boundary & $N$ & children\% & young\% & adults\% & elderly\% & $\nu$ & $(T_{\mathrm{lat}},T_{\mathrm{lon}})$ & $l$ (m) \\
         \midrule
         \textbf{Florence} & City  & 363060 & $15.1\%$ & $16.9\%$ & $43.1\%$ & $24.9\%$ & $2.08$ & $(15,12)$  & $1000$ \\ 
         \textbf{Viterbo}  & Municipality  & 66598 & $16.0\%$ & $17.8\%$ & $44.2\%$ & $22.0\%$  & $2.08$  & $(16,20)$  &  $1000$ \\
         \textbf{Sabaudia} & Municipality  & 21274 & $16.8\%$ & $18.1\%$ & $45.6\%$ & $19.5\%$  & $2.07$ & $ (26,32)$  &  $1000$ \\
         \bottomrule
    \end{tabular}
\end{table}

In this Section, we focus on a few basic properties of the network and on the design choices needed to achieve them.
As a side result, we show that any attempt to significantly simplify our network model downgrades the resulting social graph, especially in terms of heterogeneity and connectivity.
It is worth to underline that, to account for variance, hereafter any configuration is evaluated by considering the results of 10 independent runs.

\subsection{Safeguarding connectivity}
\label{sec:connectivity}

There are two main factors that impact on the overall connectivity of the graph: the average number of friends $\mu$ and the reconstructed households structure. 
In Section~\ref{sec:friendship} we found that our efforts to introduce heterogeneity in the model while preserving certain structural properties inferred from contact data yield an upper bound for $\mu$.
Albeit there is no analogous strict lower bound, $\mu$ must be chosen large enough to guarantee that the graph has the expected level of connectivity. 
To provide insights into the choice of $\mu$, we compare the graphs obtained with $\mu=1$ and $\mu=5$, measuring the percentage of nodes belonging to the giant component for different configurations.
We remind that the household layer $G_H$ is composed of a set of cliques and has average degree $\nu\approx2$ for all cities -- the average degree of the whole graph being instead $K = \nu + \mu$.
In the following, we will focus on the friendship layer $G_F$ alone and on the entire urban social network $G$.

To prevent the results from being influenced by the composition of the population, we:
(i) set the matrix $S$ to a constant, to force homogeneous mixing between the different age groups (see Appendix~\ref{app:social_contact_data});
(ii) ignore the data-driven population density, considering equally frequent age groups and distributing the individuals uniformly at random in the given territory\footnote{Actually, we map individuals only to tiles that are non-empty according to real data. This is meant to preserve the realism of the benchmark model by preventing that people are found in uninhabitable areas such as parks, lakes, sea, etc.}.
For what concerns the parameters $f_u$ and $D$, we focus on the special configurations highlighted in Section~\ref{sec:chosen_model} -- \emph{i.e.}, $f_u\sim1+ \LN(\ln(2),0.25)$ and $D(u,v)=d(u,v)^{-\beta}$ with $\beta\in\{0.5,2\}$ -- and we additionally compare them with the choice of a constant fitness $f_u\equiv1$ for all $u$.

Ideally, we would like a graph where (almost) all nodes belong to a single giant component, in accordance with empirical findings from the literature~\cite{herrera2015anatomy}.
In Table~\ref{tab:firenzeGiant} we present the case of Florence -- similar results hold for the other cities and are listed in Appendix~\ref{app:viterbo} and Appendix~\ref{app:sabaudia}.
Let us first focus on the friendship graph $G_F$ alone (\emph{i.e.}, no households).
The case $\mu=1$ clearly shows that this graph is very poorly connected if the average degree is very low.
When the fitness is Lognormally distributed, thanks to the increased degree variability, the giant component includes $\sim 20\%$ of the network.
When the fitness is constant, the giant component is instead significantly smaller, reaching a critical $1.9\%$ in the configuration that maximizes the homogeneity of the network.
When $\mu=5$, the giant component on the other hand, includes almost the entirety of the network.
Quite interestingly, in this regime switching from a constant fitness to a Lognormal one yields a slight decrease in the coverage of the giant component of the friendship network. 
The rationale is that, in the latter case, most hubs occur in denser areas of the graph, thus increasing the internal cohesion of the giant component to the detriment of its size.
By comparing the friendship graph with the whole graph, we finally see that when $\mu=1$ the households are capable to partially compensate for the limited number of friendship edges.
When $\mu=5$, finally, the households make it possible to even improve the already almost perfect coverage of the giant component.

\begin{table}[th]
\caption{[\textbf{Florence}] Percentage of nodes of the graph that belong to the giant component, on average, for the friendship graph $G_F$ and the entire social graph $G$, as $\beta$, $f_u$ and $\mu$ vary.}
\label{tab:firenzeGiant}
\centering
\begin{tabular}{llrrrr}
\addlinespace[-\aboverulesep] 
\cmidrule[\heavyrulewidth]{3-6}
& & \multicolumn{2}{c}{$\mu=1$} & \multicolumn{2}{c}{$\mu=5$}\\
$\beta$ & $f_u$ & $G_F$ & $G$ & $G_F$ & $G$\\
\midrule
0.5 & 1 & 1.9\% &  76.6\% & 99.3\% & 99.8\% \\
0.5 & $1+\LN(\ln(2), 0.25)$ & 18.8\% &  75.8\% & 98.2\% & 99.4\%  \\
2 & 1 & 8.0\% &  76.0\% & 99.0\% & 99.7\%  \\
2 & $1+\LN(\ln(2), 0.25)$ & 20.9\% &  75.1\% & 97.9\% & 99.3\%  \\
\bottomrule
\end{tabular}
\end{table}

\subsection{Accounting for age in friendships}
\label{sec:group2group}

One fundamental design choice of our friendship graph model consists in enforcing that the proportion of intra- and inter-age group connections adheres to publicly available contact data.
This information is encoded in the age-mixing matrix $S$ and, as emerged in Section~\ref{sec:friendship}, imposing this condition comes at the cost of additional constraints to the model.
To verify the importance of having age-dependent edges, we compare three different configurations:
(i) a homogeneous population connected using a constant matrix $S$; (ii) a homogeneous population connected based on the real (\emph{i.e.}, data-driven) $S$; (ii) a real (\emph{i.e.}, data-driven) population connected based on the real $S$.
With homogeneous population we mean that both the age distribution and the spatial density are taken to be uniform.
The constant matrix $S$ corresponds to age-homogeneous mixing, as explained in Appendix~\ref{app:social_contact_data}. 
Again, we consider the combinations of parameters selected in Section~\ref{sec:chosen_model}, plus a fitness-neutral model with $f_u\equiv 1$ for all $u$.
Finally, to guarantee that age-dependent preferences and patterns in the friendship edge distribution are well visible, we only consider our friendship graph and we fix $\mu=5$, to obtain a network that is well connected but not too dense.

\begin{figure}[htbp]
    \centering
    \includegraphics[width=.95\textwidth]{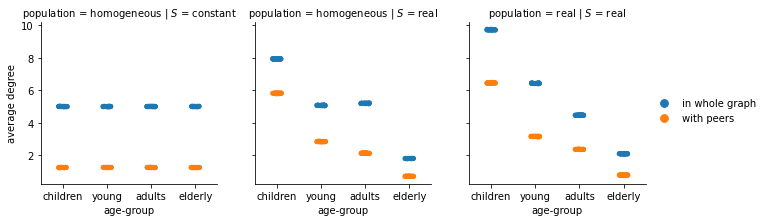}
    \caption{[\textbf{Florence}] Average degree of the individuals of each age group, in the whole friendship graph $G_F$ and with their peers, under different configurations all with $\mu=5$.}
    \label{Figure:deg_firenze_avg5}
\end{figure}

For each of the four age groups, we computed both the average degree of the members of each group in the whole friendship graph $G_F$, and their average degree considering friendship connections with their peers only, that is, within the subgraph of $G_F$ induced by the set of individuals belonging to that group.
In Figure~\ref{Figure:deg_firenze_avg5} we report these data for the city of Florence -- the other cities show analogous patterns, see Appendices~\ref{app:viterbo} and~\ref{app:sabaudia} for details.
With homogeneous population and constant $S$ (left panel) all age groups are, \emph{de facto}, equivalent and the age-stratification is irrelevant for friendships.
Introducing real nonuniform age-mixing coefficients $s_{i,j}$, even with a homogeneous population (central panel), ensures that the data-driven internal cohesion of younger age groups is preserved, but adults are more likely to establish links with other age groups.
Finally, when the population and the matrix $S$ are both data-driven (right panel), we notice an increment in the overall average degree of children and young people, which are $<25\%$ in the real population, and a decrease in the overall average degree of adults, which instead constitute $\approx43\%$ of the population of Florence.
Significantly, Figure~\ref{Figure:deg_firenze_avg5} confirms that neither $f_u$ nor $\beta$ have any impact on the age-mixing patterns of the graph, as imposed by construction.

\begin{figure}[htbp]
\centering
    \begin{subfigure}[b]{.35\textwidth}
         \includegraphics[width=\textwidth,trim={50 50 50 50}, clip]{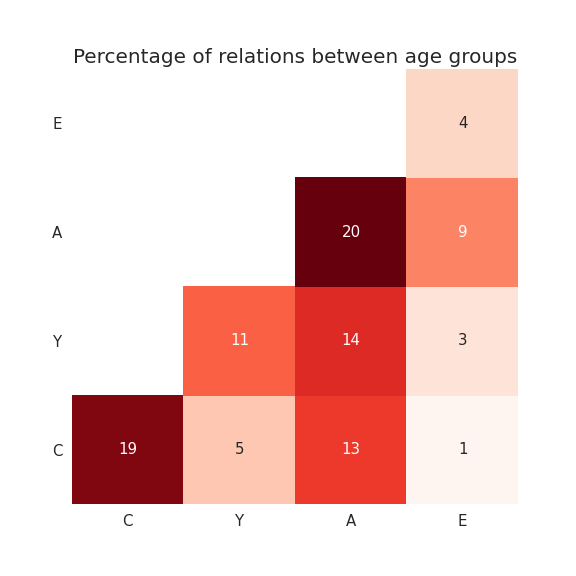}
         \caption{Friendship network.}
         \label{fig:firenze-g2g-hhFalse}
    \end{subfigure}
    \hspace{20pt}
    \begin{subfigure}[b]{.35\textwidth}
         \includegraphics[width=\textwidth,trim={50 50 50 50}, clip]{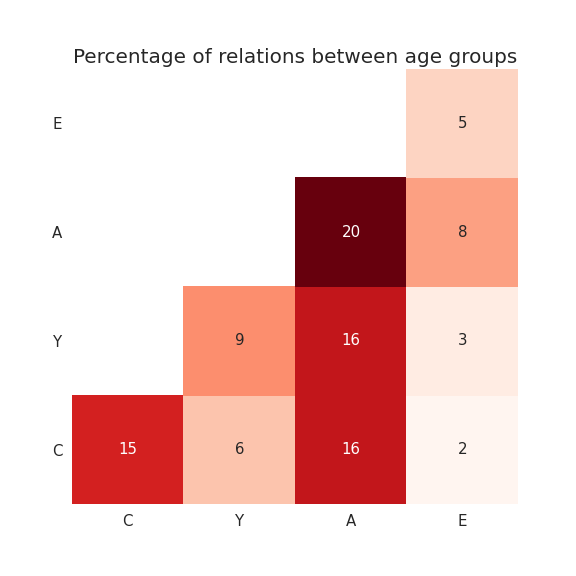}
         \caption{Entire social network.}
         \label{fig:firenze-g2g-hhTrue}
    \end{subfigure}
    \caption{[\textbf{Florence}] Percentage of edges between age groups in a configuration with data-driven population and age-based mixing, for $f_u\sim\fit$, $\mu=5$ and $D(u,v)=d(u,v)^{-2}$.}
    \label{fig:g2g_matrix_firenze}
\end{figure}

For the sake of completeness, Figure~\ref{fig:g2g_matrix_firenze} shows the percentage of the total number of edges that link each age group with all others, for the friendship graph and for the entire social graph obtained for the city of Florence using the real population and real $S$ -- see the Appendix, Figures~\ref{fig:g2g_matrix_viterbo} and~\ref{fig:g2g_sabaudia}, for Viterbo and Sabaudia.
Figure~\ref{fig:firenze-g2g-hhFalse} is essentially a visualization of expression (\ref{eq:condition_groups}), and the value in each cell is thus proportional to the size of the two groups and to their mixing coefficient $s_{i,j}$.
We see that, albeit the adults have a lower average number of friends than the younger groups, they are still involved, in total, in the majority of friendship edges.

\subsection{Using real population density and distances} 
\label{sec:densityDistance}

One of the main elements of novelty of our model resides in the use of a data-driven population, in contrast with a body of research work on spatial networks that assumes a certain regularity in the spatial distribution of the population (\emph{e.g.}, uniform or Gaussian).
The positioning of the individuals of the population and the level of penalization imposed to geographically long edges impact on the final shape of our social graph at multiple levels.
To assess the relevance of using real data for the spatial density of the population, we study the friendship network $G_F$ (\emph{i.e.}, no household edges, $\nu=0$ and $K=\mu$) obtained with either a data-driven population or a homogeneous population where individuals are distributed uniformly at random in the given territory, but fixing in both cases the size of the groups to be uniform, \emph{i.e.}, $|V_i|=\frac{N}{n}$ for all age groups.

\begin{table}[htbp]
\caption{[\textbf{Florence}] Main features of the friendship graph $G_F$ as the population type, $D(u,v)$ and $\mu$ vary, for constant $f_u$ and $S$. $\dist$ is the average path length, $C$ is the global clustering coefficient, $\rho$ is the degree assortativity, ``\# comp.'' denotes the number of connected components, ``giant \%'' denotes the percentage of nodes in the giant component.}
\label{tab:FirenzeGeoDist}
\centering
\begin{subtable}{\textwidth}
\centering
\caption{Friendship network for $\mu=5$. Expected values for an ER graph with $\mu=5$: $\dist \approx 7.95 $, $C=1.38\mathbf{e-}05$.}
\label{tab:FirenzeGeoDist_5}
\begin{tabular}{llrrrrr}
\addlinespace[-\aboverulesep] 
\cmidrule[\heavyrulewidth]{3-7}
& & \multicolumn{5}{c}{$\mu=5$} \\
Population & $D(u,v)$ & $\dist$ & 
$C$ & $\rho$ & \# comp.  & giant \% \\
\midrule
homogeneous & 1               &  8.13  &  1.5e-05  &   2.5e-04    &  2497.5   &  99.3\%  \\ 
homogeneous & $d(u,v)^{-0.5}$ &  8.11  &  1.4e-05  &   0.0018     &  2660.4   &  99.3\% \\
homogeneous & $d(u,v)^{-2}$   &  8.08  &  7.2e-05  &   0.05       &  3385.9   &  99.0\% \\ 
real & 1                      &  8.13  &  1.3e-05  &   -3.8e-04   &  2504.3   &  99.3\% \\
real & $d(u,v)^{-0.5}$        &  8.09  &  1.3e-05  &   0.0044     &  2914.1   &  99.2\% \\
real & $d(u,v)^{-2}$          &  7.83  &  5.6e-05  &   0.12       &  12443.8  &  96.3\%  \\
\bottomrule
\end{tabular}
\end{subtable}

\vspace{10pt}

\begin{subtable}{\textwidth}
\centering
\caption{Friendship network for $\mu=10$. Expected values for an ER graph with $\mu=10$: $\dist \approx 5.56 $, $C=2.75\mathbf{e-}05$.}
\label{tab:FirenzeGeoDist_10}
\begin{tabular}{llrrrrr}
\addlinespace[-\aboverulesep] 
\cmidrule[\heavyrulewidth]{3-7}
& & \multicolumn{5}{c}{$\mu=10$}\\
Population & $D(u,v)$ & $\dist$ & 
$C$ & $\rho$ & \# comp.  & giant \% \\
\midrule
homogeneous & 1               &  5.81   &    2.7e-05   &  2.9e-04   &  17.6  &  100.0\% \\ 
homogeneous & $d(u,v)^{-0.5}$ &  5.80    &    2.8e-05   &  0.0042    &  23.5  &  100.0\% \\
homogeneous & $d(u,v)^{-2}$   &  5.84   &    1.5e-04   &  0.09      &  53.7  &  100.0\%   \\ 
real & 1                      &  5.81   &    2.8e-05   &  1.7e-04   &  16.1  &  100.0\%   \\
real & $d(u,v)^{-0.5}$        &  5.80    &    2.9e-05   &  0.0088    &  32.5  &  100.0\%    \\
real & $d(u,v)^{-2}$          &  5.79   &    1.1e-04   &  0.2       &  2317.6 & 99.3\% \\
\bottomrule
\end{tabular}
\end{subtable}
\end{table}

We consider three different options for $D(u,v)$: $D(u,v)= 1$ for all $u,v$, $D(u,v)=d(u,v)^{-0.5}$ and $D(u,v)=d(u,v)^{-2}$.
Since one of the criteria for comparison will be the connectivity of the network, we let $\mu$ vary as $\mu=5$ or $\mu=10$.
Conversely, to isolate the dependency of the edge distribution upon the spatial density, we take both $S$ and $f_u$ to be constant and equal to 1.

\begin{figure}[htbp]
    \centering
    \includegraphics[width=\textwidth]{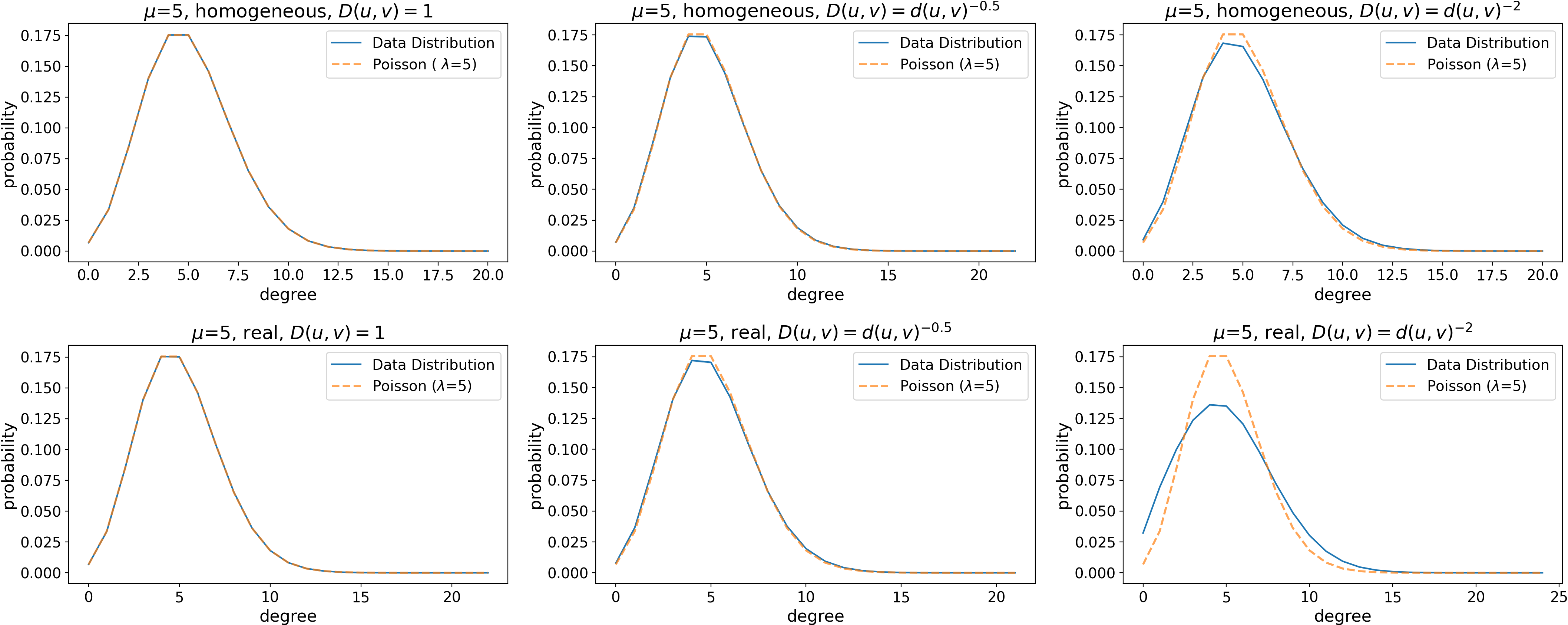}
    \caption{[\textbf{Florence}] Degree distributions of friendship graphs with $\mu=5$ and corresponding Poisson distributions with $\lambda=5$.
    The data distribution lines show the mean, whereas the shaded areas  around the lines represent the 95\% confidence interval.
    The Kullback–Leibler divergence between the Poisson and the empirical distribution are (left to right, top to bottom):
    1.6e-07,
    1.1e-05,
    0.0002,
    2.87e-07,
    5.2e-05,
    0.004.
    }
    \label{Figure:FirenzeDegDist5}
\end{figure}

\begin{figure}[htbp]
    \centering
    \includegraphics[width=\textwidth]{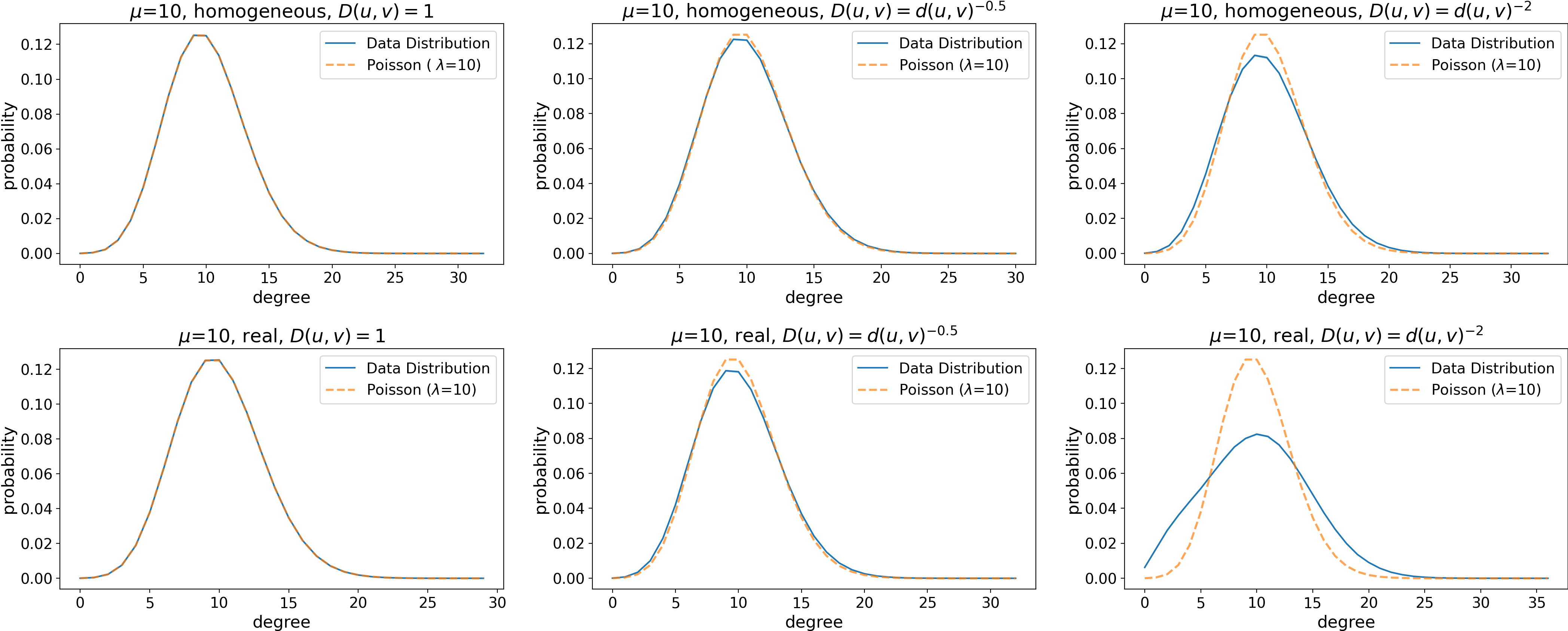}
    \caption{[\textbf{Florence}] Degree distributions of friendship graphs with $\mu=10$ and corresponding Poisson distributions with $\lambda=10$.
    The data distribution lines show the mean, whereas the shaded areas  around the lines represent the 95\% confidence interval.
    The Kullback–Leibler divergence between the Poisson and the empirical distribution are (left to right, top to bottom): 
    2.5e-07,
    3.2e-05,
    0.0005,
    2.01e-07,
    0.0001,
    0.009.
    }
    \label{Figure:FirenzeDegDist10}
\end{figure}

\begin{figure}[htbp]
    \centering
    \includegraphics[width=\textwidth]{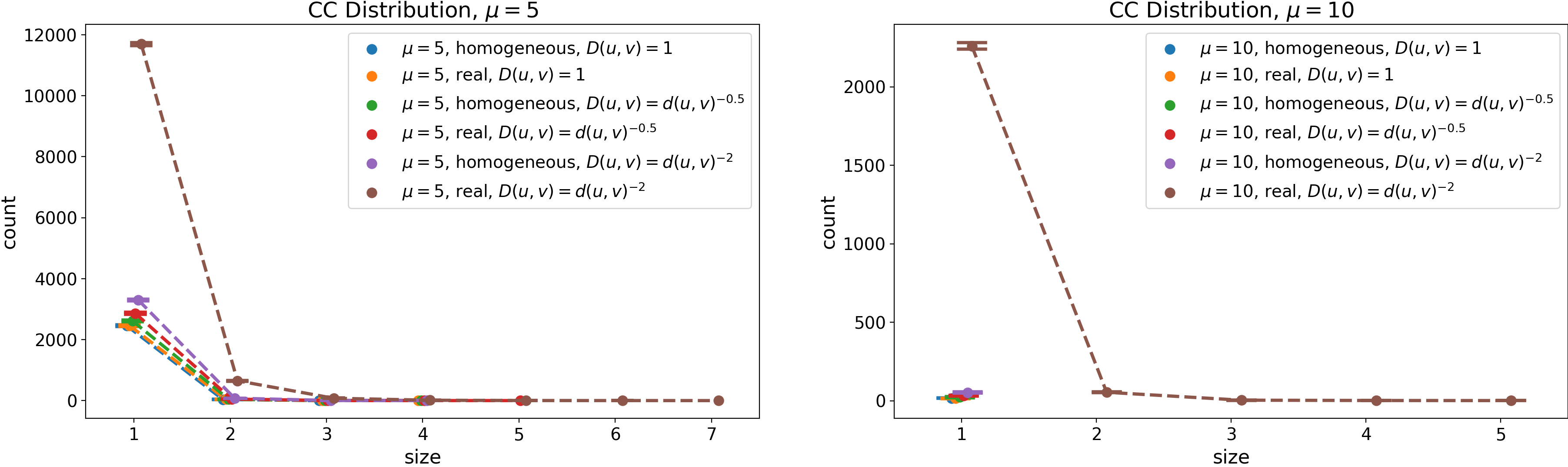}
    \caption{[\textbf{Florence}] Distribution of the size of the connected components other than the giant for friendship graphs with $\mu=5$ and $\mu=10$.
    The lines show the mean, whereas the error bars (with ``caps'') represent the 95\% confidence interval computed over each set of runs.}
    \label{Figure:FirenzeConnComp}
\end{figure}

Table~\ref{tab:FirenzeGeoDist} and Figures~\ref{Figure:FirenzeDegDist5},~\ref{Figure:FirenzeDegDist10},~\ref{Figure:FirenzeConnComp} show the results of our tests for the city of Florence -- for the other two cities, see Appendices~\ref{app:viterbo} and~\ref{app:sabaudia}. 
We repeated the construction of the social graph 10 times for each configuration.
Since the variance is negligible and adds no valuable element of discussion, for the sake of clarity we only included in Table~\ref{tab:FirenzeGeoDist} the average value for each metrics over each set of runs.
Figures~\ref{Figure:FirenzeDegDist5},~\ref{Figure:FirenzeDegDist10} instead report the mean distributions over each set of runs with a 95\% confidence interval.
To measure the distance between the data distributions and the Poisson distribution, we compute the relative entropy or Kullback–Leibler (KL) distance -- the values are reported in the caption of the figures.
The KL distance $D(p||q)$ between distribution $p$ and $q$ measures the amount of information lost when the distribution $q$ is used to approximate the distribution $p$~\cite{cover2012elements}.

We can divide our tests in three main categories, based on the used combination of parameters:
\begin{enumerate}
    \item \textbf{Homogeneous/real population and $D(u,v) = 1$}: 
    Geographical distances and the distribution of the population over the territory do not affect the graph construction if we set $D(u,v) = 1$ and we impose a uniform size for all age groups. 
    In this case, we expect to create an ER graph, because we are actually fixing the edge probability $Pr[u,v]$ to be the same for any node pair $u, v$. In an ER graph~\cite{albert2002statistical,newman2010networks} the expected clustering coefficient is $C = \frac{\mu}{N-1}$, where $\mu$ is the mean degree, the degree distribution is a Poisson distribution (for large $N$), and the expected average shortest path is $\dist \approx \frac{\ln N}{\ln \mu}$. The graphs of our tests have 363060 nodes (the population of Florence), thus we expect the following values: $C=1.38\mathbf{e-}0$ for $\mu=5$, $C=2.75\mathbf{e-}05$ for $\mu=10$,  $\dist \approx 7.95$ for $\mu=5$ and $\dist \approx 5.56$ for $\mu=10$. 
    Our results (see Table~\ref{tab:FirenzeGeoDist}, Figures~\ref{Figure:FirenzeDegDist5} and~\ref{Figure:FirenzeDegDist10}) confirm the expected values and degree distributions. 
    Therefore, if we do not consider real population density and distances we are \emph{de facto} building an ER graph.
    
    \item \textbf{Homogeneous population and $D(u,v)=d(u,v)^{-0.5}$/$D(u,v)=d(u,v)^{-2}$}: Geographical distances do affect the graph construction, under both regimes of $\beta\in\{0.5,2\}$, even if we impose a uniform distribution of the population over the territory, other than equal size age groups. 
    Our results (see Table~\ref{tab:FirenzeGeoDist}, Figures~\ref{Figure:FirenzeDegDist5} and~\ref{Figure:FirenzeDegDist10}) show that $\beta=0.5$ has a major effect on the assortativity of the network, whereas $\beta=2$ affects both the global clustering coefficient and the assortativity.
    The latter is always positive and for $\beta=2$ the global clustering coefficient is about 5 times greater than in all graphs built in the previous cases.
    Again, these are desirable features in a social network model~\cite{newman2003social,wong2006spatial}. Geographical distances have also an effect on the degree distribution, but it is appreciable only for $\beta=2$ (see Figures~\ref{Figure:FirenzeDegDist5},~\ref{Figure:FirenzeDegDist10}).
    
    \item \textbf{Real population and $D(u,v)=d(u,v)^{-0.5}$/$D(u,v)=d(u,v)^{-2}$}: Finally, the combined use of the real population and of a penalization for long edges is clearly visible in our simulation, for both values of $\beta$, but especially when $\beta=2$.
    Real density and distances mainly influence the number of connected components and the assortativity of the network. Notice that the number of connected components is also affected by the average degree of the network: the higher the degree, the lower is the number of connected components. Specifically, the use of real data seems to favor higher assortativity values and a higher number of connected components. 
    While a uniform population density creates a giant component whose size is in line with the giant component of an ER graph, a real density creates a smaller giant component and a higher number of connected components (see Figure~\ref{Figure:FirenzeConnComp}). 
    For uniform densities the size of the connected components is almost always 1 (the same holds when we set $D(u,v) = 1$), instead for real densities we have more components whose size is more variable.
    Finally, as shown in Figures~\ref{Figure:FirenzeDegDist5} and \ref{Figure:FirenzeDegDist10}, the degree distribution is also affected  by real densities and distances.
    Preventing a Poisson degree distribution is a further step toward the definition of a suitable model, because social networks are usually characterized by skewed degree distributions~\cite{wong2006spatial}.

\end{enumerate}

Summarizing, our tests show that  penalizing long edges in (\ref{eq:edge_prob}) while using the real population density allows to construct a friendship graph deprived of the typical characteristics of ER-like homogeneous graphs.
Even more, using real data we are able to simulate some of the features usually observed in social networks.
These findings support the rationale that a data driven approach is important: real population density and distances must be taken into consideration for building urban social graphs.
Finally, we observe that the impact on graph construction of geographical distances, when $\beta=0.5$, is negligible if not combined with real densities, as shown by the degree distributions (Figures~\ref{Figure:FirenzeDegDist5},~\ref{Figure:FirenzeDegDist10}) and values in Table~\ref{tab:FirenzeGeoDist}.




\section{Characterization of the Urban Social Network}
\label{sec:USG}

In Section \ref{sec:parameters} we verified that each and every parameter of our model has a role in the definition of the resulting urban social graph, and that making proper use of the available data is paramount in order not to end up with a too simple network model (\emph{e.g.}, a ER-like graph).
In the following, we therefore consider the fully data-driven version of our model: the synthetic population is constructed as described in Section~\ref{sec:population} based on real age-stratification data, real spatial density estimates, and real household composition data; the dependence of the friendship edges upon the age of the individuals is inferred from publicly available contact matrices.
We study the topological features of the graph and their relation with the imposed geographical constraints, as the remaining parameters vary as follows: (i) $\mu=5$ or $\mu=10$; (ii) $D(u,v) = d(u,v)^{-\beta}$ with $\beta=0.5$ or $\beta=2$; (iii) $f_u \equiv 1$ for all $u$ or $f_u\sim 1+\LN(\ln(2),0.25)$.
Notably, we already know from Section~\ref{sec:parameters} that all these configurations yield, albeit to different extents, desirable properties of social networks, including good connectivity, decent network assortativity and global clustering coefficient, and a non-Poisson degree distribution.

\subsection{Global metrics}
\label{sec:global_metrics}

We assessed the impact of using the real population density and an inverse-power-law distance penalization in Section \ref{sec:densityDistance}.
In this section, we focus again on the main global features of the graph, but considering the fully data-driven model that includes the household layer $G_H$ and the age-based social mixing imposed through the matrix $S$ (see Appendix~\ref{app:social_contact_data}).  

By using the data-driven values of $S$, we are adding to the model an age-specific homophily and we expect an increase in the assortativity of the resulting graph. 
Indeed, for the city of Florence, considering the other parameters as in Section~\ref{sec:densityDistance} with $\beta = 2$ and $\mu=10$, the assortativity grows from $0.2$ to $0.48$. 
Moreover, by connecting with higher probability nodes belonging to specific subsets (\emph{i.e.}, those in the same age groups) we report a four-fold increase of the global clustering coefficient from $0.00012$ to $0.0004$.
On the other hand, the layer $G_H$ introduces, by construction, a large amount of cliques -- albeit small -- so we can expect a great impact on the clustering coefficient (\emph{i.e.}, the number of triangles). 
In fact, considering again the city of Florence, with the same parameters of the previous example and adding the households, the global clustering coefficient increases $\sim 40$ times from $0.0004$ to $0.0151$.

An overview of the resulting graph for the city of Florence with the parameter chosen as described at the beginning of Section~\ref{sec:USG} is reported in Table~\ref{tab:Metrics} -- for Viterbo and Sabaudia see Appendices~\ref{app:viterbo} and~\ref{app:sabaudia}. 
For each set of parameters, we repeated the construction of the social graph 10 times. The values reported in the table are the mean computed over each set of runs, the variance is negligible for all the values and thus omitted.
By looking at Table~\ref{tab:Metrics} we see that the contribution of the fitness to the clustering coefficient, the assortativity and the connectivity is only marginal compared to $S$ and to the households (for a direct comparison with the configurations considered in Section~\ref{sec:densityDistance}, see Table~\ref{tab:FirenzeGeoDist}).
The value of the global clustering is several orders of magnitude greater than the corresponding ER graphs, it shows a small decrease with the fitness and for higher values of $\beta$ (and $\mu$).
Both circumstances favor the creation of hubs that attract edges from the rest of the graph thus making ``local'' triangles less likely to occur. 
Besides the presence of hubs, there will be a higher number of nodes at a lower degree and therefore a higher probability of having many connected components.
The degree heterogeneity also penalizes the assortativity which is consistently lower for the run with the fitness but increases with $\beta$. It is likely that, on average, the higher value of $\beta$ reinforces the group-to-group assortativity of $S$.
On the other hand, the average local clustering has the opposite behaviour and constantly increases with the fitness, $\beta$ and $\mu$. Local connections are favored by a stronger dependence on distance and by the presence of local hubs.
Almost regardless of the configuration, the average shortest path length is comparable to $\frac{\ln(N)}{\ln(K)}$, where $K=\mu+\nu$ is the average degree of the graph $G$.
This value is typical for small world networks.  


\begin{table}[htbp]
\caption{[\textbf{Florence}] Main features of the social graph as $f_u$, $\beta$ and $\mu$ vary, for data-driven population and $S$. $K=\mu+\nu$ is the average degree, $\dist$ is the average path length, $C$ and $C_{\mathrm{loc}}$ are the global and average local clustering coefficients, $\rho$ is the degree assortativity, ``\# comp.'' denotes the number of connected components, ``giant \%'' denotes the percentage of nodes in the giant component.}
\label{tab:Metrics}
\centering
\begin{subtable}{\textwidth}
\centering
\caption{Social graph for $\mu=5$.}
\label{tab:Metrics_Firenze_5}
\begin{tabular}{llrrrrrrr}
\addlinespace[-\aboverulesep] 
\cmidrule[\heavyrulewidth]{3-9}
& & \multicolumn{7}{c}{$\mu=5$} \\
$f_u$ & $\beta$ & $K$ & $\dist$ & $C$ & $C_{\mathrm{loc}}$ & $\rho$ & \# comp. & giant \% \\
\midrule
1      & 0.5 & \phantom{0}6.82 & 6.72 & 0.053 & 0.074 & 0.261 & 4071.1 & 98.8\% \\
1      & 2   & \phantom{0}6.81 & 6.72 & 0.050 & 0.089 & 0.334 & 6824.8 & 97.7\%  \\
$\fit$ & 0.5 & \phantom{0}6.82 & 6.57 & 0.049 & 0.085 & 0.203 & 5357.4 & 98.4\%  \\
$\fit$ & 2   & \phantom{0}6.81 & 6.58 & 0.046 & 0.100 & 0.258 & 8109.3 & 97.3\%  \\
\bottomrule
\end{tabular}
\end{subtable}

\vspace{10pt}

\begin{subtable}{\textwidth}
\centering
\caption{Social graph for $\mu=10$.}
\label{tab:Metrics_Firenze_10}
\begin{tabular}{llrrrrrrr}
\addlinespace[-\aboverulesep] 
\cmidrule[\heavyrulewidth]{3-9}
& & \multicolumn{7}{c}{$\mu=10$} \\
$f_u$ & $\beta$ & $K$ & $\dist$ & 
$C$ & $C_{\mathrm{loc}}$ & $\rho$ & \# comp.  & giant \% \\
\midrule
1      & 0.5 & 11.81 & 5.36 & 0.017 & 0.027 & 0.317 & 463.1  & 99.9\% \\
1      & 2   & 11.81 & 5.41 & 0.016 & 0.037 & 0.383 & 1726.9 & 99.5\% \\
$\fit$ & 0.5 & 11.81 & 5.26 & 0.016 & 0.033 & 0.211 & 924.9  & 99.7\% \\
$\fit$ & 2   & 11.82 & 5.32 & 0.015 & 0.044 & 0.260 & 2333.1 & 99.3\% \\
\bottomrule
\end{tabular}
\end{subtable}
\end{table}

\subsection{Connectivity, communities, degree}
\label{sec:conn_clust_deg}

Hereafter, we focus on the graph we obtained for the city of Florence, for which we provide an overview of the main characteristics in Figure~\ref{fig:firenze}.
Similar results hold for the other two cities and are reported in Appendix~\ref{app:viterbo} and Appendix~\ref{app:sabaudia}.

\begin{figure}[htbp]
    \centering
    \begin{subfigure}[b]{.48\textwidth}
         \centering
         \includegraphics[width=\textwidth]{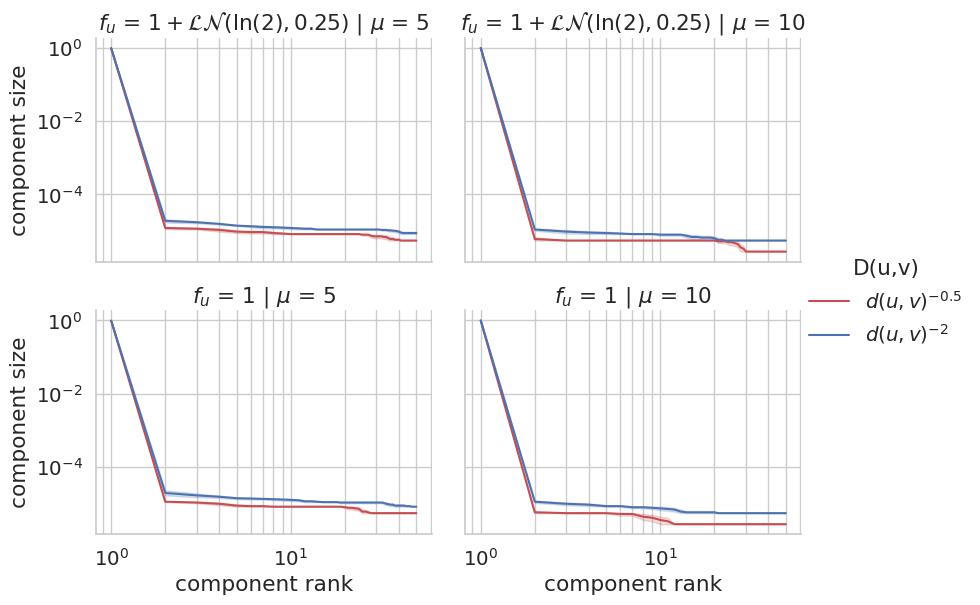}
         \caption{Size of the largest 50 connected components.}
         \label{fig:components_firenze}
    \end{subfigure}
    \hfill
    \begin{subfigure}[b]{.48\textwidth}
         \centering
         \includegraphics[width=\textwidth]{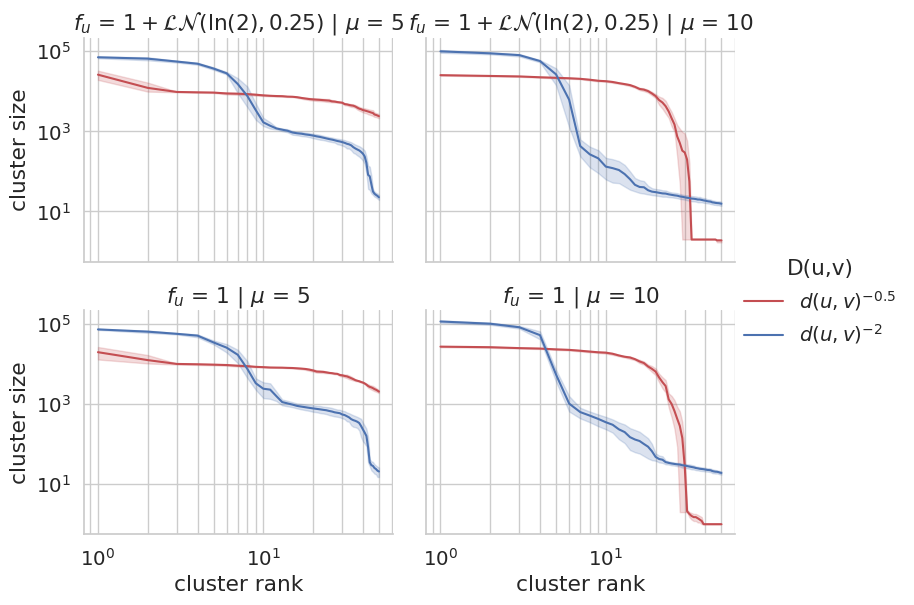}
         \caption{Size of the largest 50 clusters (Louvain).}
         \label{fig:cluster_size_firenze}
    \end{subfigure}
    
    \begin{subfigure}[b]{.48\textwidth}
         \centering
         \includegraphics[width=\textwidth]{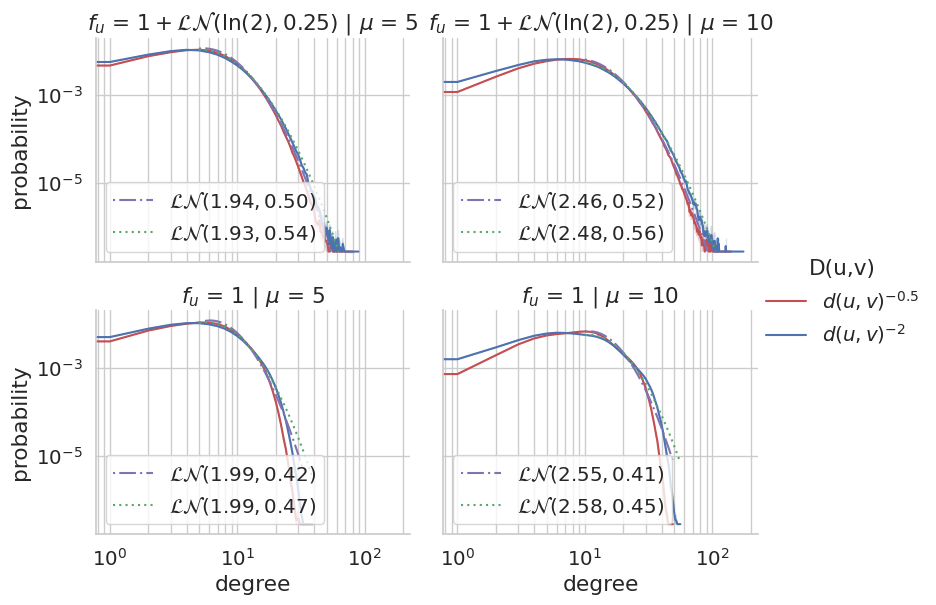}
         \caption{Degree distribution with Lognormal fit.}
         \label{fig:degree_distribution_firenze}
    \end{subfigure}
    \hfill
    \begin{subfigure}[b]{.48\textwidth}
         \centering
         \includegraphics[width=\textwidth]{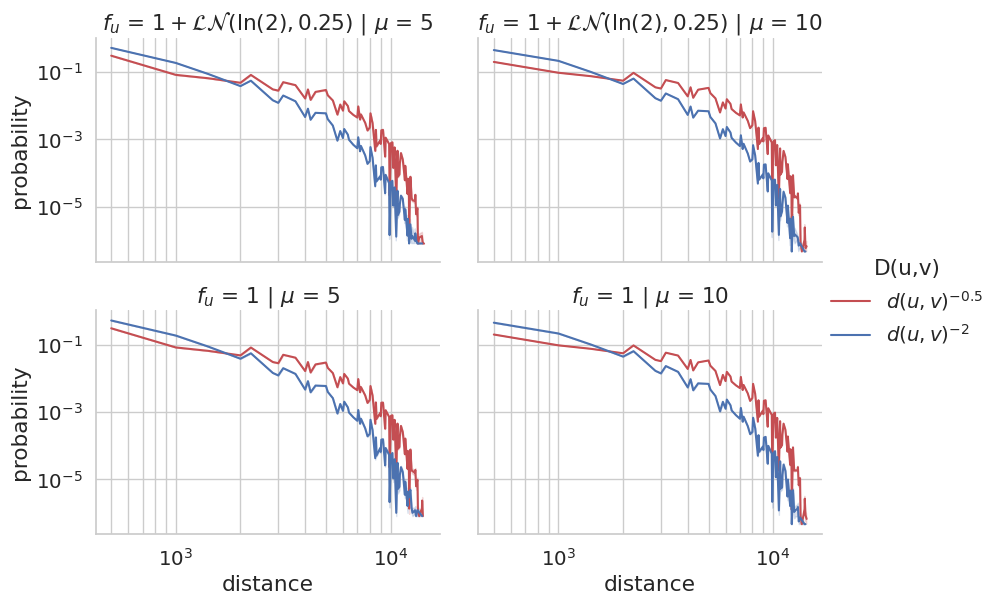}
         \caption{Geographical distance between adjacent vertices.}
         \label{fig:distance_distribution_firenze}
    \end{subfigure}
    
    \begin{subfigure}[b]{.48\textwidth}
         \centering
         \includegraphics[width=\textwidth]{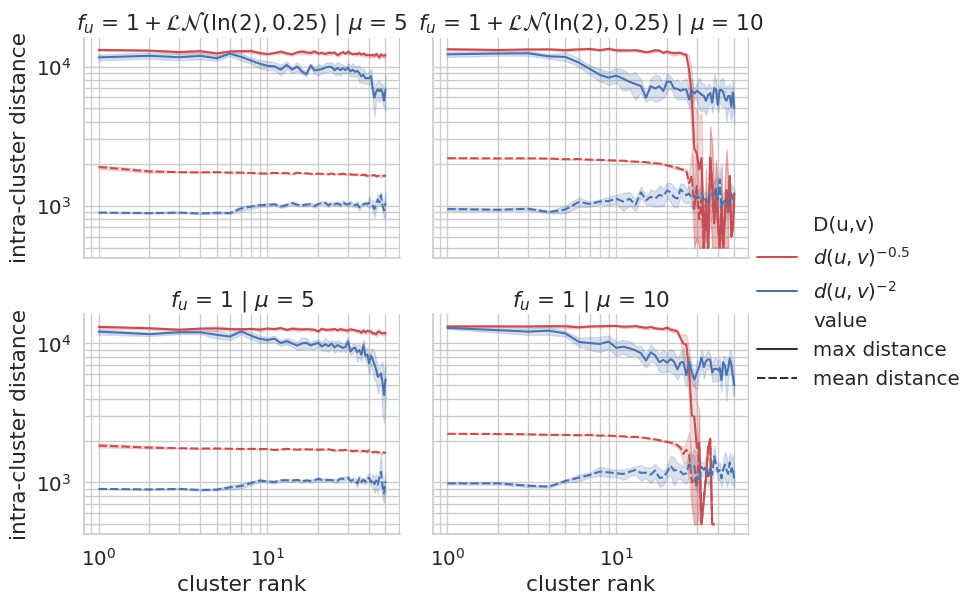}
         \caption{Mean and max intra-cluster geographical distances.}
         \label{fig:cluster_distance_firenze}
    \end{subfigure}
    \hfill
    \begin{subfigure}[b]{.48\textwidth}
         \centering
         \includegraphics[width=\textwidth]{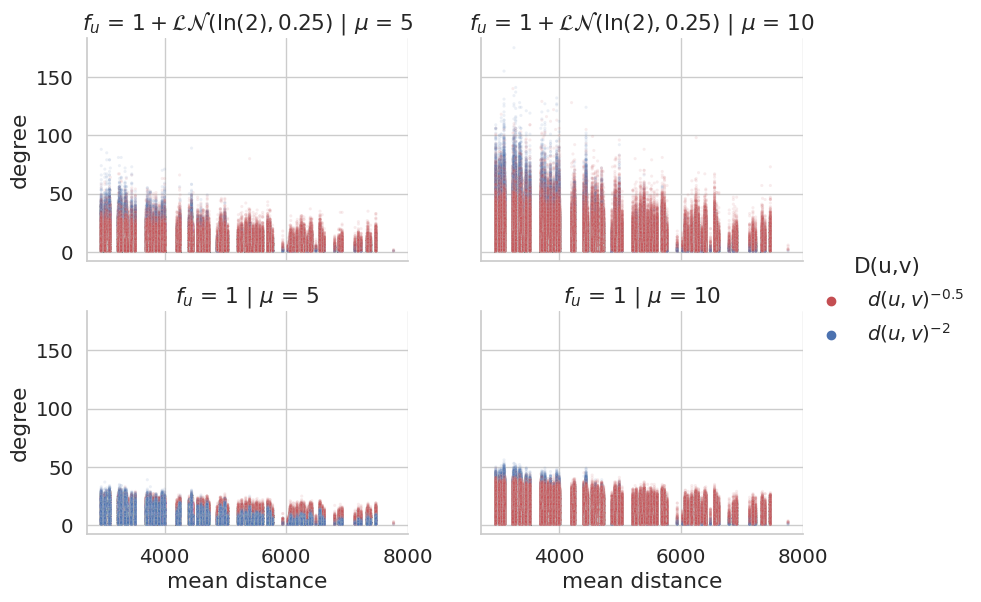}
         \caption{Mean distance to all other vertices \emph{vs.} Degree.}
         \label{fig:degree_distance_firenze}
    \end{subfigure}
    \caption{[\textbf{Florence}] Overview of the urban social graph. Each plot shows the average with confidence interval for 10 independent runs with the same configuration.}
    \label{fig:firenze}
\end{figure}

\begin{figure}[htbp]
    \centering
    \includegraphics[width=.6\textwidth]{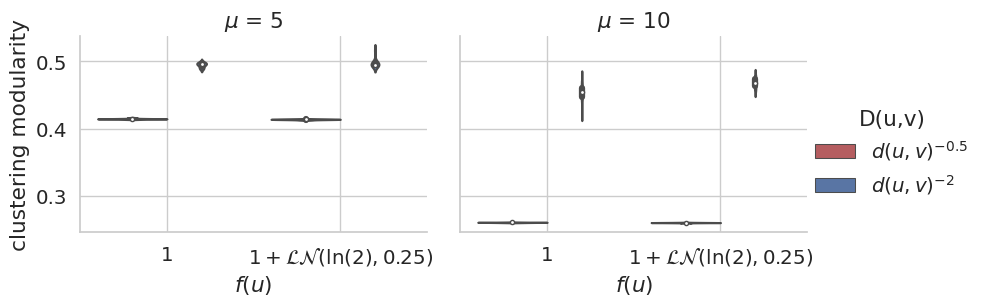}
    \caption{[\textbf{Florence}] Modularity of the obtained clustering structure -- distribution for 10 independent runs per configuration.}
    \label{fig:modularity_firenze}
\end{figure}

First of all, by looking at Figure~\ref{fig:components_firenze} we notice that the fragmentation of the graph into connected components is very stable across different simulations and different configurations.
For all eight considered combinations of parameters, a single giant component exists, covering at least 97.3\% of the whole graph, with a few additional components having tenths to hundreds of nodes and a constellation of isolated vertices.

We then focus on the modularity-based clustering obtained by applying to our network the well-known Louvain algorithm.
From Figure~\ref{fig:cluster_size_firenze}, we see that some parameters do have an impact on the organization of the graph into densely connected communities.
In particular, a steeper cluster size distribution is associated with a greater average degree and with a stronger dependence of friendships on distance.
These two conditions, in fact, favor the emergence of just a few giant clusters and of a multitude of clusters of variable size.
When $\mu=10$ and $\beta=0.5$ we notice a sudden drop of the cluster size, with no clusters of size $\approx 100$.
Quite surprisingly, taking $f_u\sim 1+\LN(\ln(2),0.25)$ seems to have a minor impact from this point of view, albeit this choice is supposed to guarantee the existence of large hubs.
For the sake of completeness, in Figure~\ref{fig:modularity_firenze} we compare the modularity of the obtained clustering structure for different configurations.
We see that the modularity ranges from less than $0.3$, when $\mu=10$ and $\beta=0.5$, to $\approx0.5$ when $\mu=5$ and $\beta=2$.
Not surprisingly and in line with previous findings, the density of the network has a major impact on the quality of the obtained communities.
Especially with denser networks, the penalty applied to long edges also has a visible effect, contrarily to the fitness score.

In Figure~\ref{fig:degree_distribution_firenze} we finally show the degree distribution of the graph (we remind that household edges are included) for all configurations, in a log-log scale.
At least when $f_u\sim 1+\LN(\ln(2),0.25)$, we expect the right tail of the degree distribution of the graph to be heavy but not fat (\emph{i.e.}, subexponential but not power-law).
To verify this insight, we included in the plot the result of a Lognormal fit of the portion of the distribution corresponding to degrees $\geq\mu$.
The fit looks indeed accurate when $f_u\sim 1+\LN(\ln(2),0.25)$, whereas when $f_u\equiv1$ the tail of the distribution is much shorter.
The impact of the exponent $\beta$ is slightly but consistently visible in all plots: when $D= d^{-2}$ the distribution is more skewed than when $D= d^{-0.5}$, with a larger portion of loosely connected vertices compensated by the presence of greater hubs.
This phenomenon is more visible in the other cities (see Appendices~\ref{app:viterbo} and~\ref{app:sabaudia}), where the territory is proportionally wider. 
The rationale is that a weaker dependence on the distance pushes individuals living in central and denser areas to connect to peripheral vertices, that would otherwise remain isolated.

\subsection{Geography of the graph}

Since our model includes a factor purposely designed to penalize long friendship edges, it is especially worth assessing the correlation between topological properties and population density.
To this end, we first show in Figure~\ref{fig:distance_distribution_firenze} the distribution of the edges' physical length.
We immediately see that, as expected, the fitness function and the average degree appear to have a negligible impact on the distribution, contrary to the distance function.
Our graph model indeed guarantees that both $\mu$ and $f_u$ only impact on the number of friends of $u$, whereas, regardless of $\mu$ and $f_u$, $D(u,v)$ is what ultimately determines the ratio of friends $v$ that $u$ will have at any given geographical distance.
Of course, this works as long as the fitness is distributed independently of the location, which is true by design on a probability base.
For what concerns $D$, with respect to $\beta=2$, setting $\beta=0.5$ significantly favors the creation of long edges at the expenses of the very short ones.

We now look into the obtained communities to see if they are geographically concentrated and/or bounded.
To this end, Figure~\ref{fig:cluster_distance_firenze} shows the mean and max intra-cluster distance for the first 50 clusters of the graph.
By comparing Figure~\ref{fig:cluster_distance_firenze} with Figure~\ref{fig:cluster_size_firenze}, we realize that, albeit long intra-cluster edges tend to disappear as the cluster size gets smaller, only clusters that are at least 3 orders of magnitude smaller than the whole graph are geographically bounded -- a phenomenon that is especially visible when $\mu=10$, that is, when the cluster size distribution is steeper.
This is partially in line with a previous work showing a sudden rise in the geographical extension of the communities of empirical networks~\cite{onnela2011geographic}.
All sufficiently large clusters behave very similarly to each other and to the whole graph: most intra-cluster edges connect nearby vertices, yet very long edges do exist in each cluster.
This plot underlines that the parameter $\beta$ has a paramount role even in the formation of clusters, with both the mean and max intra-cluster distance being consistently greater when $\beta=0.5$ compared to the case $\beta=2$.
In particular, the mean distance when $\beta=2$ is often less than the tile side $l$ (set to 1 km as per Section~\ref{sec:population}), meaning that most adjacent vertices are at one tile of distance or less; the mean distance is instead between $2l$ and $3l$ when $\beta=0.5$.

Finally, in Figure~\ref{fig:degree_distance_firenze} we plotted the average geographical distance of a vertex $u$ from all others  against the degree of the vertex $u$.
Since we introduced a penalization for long edges, it is reasonable to expect that vertices that occupy a favorable position, closer, on average, to the other vertices, will generally have a greater degree.
However, setting $D= d^{-0.5}$, seems to be enough to significantly dissipate this effect, which is instead clearly visible when $D= d^{-2}$.
Moreover, we also see that all large hubs have a very small average distance from all other vertices, regardless of the specific configuration.
In particular, when $f_u\sim 1+\LN(\ln(2),0.25)$ the greatest hubs correspond always to vertices having both a large $f_u$ and a favorable position in the territory.
Albeit the introduction of a social fitness in the model allows to have medium-large hubs in sparsely populated regions, the greater prevalence, on average, of hubs in densely populated areas may exacerbate the tendency of other vertices to establish links with individuals in those areas.

\subsection{Network adjacency matrices and degree density plot}
\label{sec:adjacency}
\begin{figure}[htbp]
\centering
    \begin{subfigure}[b]{.35\textwidth}
         \includegraphics[width=\textwidth,trim={100 100 100 100}, clip]{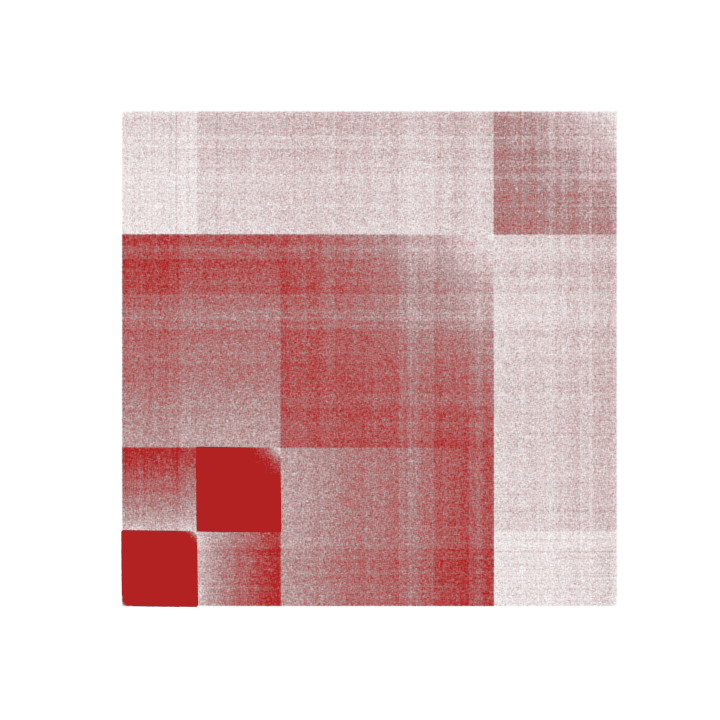}
         \caption{$\beta=0.5$.}
         \label{fig:firenze-adj-0510}
    \end{subfigure}
    \hspace{20pt}
    \begin{subfigure}[b]{.35\textwidth}
         \includegraphics[width=\textwidth,trim={100 100 100 100}, clip]{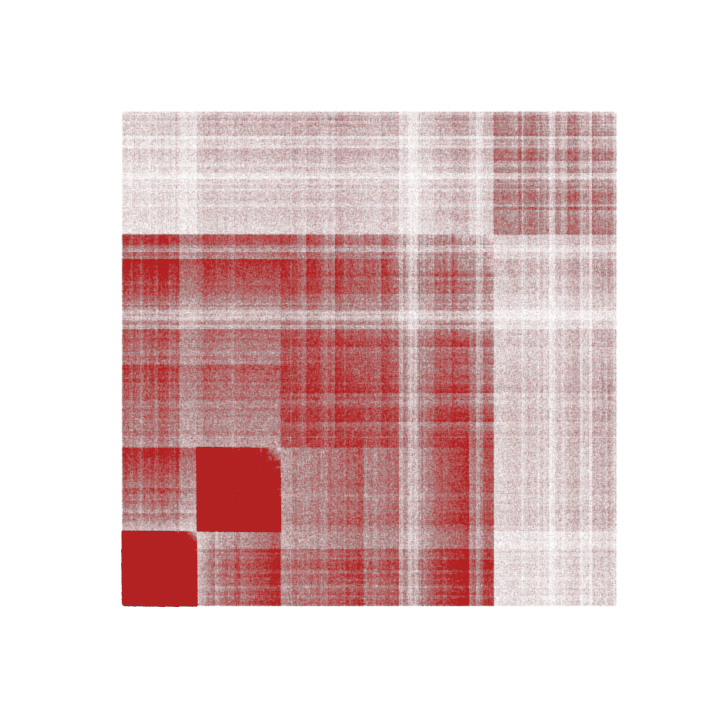}
         \caption{$\beta=2$.}
         \label{fig:firenze-adj-210}
    \end{subfigure}
    \caption{[\textbf{Florence}] Adjacency matrix of the social graph with nodes (people) ordered by age-group.
    In both cases $f_u\sim\fit$, $\mu=10$ and $D(u,v)=d(u,v)^{-\beta}$, but $\beta$ varies between the two figures.}
    \label{fig:adj_matrix_firenze}
\end{figure}

In Figure \ref{fig:adj_matrix_firenze} for Florence (and in Figures~\ref{fig:adj_matrix_viterbo} and \ref{fig:adj_matrix_sabaudia} in the Appendix for the other cities) we report the scatter plot of the adjacency matrices relative to the obtained social graph.
In the plots, nodes are ordered by their age and, within each age-group, by tile.

In all matrices the prevalence of intra-groups edges (assortativity by age) over inter-groups edges is clearly visible and in qualitative agreement with previous works on social mixing patterns \cite{delvalle2007episims, AJELLI20171, klepac2020contacts, read2014china, mistry2020inferring}. 
The white stripes, clearly visible for $\beta = 2$ (Figure~\ref{fig:firenze-adj-210}),  are a consequence of the sorting by tile (and of the use of tile-to-tile distances) and show the impact of distance on connection patterns.
In \cite{delvalle2007episims, mistry2020inferring} the authors also identified sub-diagonals which account for parent-children contacts. These sub-structures cannot be seen in our matrices because we used only four age-groups, while the aforementioned studies relied on a stratification of the population into 5-year segments. In contrast, we see that the adult group, which is the largest group in the population, dominates the inter-groups contacts.

\begin{figure}[htbp]
    \begin{subfigure}[b]{.33\textwidth}
         \centering
         \includegraphics[width=\textwidth]{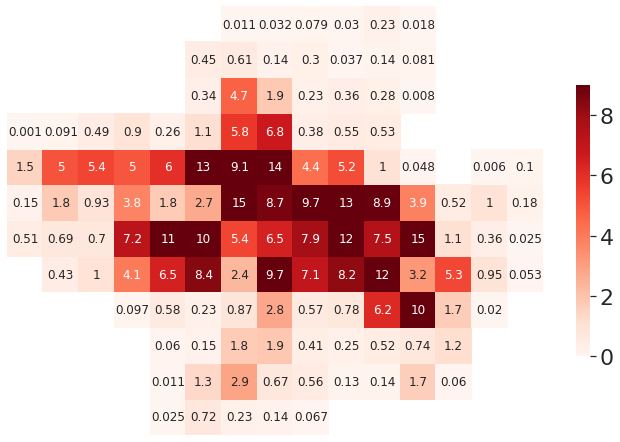}
         \caption{Population in thousands.}
         \label{fig:firenze-pop-heatmap}
    \end{subfigure}
    \begin{subfigure}[b]{.33\textwidth}
         \centering
         \includegraphics[width=\textwidth]{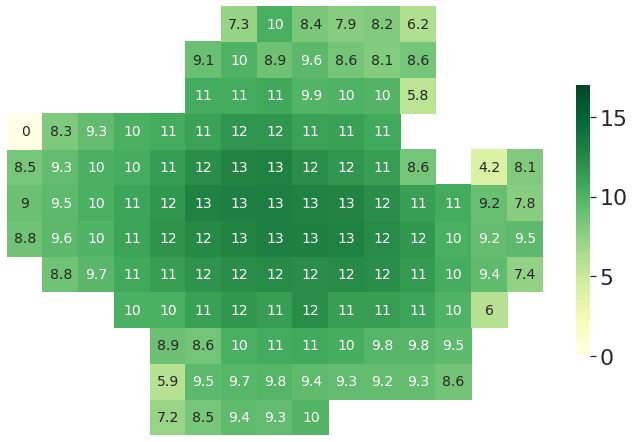}
         \caption{Mean degree, $\beta=0.5$.}
         \label{fig:firenze-heatmap-0510}
    \end{subfigure}
    \begin{subfigure}[b]{.33\textwidth}
         \centering
         \includegraphics[width=\textwidth]{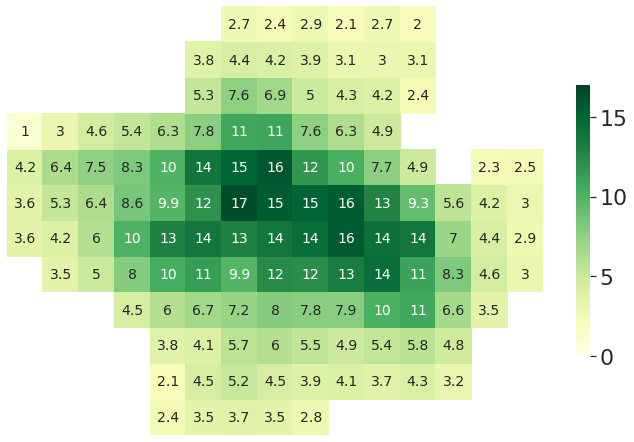}
         \caption{Mean degree, $\beta=2$.}
         \label{fig:firenze-heatmap-210}
    \end{subfigure}

    \caption{[\textbf{Florence}] Heatmaps of the population and of the average degree per tile of the considered territory.
    The average degree is obtained for a graph with $f_u\sim\fit$, $\mu=10$ and $D(u,v)=d(u,v)^{-\beta}$, for both $\beta\in\{0.5,2\}$.
    }
    \label{fig:heatmap_firenze}
\end{figure}

To further investigate the role of population density in shaping the graph, in Figure~\ref{fig:heatmap_firenze} (and \ref{fig:heatmap_viterbo}, \ref{fig:heatmap_sabaudia} in the Appendix) we show the number of individuals living in each tile and their average degree, for two selected configurations both with $\mu=10$, so that the average degree of the whole graph is $K\approx 12$.
It is apparent that the average degree per tile is strongly influenced by the choice of the distance function.
While in Figure~\ref{fig:firenze-heatmap-0510} the situation is mostly homogeneous with few tiles above average, in Figure~\ref{fig:firenze-heatmap-210} most tiles are far below average while the tiles surrounded by a densely populated area have a high average degree.
The increase of $\beta$ has the effect of decreasing the probability of long contacts, a condition that favors the concentration of hubs in high density areas.

\begin{figure}[htbp]
    \begin{subfigure}[b]{.48\textwidth}
         \centering
         \includegraphics[width=\textwidth]{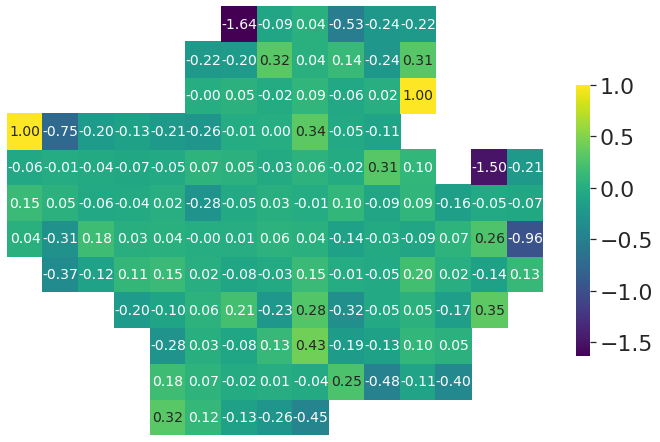}
         \caption{Difference of the \emph{mean} degree of each tile between the configuration with $f_u=\fit$ and the configuration with $f_u\equiv 1$.}
         \label{fig:firenze_degmean}
    \end{subfigure}
    \hfill
    \begin{subfigure}[b]{.46\textwidth}
         \centering
         \includegraphics[width=\textwidth]{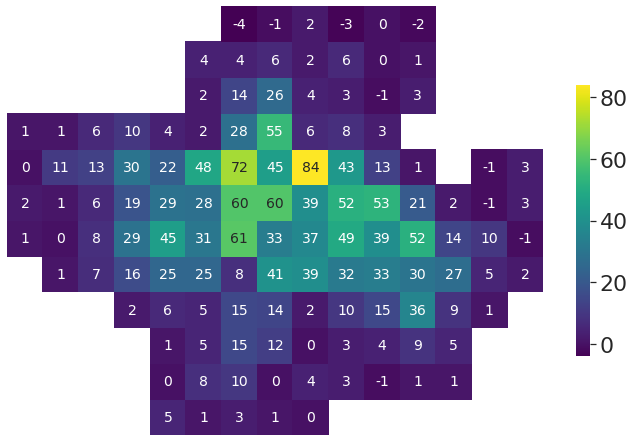}
         \caption{Difference of the \emph{maximum} degree of each tile between the configuration with $f_u=\fit$ and the configuration with $f_u\equiv 1$.}
         \label{fig:firenze_degmax}
    \end{subfigure}

    \caption{[\textbf{Florence}] Impact of switching from $f_u \equiv 1$ to  $f_u\sim\fit$ on the degree distribution of each tile.
    In both cases, $\mu=10$ and $D(u,v)=d(u,v)^{-2}$.}
    \label{fig:heatmaps_degmax}
\end{figure}

Finally, to assess the impact of $f_u$, we consider two different configurations having both $\mu=10$ and $\beta=2$, but one with $f_u\sim\fit$ and the other with $f_u\equiv 1$.
Figure \ref{fig:firenze_degmean} shows the difference of the mean degree per tile between these two configurations, which results being negligible across the whole city.
Figure \ref{fig:firenze_degmax} instead shows the difference of the maximum degree per tile, which is huge for central and densely populated areas.
In essence, for fixed $\mu$, the distance function $D$ governs the average degree of each tile, whereas the fitness scores $f_u$ has a major impact on its variance, as already emerged from Figure~\ref{fig:degree_distribution_firenze} for the whole graph.
If $f_u\equiv1$ the degree distribution within a single tile is almost flat, except for the age-based variance induced by the social mixing matrix $S$.
With a long-tailed distribution for $f_u$ we instead obtain individuals with different sociability inside each tile.

\section{Discussion and Conclusions} 
\label{sec:discussion}

We have defined and implemented a probabilistic model of the strong social ties binding the population of a given territory, organized into households.
Despite the abundance of related empirical and modeling studies, there is no general agreement concerning the mechanisms behind the formation of urban social networks.
Our model is therefore designed on top of just a few clear assumptions: (i) not all individuals are equally sociable; (ii) the geographical distance and the age difference play a role in the probability that two individuals become friends; (iii) we shall make use of all available data, bearing in mind their shortcomings.
An overview of the main features of the model can be found in~\cite{guarino2021model}, where the potential of the proposed framework is confirmed by the results of a set of epidemic simulations on the obtained network.
In this paper we instead provided a detailed analysis and an extensive experimental validation to assess the robustness and the flexibility of our model.

The main goal of this work is permitting the recreation of synthetic social networks in the common circumstance where aggregated demographic data and some estimate of the age mixing patterns are the only available information. 
The ubiquitous use of online social networks and wearable sensors offers the opportunity to analyze networks that span the globe, but the extent to which they can be used to track real geographical networks and infer relationships is still an open question.
On the other hand, social surveys are rarely available and intrinsically limited in the size of the reconstructed network.
Data related to mobile (and/or landline) phone calls may seem a good compromise, but they are also difficult to acquire, often disaggregated among several operators, and variable in terms of pervasiveness and geographical resolution.
Our tool addresses these issues providing a way to simulate a population within an arbitrary territory, whose individuals may be positioned with (almost) arbitrary precision, and for which the social mixing patterns can be inherited from any already existing dataset.

We extensively evaluated the resulting social network and its dependence on system parameters, considering three Italian cities that differ for both the size of the population and the geography of the territory.
We found that setting the average number of friends to 5 is sufficient to obtain a giant component that almost spans the entire network, even in the absence of household edges.
Age and proximity based homophily do provide the intended benefits: imposing data-driven age-based mixing patterns is critical to guarantee the internal cohesion of single age-groups -- in particular, of young people; using the real population density and penalizing physically long edges prevents a Poisson-like degree distribution; both improve the assortativity and the transitivity of the network.
The clustering coefficient, however, is rather low in the friendship layer of our graph, albeit the whole social network has strong clustering thanks to the complete subgraphs used to represent data-driven households.
If we introduce a variable (specifically, Lognormally distributed) sociability we obtain the often desired heavy-tailed degree distribution.
This helps connecting peripheral areas to the core of the network, especially when the average degree is small.
However, sociable hubs tend to concentrate in densely populated areas, with the combined effect of exacerbating the correlation between favorable positioning and degree. 
The configurations that favor the rise of large hubs (\emph{i.e.}, Lognormal social fitness and strong penalization of long edges) slightly worsen the global transitivity, but they improve the average local transitivity.
If we increase the average number of friends, most of the network distributes in just a few giant communities, a phenomenon that is amplified by a weak penalization of long edges and not affected by the distribution of the social fitness. 
Almost regardless of their size, the communities tend to have a large spatial extension, even though the average distance of their members is small.
Finally, by controlling the penalization of long edges we can not only control the distribution of the physical distance between adjacent nodes, but also the dependence of the average degree in a tile upon the position of the tile.
The variability of the degree internal to a tile is however entirely controlled by the social fitness.

Our model is intrinsically conditioned by the meaning we give to the word ``friendship'', which is not only related to the value we assign to the average number of friends $\mu$.
For instance, the experimentally measured dependence of $u$'s average degree upon the density of vertices at different distances from $u$ was already predicted by (\ref{eq:average_degree_u}). 
However, we mostly designed this simulator as a tool to study and possibly forecast interactions, \emph{e.g.} for use in computational epidemiology, even when accurate data upon such interactions are not available.
When the friendship network is used to infer the likelihood of physical contacts, it may be reasonable to assume that living in a high density area implies being part of a dense subnetwork of friends that have a greater chance to physically meet. 
In any case, (\ref{eq:average_degree_u}) also suggests that corrections to our model are possible to make it less sensitive to population density, for instance by imposing a limit to the total sociability of each tile or using a density-aware $D$ such as the rank-based model defined in~\cite{liben2005geographic}.
Another related element of discussion is the identification of the most suitable distribution for the social fitness parameter, or even whether this parameter should be used in the first place or not.
While a heavy-tailed degree distribution characterizes many real-world networks~\cite{newman2010networks}, it is widely acknowledged that even sociable human beings can establish a limited amount of strong relationships~\cite{duck1991friends, krackhardt2003strength}.
In other words, since our urban social graph models strong ties, whether a heavy-tail effect should be present and to what extent, is open to interpretation.
A Lognormal fitness seems to yield the typical degree distribution of many real-world networks, but setting $f_u\equiv1$ still guarantees that the maximum degree of the network is 5 to 6 times greater than the mode of the distribution, which may be preferable in many practical cases.
Finally, by not incorporating any preferential attachment mechanism other than proximity and age-based homophily, our friendship graph model may fail to capture some of the typical features of friendship networks.
In particular, our simulations highlight a limited tendency of friends to create triangles, which can be only partially ascribed to the low average degree considered.
In this sense, the model may benefit from the usage of a fine-grained age-stratification, to intensify the internal cohesion of all age groups, or from the definition of age-specific penalties for long edges, to foster triangles in certain (\emph{e.g.}, school age) groups.
Exploring all possible corrections to the model and providing a final answer to these and other similar questions is way beyond the scope of the present work.
By making the simulator parametric and releasing it to the public as open source software, we hope to stimulate the interest of a wide audience of users which may adjust the simulator to their needs and possibly contribute to its further development. 


\section*{Availability of data and materials}
The code and data used in this paper for the creation of the urban social network are publicly available under the GPL v3 at \href{https://gitlab.com/cranic-group/usn}{gitlab.com/cranic-group/usn}.

\section*{Acknowledgment}
The authors thank the municipality of Florence for the kind support provided and Francesca Colaiori for useful discussions.

\bibliographystyle{unsrt}
\bibliography{biblio}


\appendix
\section{Population Synthesis: Territory and Households}\label{app:population}

\begin{figure}[htbp]
    \centering
    \begin{subfigure}{.35\textwidth}
         \centering
         \includegraphics[width=\textwidth]{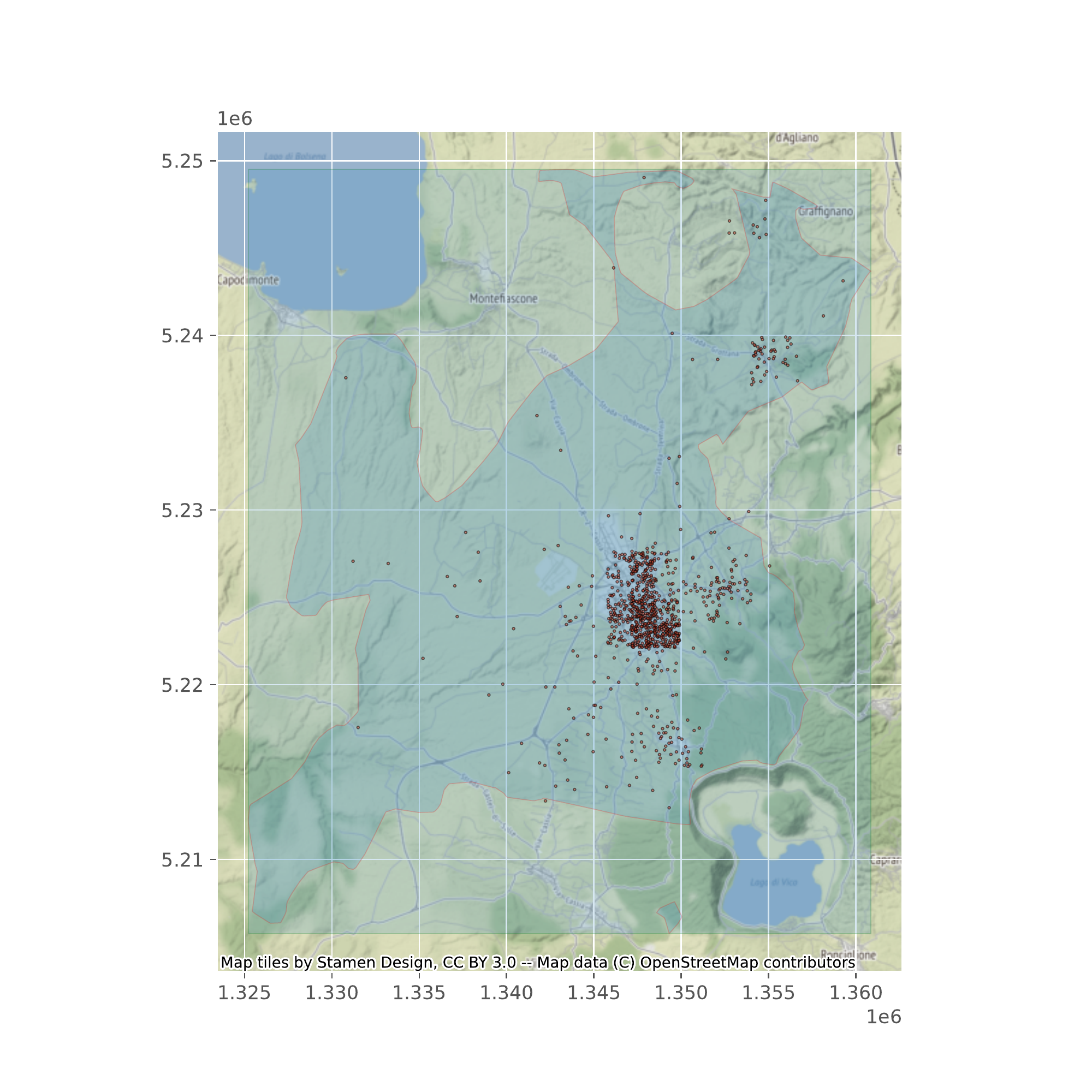}
         \caption{Random sample of 1000 people on the territory of Viterbo.}
         \label{fig:viterbo_shape}
    \end{subfigure}
    \hspace{20pt}
    \begin{subfigure}{.5\textwidth}
         \centering
         \includegraphics[width=\textwidth]{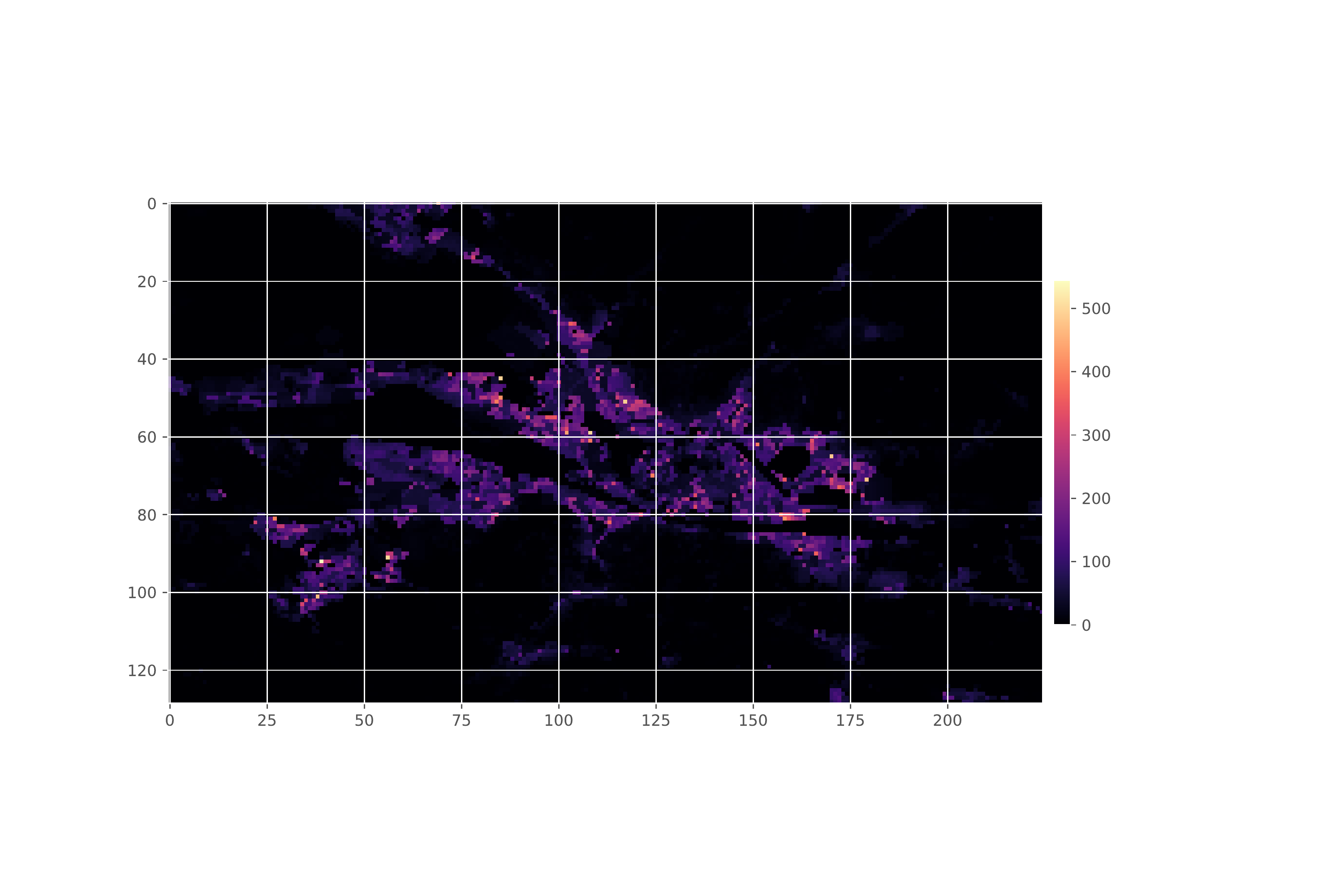}
         \caption{Population density for the city of Florence.}
    \label{fig:firenze_density}
    \end{subfigure}
    
    \caption{A graphical representation of the Territory of \textbf{Viterbo} (a) and of the population of \textbf{Florence} (b).}
    \label{fig:city_shape_app}
\end{figure}

The first step to building a synthetic, but realistic, population for a territory of interest consists in the definition of the territory itself.
We resorted to the well known OpenStreetMap database by means of the \verb|overpass| API: for each city/municipality of interest we download the shape file with its boundary.
This is used, at first, to find out the territory grid as the minimal rectangle that contains the shape and, later, to select only the tiles of the grid whose center actually falls inside the shape.
This simple technique allows to reproduce with a high accuracy the actual number of people living in the selected area.
Comparing our reconstructed population with the ISTAT data we observed a difference of $\approx 1\%$ for all the three cities of Florence, Sabaudia and Viterbo.
Figure \ref{fig:viterbo_shape} is the analogous to Figures~\ref{fig:firenze_shape} and~\ref{fig:sabaudia_shape} for Viterbo and shows the selected area with the city shape and a sample population of 1000 individuals.

As mentioned in Section \ref{sec:population}, we aim at recreating a synthetic population that is statistically indistinguishable from the real one.
To this end, we selected the following sources of information: (i) population density data from the WorldPop project~\cite{worldpop}; (ii) ISTAT age distribution data aggregated at the provincial level (for Sabaudia, we used the Province of Latina, to which the city belongs); (iii) due to the absence of analogous data at any greater resolution, we use ISTAT household composition and frequency data aggregated at the national level.
The reconstructed population density for the city of Florence is depicted in Figure~\ref{fig:firenze_density}.
The set of available household types and roles based on ISTAT data are reported in Table~\ref{tab:households}, where ``various'' comprehends all possible cases not included in the previous instances (\emph{e.g.}, non-partner adults living together).
It is worth to underline that acquiring all data from the same source was unfortunately impossible.
In particular, albeit WorldPop makes available an age-stratified population density, we verified that using those data induces an age distribution that is inconsistent with ISTAT data on household composition.

Based on the collected data we extract the tile label $t_u$ and the age label $g_u$ for each individual $u$ of the population, as defined in Section~\ref{sec:population}.
We also assign a role to each agent based on the joint-distribution of age and household roles provided by ISTAT, \emph{i.e.}, conditioning on $g_u$ when we draw $u$'s role $r_u$.
We then generate the households with the algorithm described in Section~\ref{sec:households}.

\begin{figure}[htbp]
    \centering
    \begin{subfigure}[b]{.49\textwidth}
         \centering         \includegraphics[width=\textwidth]{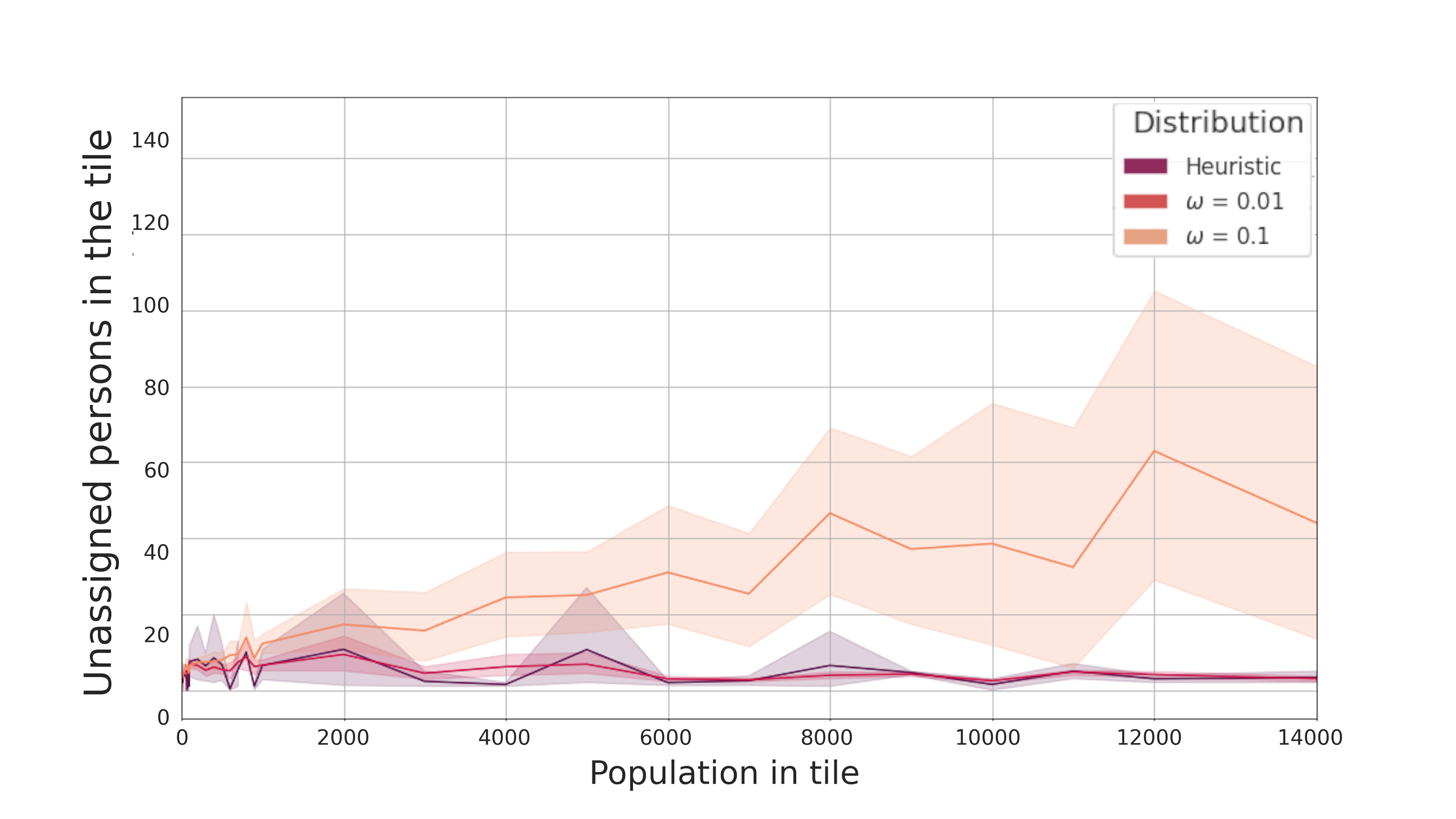}
         \caption{Number of unassigned people \emph{vs.} population size.}
    \label{fig:undone_pop}
    \end{subfigure}
    \hfill
    \begin{subfigure}[b]{.49\textwidth}
         \centering         \includegraphics[width=\textwidth]{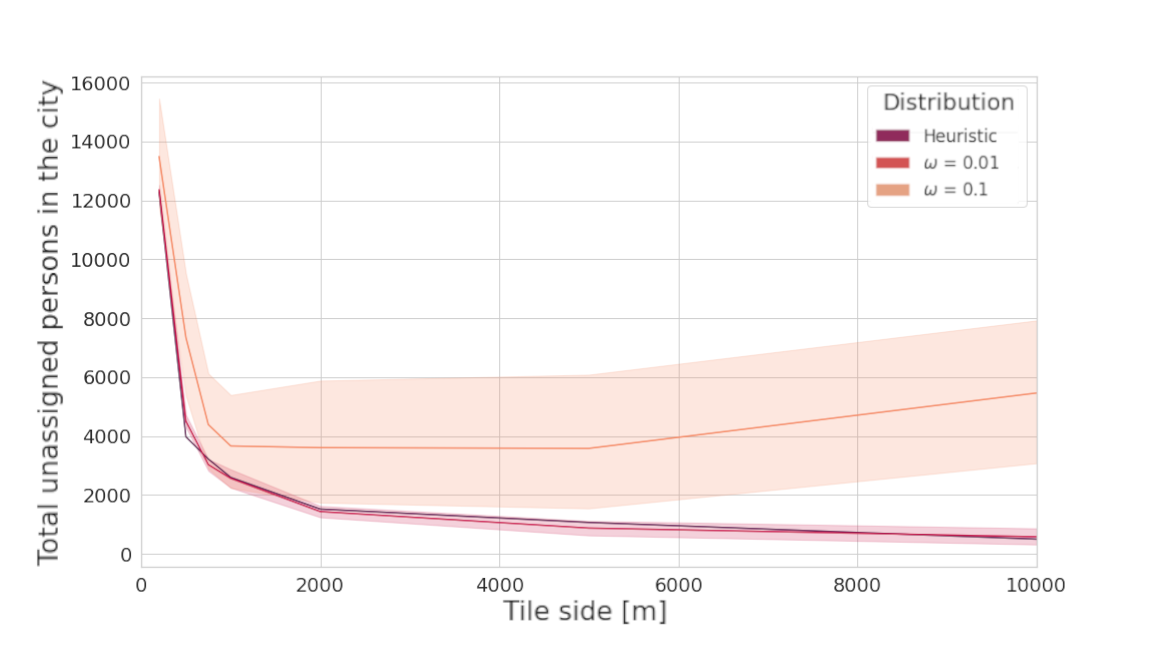}
         \caption{Number of unassigned people \emph{vs.} tile size (Florence).}
    \label{fig:undone_tile}
    \end{subfigure}
    \caption{[\textbf{Florence}] Number of people not assigned to any household in different simulations, as the size of the population, the side of each tile and the noise added to the age distribution vary.}
    \label{fig:undone}
\end{figure}

We assess the robustness of the algorithm under perturbations to the age distribution, by considering independent relative variations -- either positive or negative -- to the probability of each age-group, controlled by a real-valued parameter $\omega$.
Formally, if $\pi_i=\frac{|V_i|}{N}$ is the data-driven probability of age group $i$, the perturbed value $\pi_i^*$ is obtained as follows: we draw $\epsilon\sim\frac{1}{2}\mathcal{N}(\omega,\omega^2)+\frac{1}{2}\mathcal{N}(-\omega,\omega^2)$, where $\mathcal{N}(\lambda,\sigma^2)$ denotes the normal distribution with mean $\mu$ and variance $\sigma^2$; we set $\pi_i^*=\pi_i\cdot(1+\epsilon)$; we normalize $\pi_i^*$ so that $\sum_{i=1}^n\pi_i^*=1$.
For both $\omega=0.01$ and $\omega=0.1$ we pick $20$ different perturbed age distributions and we evaluate the quality of the resulting households under perturbation to the input data.

First, we address the problem that our heuristics does not guarantee that all individuals of the population are assigned to some household.
For instance, this may happen if in any of the tiles the number of ``single parents'' exceeds the number of ``children of a single parent''.
We therefore count the number of unassigned people in two different sets of simulations: (i) we consider a single tile and vary the number of individuals in the tile (see Figure~\ref{fig:undone_pop}, with the range for the population size taken from the city of Florence); or (ii) we consider the whole city and vary the side of the tiles used to define the grid (see Figure~\ref{fig:undone_tile}, again for the case of Florence).
Figure~\ref{fig:undone_pop} shows that the number of people left out of all households is negligible, provided that the number of individuals in each tile is large enough to guarantee that all roles are sufficiently represented.
Accordingly, Figure~\ref{fig:undone_tile} shows that the number of unassigned people in the city drops very fast as soon as the tiles are large enough to contain, on average, a sufficient number of people; this number approaches zero as the tile size grows, \emph{i.e.}, as the population of each tile approaches the total population $N$.
The quality of the heuristics decreases when the noise volume $\omega$ grows, thus highlighting the importance of internal consistency in the data.
However, the plots show that the number of unassigned people stays safely below 1\% of the total population even in the presence of (reasonable) fluctuations in the input data.

\begin{figure}[htbp]
    \centering
    \begin{subfigure}[b]{.43\textwidth}
         \centering         \includegraphics[width=\textwidth]{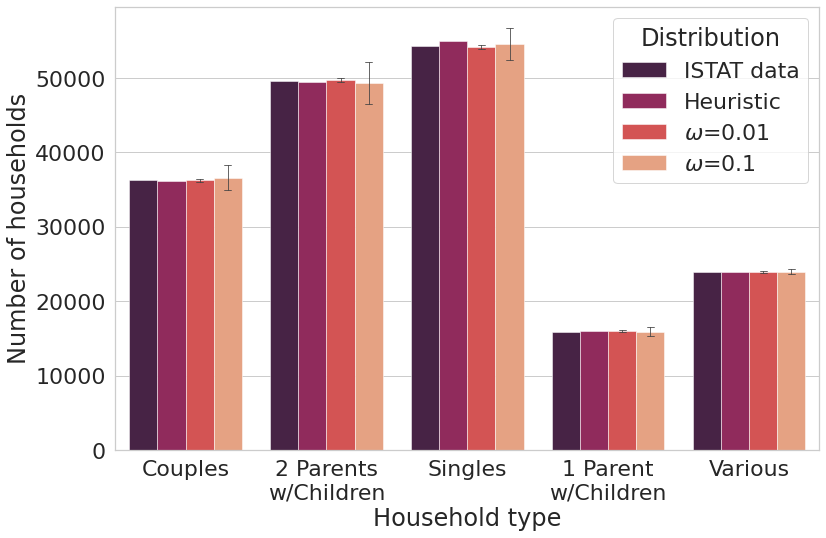}
         \caption{Household distribution per type}
    \label{fig:familydis_type}
    \end{subfigure}
    \hfill
    \begin{subfigure}[b]{.52\textwidth}
         \centering         \includegraphics[width=\textwidth]{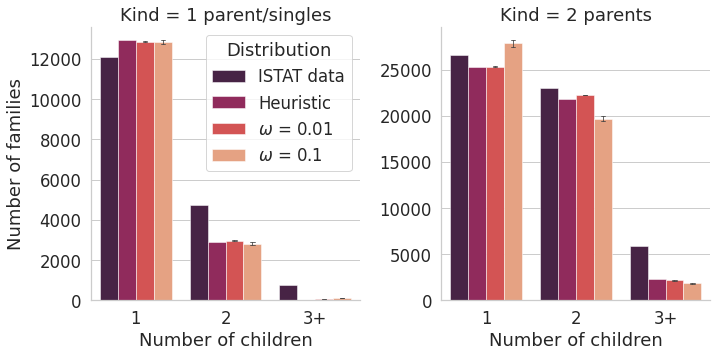}
         \caption{Distribution of the number of children in families.}
    \label{fig:familydis_num}
    \end{subfigure}
    
    \caption{[\textbf{Florence}] Distribution of household types and number of children obtained with our heuristic, with and without noise on the input data, compared with ISTAT data.}
    \label{fig:familydis}
\end{figure}

Next, we focus on the distribution of households by type and by number of children.
In Figure~\ref{fig:familydis_type} we plot the number of households of each type obtained for the city of Florence with our heuristics, on both clean and noisy data, compared with an estimate based on ISTAT data.
In the plot, ``ISTAT data'' refers to the weighted average of the conditional probability of being the head of a specific household type given the age, computed through the law of total probability from ISTAT data.
For instance, the number of households of type ``1 Parent w/Children'' is computed as the weighted average of the role (single-parent, parent).
This approach allows to gain an estimate based on ISTAT data that takes into account the specific age distribution of Florence.
The figure shows that our results do match the expected distribution almost perfectly and that the heuristics is very robust with respect to noise.
In Figure~\ref{fig:familydis_num} we instead show the distribution of the number of children for the types of household composed of parents and children.
In this case, ``ISTAT data'' is obtained from the national aggregate distribution of the number of children per family; to the best of our knowledge, equivalent data are unfortunately unavailable for the province of Florence.
The two distributions might significantly differ in light of the demographic differences between the residents of the city and the entirety of the Italian population.
This difference may explain, at least in part, the divergence between the ISTAT data and the results of our simulations.
In any case, the results look reasonable and are, again, stable with respect to perturbations in the input data.

\section{Social Contact Data}\label{app:social_contact_data}

In our urban social graph, the probability $\Pr[u,v]$ of an edge connecting individuals $u$ and $v$ is given by (\ref{eq:edge_prob}).
The dependence of $\Pr[u,v]$ upon the age groups $g_u$ and $g_v$ is determined by the matrix $S$ which is defined by using the available social contact data as follows.

\paragraph{Initial data}
We assume that social contact patterns among individuals of our age-stratified population are available in the form of a $n\times n$ social contact matrix $\Gamma=\{\gamma_{i,j}\}_{i,j\in\{0,\ldots,n-1\}}$, where $n$ is the number of considered age groups.
Following a systematic literature review, the project SOCRATES~\cite{willem2020socrates} recently released an online tool\footnote{\url{https://lwillem.shinyapps.io/socrates_rshiny/}} to extract and analyze social contact patterns for a wide range of countries based on the best publicly available survey datasets.
The tool allows to select a number of parameters upon which social contact matrices are usually dependent, such as age breaks, gender, day of the week, duration or location of the contact.
Once the preferred dataset and the aforementioned parameters are selected, the tool produces the desired square matrix with $\gamma_{i,j}$ measuring the average number of daily contacts. 
For the scope of this paper, we relied upon the SOCRATES tool (v1.32), using the Polymod~\cite{Mossong-2008} dataset for Italy and the following set of parameters:
\begin{itemize}
    \item Age breaks: 0,18,35,65 (the choice was dictated by the available data from ISTAT to build households);
    \item Type of day: all contacts;
    \item Contact duration: more than 15 minutes;
    \item Contact intensity: all contacts;
    \item Gender: all;
    \item Reciprocity: yes;
    \item Weigh by age: yes;
    \item Weigh by week/weekend: yes;
    \item Locations: all except Home (because we use households for home contacts).
\end{itemize}
While we have no particular reason to recommend the use of the SOCRATES tool to determine $\Gamma$, we believe it is a valuable resource that perfectly fits our needs.
More generally, the construction of a social contact matrix $\Gamma$ is a typical problem in computational social science and our simulator simply assumes that $\gamma_{i,j}$ is an estimate of the volume of interactions/relationships that an individual of age group $i$ has with any other individual of age group $j$ within a given time span.
It goes without saying that the type of contacts described by $\Gamma$ is partially reflected on the final structure of the graph.

\paragraph{Data preprocessing}
In general the matrix $\Gamma$ may be not symmetric, because $\gamma_{i,j}$ depends upon the habits of the individuals of group $i$, whereas $\gamma_{j,i}$ on the habits of group $j$.
Using graph terminology, in the bipartite graph composed of individuals of groups $i$ and $j$, $\gamma_{i,j}$ and $\gamma_{j,i}$ are an estimate of the average degree of vertices of type $i$ and $j$, respectively. 
Even if asymmetric, the matrix should be consistent, in the sense that its entries should guarantee the \emph{reciprocity} of contacts.
At the population level, reciprocity means that in the bipartite graph the total number of edges that ``exit'' group $i$ must be equal to the total number of edges that ``enter'' group $j$, \emph{i.e.}, that $|V_i|\cdot \gamma_{i,j} = |V_j|\cdot \gamma_{j,i}$.
Since survey data rarely meet this requirement, it is standard practice to introduce a reciprocity correction~\cite{10.1093/aje/kwj317,klepac2020contacts} by taking the arithmetic mean $\frac{1}{2}\left(\gamma_{i,j}\cdot|V_i| + \gamma_{j,i}\cdot|V_j|\right)$ as an estimate of the total number of edges in the bipartite graph.
When $i=j$, instead, the total number of intra-group contacts for group $i$ is given by $\frac{1}{2}\left(\gamma_{i,i}\cdot |V_i|\right)$, because the graph is \emph{not} bipartite.
This leads to a group-adjacency matrix $A=\{\alpha_{i,j}\}$ defined as follows:
\begin{equation}\label{eq:adjacency}
\alpha_{i,j} = \begin{cases} \frac{1}{2}\left(\gamma_{i,j}\cdot|V_i| + \gamma_{j,i}\cdot |V_j|\right) &\text{if } i\neq j \\ 
\frac{1}{2}\left(\gamma_{i,i}\cdot |V_i|\right) &\text{if } i=j\end{cases}
\end{equation}
Finally, we divide each $\alpha_{i,j}$ by the total number of potential edges $m_{i,j}$ (see (\ref{eq:m_ij}) in Section~\ref{sec:friendship}) to obtain the matrix $S=\{s_{i,j}\}$ of inter-group edge frequencies:
$$s_{i,j} = \frac{\alpha_{i,j}}{m_{i,j}} = \begin{cases} \frac{1}{2}\left(\frac{\gamma_{i,j}}{|V_j|} + \frac{\gamma_{j,i}}{|V_i|}\right) &\text{if } i\neq j \\ 
\frac{\gamma_{i,i}}{|V_i|-1} &\text{if } i=j\end{cases}$$

It is worth underlining that different sub-territories of the same country may have different age-group ratios and therefore require different corrections.
The reciprocity correction, implicit in  (\ref{eq:adjacency}), is therefore necessary regardless of whether the matrix $\Gamma$ had already been corrected beforehand.
For instance, the SOCRATES tool implements the reciprocity correction but with national age-group statistics.
If the local and national age-group statistics are identical, it is easy to verify that a double correction is useless yet harmless.

\paragraph{Age-homogeneous mixing}
In the absence of any age-based homophily, we have $\gamma_{i,j} = h_i\cdot \frac{|V_j|}{N-1}$ if $i\neq j$ and $\gamma_{i,i} = h_i\cdot \frac{|V_i|-1}{N-1}$ if $i=j$, where $h_i$ is the average number of (\emph{e.g.}, daily) contacts of the individuals of group $i$.
This leads to $s_{i,j} = \frac{h_i}{N-1}$ for all $j$.
If, additionally, all age-groups are ``equally sociable'', we have $h_i\equiv h$ for all $i$, leading to a constant $s_{i,j} = \frac{h}{N-1}$ for all $i,j$.
This scenario represents a condition of \emph{age-homogeneous mixing}, in which the age groups have no impact whatsoever on the edge probability -- this condition is used as a benchmark in part of the analysis presented in Section~\ref{sec:parameters}.

As highlighted in Section~\ref{sec:friendship}, our model is invariant under multiplication of $S$ by any positive constant.
This means that the value $h$ is irrelevant and that age-homogeneous mixing can be obtained by taking $S$ to be any constant matrix, such as $s_{i,j}\equiv1$.
If we choose to have $s_{i,j}$ represent the probability that a randomly chosen edge of the graph connects groups $i$ and $j$, age-homogeneous mixing corresponds to $s_{i,j}=\frac{1}{n^2}$ for all $i,j$, so that $\sum_{i,j}s_{i,j}=1$.

\section{Results for the City of Viterbo}\label{app:viterbo}

The same analysis discussed in Sections~\ref{sec:parameters} and~\ref{sec:USG} for the city of Florence has been carried out for the cities of Viterbo and Sabaudia.
Except for a few details, mostly ascribable to differences in the demographics and in the geography of the considered territories, the same observations and conclusions drawn for the case of Florence apply to these other two cities, as confirmed by the results shown in this and the next appendix.
In the following, we report all plots and tables for the city of Viterbo, which are by all means analogous to those already presented for Florence.

\begin{table}[htbp]
\caption{[\textbf{Viterbo}] Percentage of nodes of the graph that belong to the giant component, on average, for the friendship graph $G_F$ and the entire social graph $G$, as $\beta$, $f_u$ and $\mu$ vary.
}
\label{tab:viterboGiant}
\centering
\begin{tabular}{llrrrr}
\addlinespace[-\aboverulesep]
\cmidrule[\heavyrulewidth]{3-6}
& & \multicolumn{2}{c}{$\mu=1$} & \multicolumn{2}{c}{$\mu=5$}\\
$\beta$ & $f_u$ & $G_F$ & $G$ & $G_F$ & $G$ \\
\midrule
0.5 & 1                     & 3.2\%  &  76.1\% & 99.2\% & 99.7\% \\
0.5 & $1+\LN(\ln(2), 0.25)$ & 19.4\% &  75.0\% & 98.1\% & 99.4\% \\
2 & 1                       & 15.5\% &  73.1\% & 98.3\% & 99.5\% \\
2 & $1+\LN(\ln(2), 0.25)$   & 24.0\% &  72.7\% & 97.2\% & 99.0\% \\
\bottomrule
\end{tabular}
\end{table}

\begin{figure}[htbp]
    \centering
    \includegraphics[width=.95\textwidth]{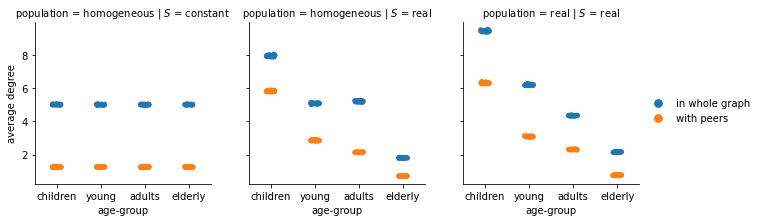}
    \caption{[\textbf{Viterbo}] Average degree of the individuals of each age group, in the whole friendship graph $G_F$ and with their peers, under different configurations all with $\mu=5$.}
    \label{Figure:deg_viterbo_avg5}
\end{figure}

\begin{figure}[htbp]
    \centering
    \begin{subfigure}[b]{.35\textwidth}
         \includegraphics[width=\textwidth, trim={50 50 50 50},clip]{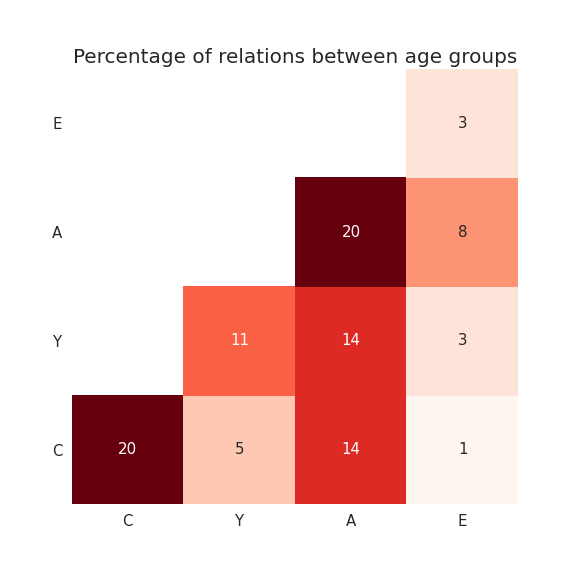}
         \caption{Without household edges.}
         \label{fig:viterbo-g2g-hhFalse}
    \end{subfigure}
    \hspace{20pt}
    \begin{subfigure}[b]{.35\textwidth}
         \includegraphics[width=\textwidth, trim={50 50 50 50},clip]{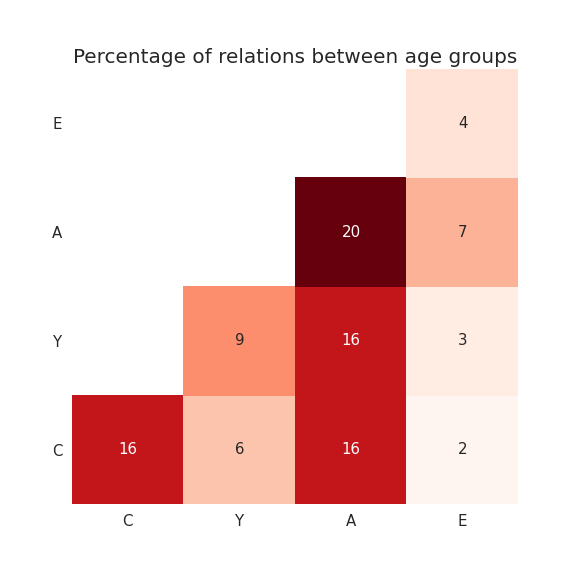}
         \caption{With household edges.}
         \label{fig:viterbo-g2g-hhTrue}
    \end{subfigure}
    \caption{[\textbf{Viterbo}] Percentage of edges between age groups in a configuration with data-driven population and age-based mixing, for $f_u\sim\fit$, $\mu=5$ and $D(u,v)=d(u,v)^{-2}$.}
    \label{fig:g2g_matrix_viterbo}
\end{figure}

\begin{table}[htbp]
\caption{[\textbf{Viterbo}] Main features of the friendship graph $G_F$ as the population type, $D(u,v)$ and $\mu$ vary, for constant $f_u$ and $S$. $\dist$ is the average path length, $C$ is the global clustering coefficient, $\rho$ is the degree assortativity, ``\# comp.'' denotes the number of connected components, ``giant \%'' denotes the percentage of nodes in the giant component.}
\label{tab:ViterboGeoDist}
\centering
\begin{subtable}{\textwidth}
\centering
\caption{Friendship network for $\mu=5$. Expected values for an ER graph with $\mu=5$: $\dist \approx 6.9$, $C=7.5\mathbf{e-}05$.}
\begin{tabular}{llrrrrr}
\addlinespace[-\aboverulesep]
\cmidrule[\heavyrulewidth]{3-7}
& & \multicolumn{5}{c}{$\mu=5$} \\
Population  & $D(u,v)$ & $\dist$ & $C$ & $\rho$ & \# comp.  & giant \% \\
\midrule
homogeneous & 1               &   7.08   &   8.2e-05     & -5.06e-04  &  458.9  & 99.3\%     \\
homogeneous & $d(u,v)^{-0.5}$ &   7.04   &   8.2e-05     &  0.005     &  546.9  & 99.2\%    \\
homogeneous & $d(u,v)^{-2}$   &   7.15   &   7.95e-04    &  0.13      &  1022.1 & 98.4\%    \\
real        & 1               &   7.08   &   8.2e-05     & -1.91e-04  &  463.1  & 99.3\%    \\
real        & $d(u,v)^{-0.5}$ &   6.94   &   9.2e-05     &  0.02      &  1053.9 & 98.4\%    \\
real        & $d(u,v)^{-2}$   &   6.44   &   0.00025     &  0.2       &  9137.3 & 84.0\%    \\
\bottomrule
\end{tabular}
\end{subtable}

\vspace{10pt}

\begin{subtable}{\textwidth}
\centering
\caption{Friendship network for $\mu=10$. Expected values for an ER graph with $\mu=10$: $\dist \approx 4.82$, $C=1.5\mathbf{e-}04$.}
\begin{tabular}{llrrrrr}
\addlinespace[-\aboverulesep]
\cmidrule[\heavyrulewidth]{3-7}
& & \multicolumn{5}{c}{$\mu=10$}\\
Population & $D(u,v)$ & $\dist$ &
$C$ & $\rho$ & \# comp.  & giant \% \\
\midrule
 homogeneous & 1               &  5.08  &   1.52e-04   &  -5.31e-04  &  4.1     &  100.0\%    \\
 homogeneous & $d(u,v)^{-0.5}$ &  5.06  &   1.56e-04   &   0.011     &  8.0     &  100.0\%    \\
 homogeneous & $d(u,v)^{-2}$   &  5.23  &   0.0016     &   0.22      &  42.9    &  99.9\%  \\
 real        & 1               &  5.08  &   1.56e-04   &  -9.15e-04  &  4.8     &  100.0\%   \\
 real        & $d(u,v)^{-0.5}$ &  5.04  &   1.69e-04   &   0.04      &  55.8    &  99.9\%   \\
 real        & $d(u,v)^{-2}$   &  5.17  &   0.0005     &   0.31      &  4461.6  &  92.6\%  \\
\bottomrule
\end{tabular}
\end{subtable}
\end{table}

\begin{figure}[htbp]
    \centering
    \includegraphics[width=\textwidth]{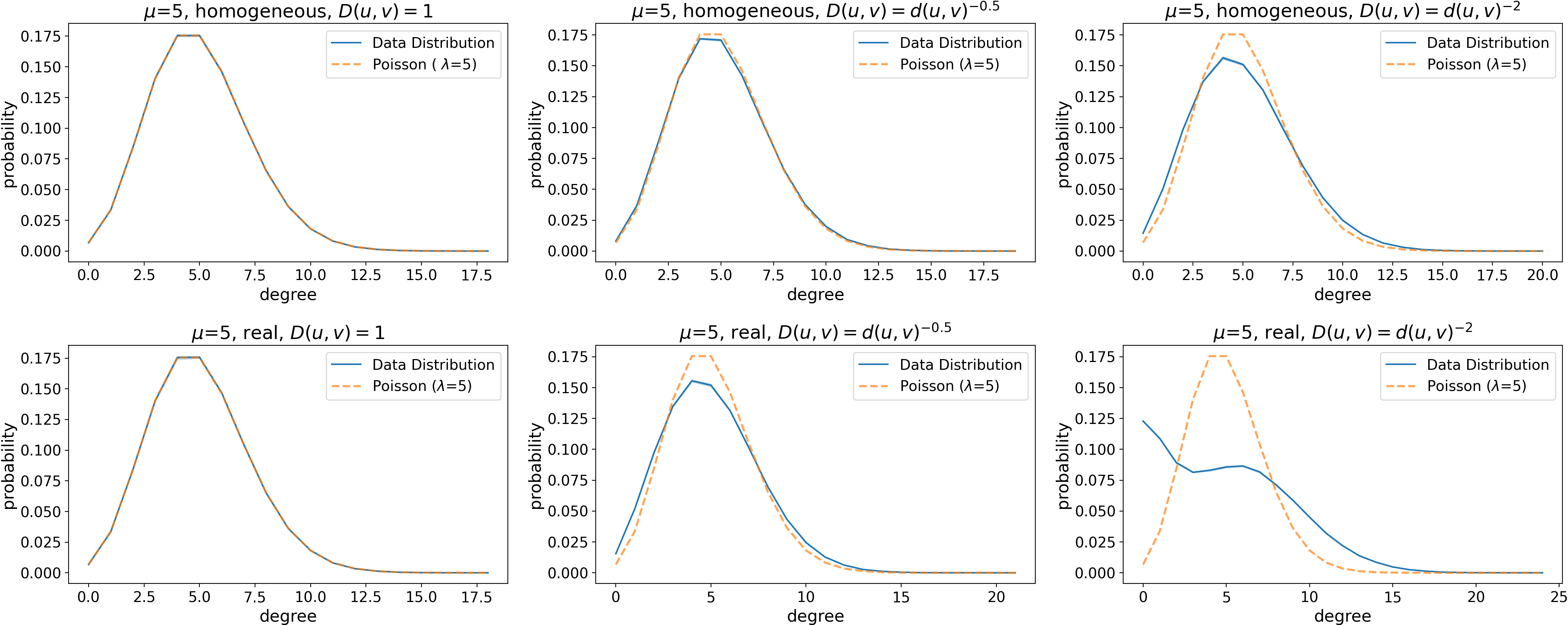}
     \caption{[\textbf{Viterbo}] Degree distributions of friendship graphs with $\mu=5$ and corresponding Poisson distributions with $\lambda=5$.
    The data distribution lines show the mean, whereas the shaded areas  around the lines represent the 95\% confidence interval.
    The Kullback–Leibler divergence between the Poisson and the empirical distribution are (top to bottom, left to right):
    8.4e-07,
    6.03e-05,
    0.001,
    7.2e-07,
    0.001,
    0.03.
    }
    \label{Figure:ViterboDegDist5}
\end{figure}

\begin{figure}[htbp]
    \centering
    \includegraphics[width=\textwidth]{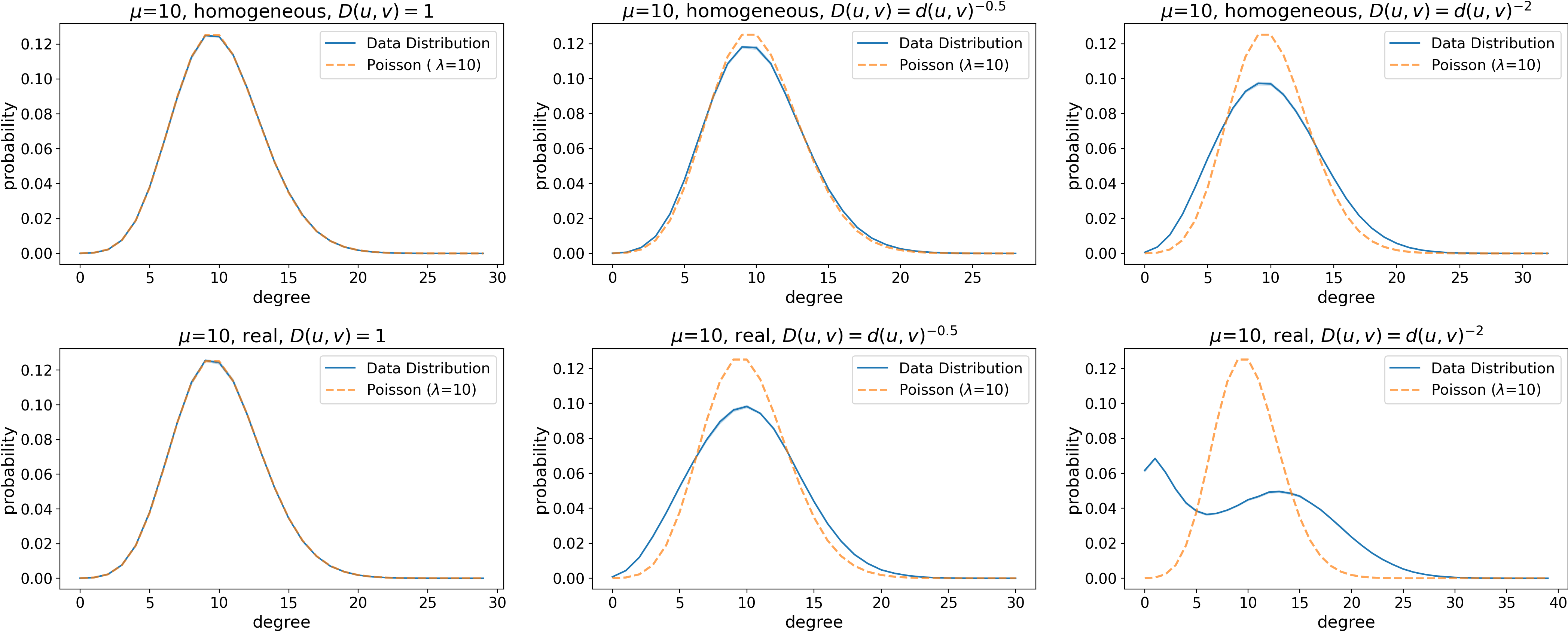}
     \caption{[\textbf{Viterbo}] Degree distributions of friendship graphs with $\mu=10$ and corresponding Poisson distributions with $\lambda=10$.
    The data distribution lines show the mean, whereas the shaded areas  around the lines represent the 95\% confidence interval.
    The Kullback–Leibler divergence between the Poisson and the empirical distribution are (left to right, top to bottom):
    1.41e-06,
    0.0002,
    0.003,
    1.003e-06,
    0.003,
    0.05.
    }
    \label{Figure:ViterboDegDist10}
\end{figure}

\begin{figure}[htbp]
    \centering
    \includegraphics[width=\textwidth]{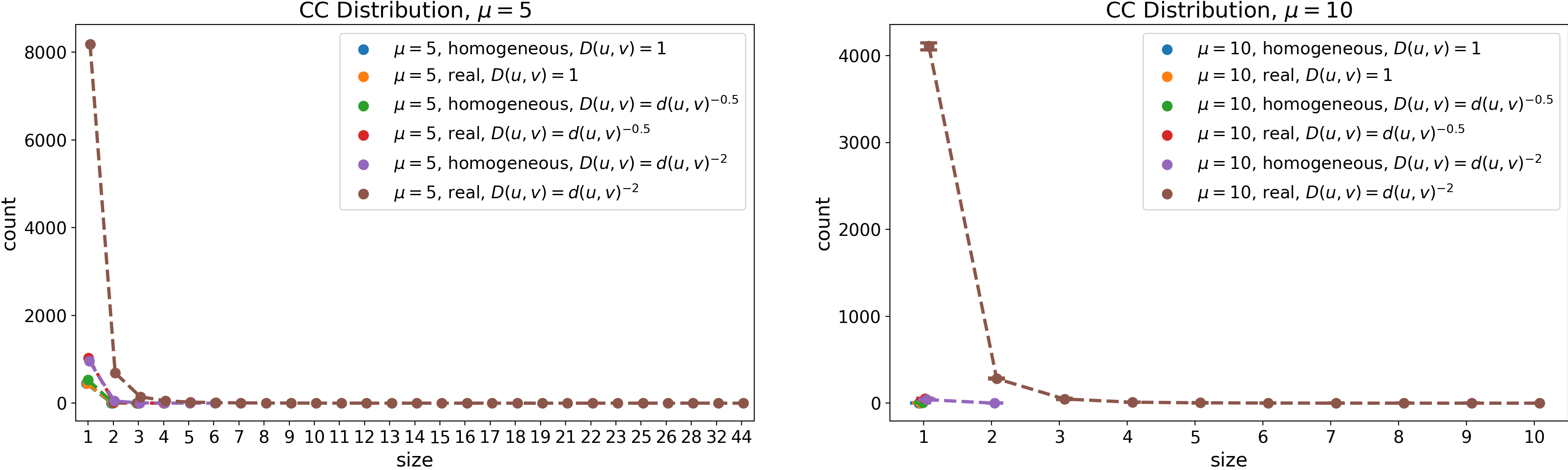}
    \caption{[\textbf{Viterbo}] Distribution of the size of the connected components other than the giant for friendship graphs with $\mu=5$ and $\mu=10$.
    The lines show the mean, whereas the error bars (with ``caps'') represent the 95\% confidence interval computed over each set of runs.}
    \label{Figure:ViterboConnComp}
\end{figure}

\begin{table}[htbp]
\caption{[\textbf{Viterbo}] Main features of the social graph as $f_u$, $\beta$ and $\mu$ vary, for data-driven population and $S$. $K=\mu+\nu$ is the average degree, $\dist$ is the average path length, $C$ and $C_{\mathrm{loc}}$ are the global and average local clustering coefficients, $\rho$ is the degree assortativity, ``\# comp.'' denotes the number of connected components, ``giant \%'' denotes the percentage of nodes in the giant component.}
\centering
\begin{subtable}{\textwidth}
\centering
\caption{Social graph for $\mu=5$.}
\label{tab:Metrics_Viterbo_5}
\begin{tabular}{llrrrrrrr}
\addlinespace[-\aboverulesep] 
\cmidrule[\heavyrulewidth]{3-9}
& & \multicolumn{7}{c}{$\mu=5$} \\
$f_u$ & $\beta$ & $K$ & $\dist$ & $C$ & $C_{\mathrm{loc}}$ & $\rho$ & \# comp.  & giant \% \\
\midrule
1      & 0.5  & 6.795 & 5.902  & 0.052 & 0.078 & 0.264  & 894.2  & 98.5 \\
1      & 2    & 6.789 & 6.081  & 0.045 & 0.130 & 0.420  & 3578.8 & 91.9 \\
$\fit$ & 0.5  & 6.788 & 5.786  & 0.048 & 0.089 & 0.207 & 1127.3  & 98.1 \\
$\fit$ & 2    & 6.781 & 5.970  & 0.041 & 0.138 & 0.320  & 3826.3 & 91.5 \\
\bottomrule
\end{tabular}
\end{subtable}

\vspace{10pt}

\begin{subtable}{\textwidth}
\centering
\caption{Social graph for $\mu=10$.}
\label{tab:Metrics_Viterbo_10}
\begin{tabular}{llrrrrrrr}
\addlinespace[-\aboverulesep] 
\cmidrule[\heavyrulewidth]{3-9}
& & \multicolumn{7}{c}{$\mu=10$} \\
$f_u$ & $\beta$ & $K$ & $\dist$ & $C$ & $C_{\mathrm{loc}}$ & $\rho$ & \# comp.  & giant \% \\
\midrule
1      & 0.5 &  11.795 & 4.723 & 0.017 & 0.029 & 0.309  & 140.2   & 99.8 \\
1      & 2   &  11.780 & 4.998 & 0.015 & 0.074 & 0.452  & 1947.3 & 96.0 \\
$\fit$ & 0.5 &  11.781 & 4.656 & 0.016 & 0.035 & 0.208  & 226.7  & 99.6 \\
$\fit$ & 2   &  11.791 & 4.924 & 0.014 & 0.080 & 0.311  & 2088.6 & 95.7 \\ 
\bottomrule
\end{tabular}
\end{subtable}
\end{table}

\begin{figure}[htbp]
    \centering
    \begin{subfigure}[b]{.48\textwidth}
         \centering
         \includegraphics[width=\textwidth]{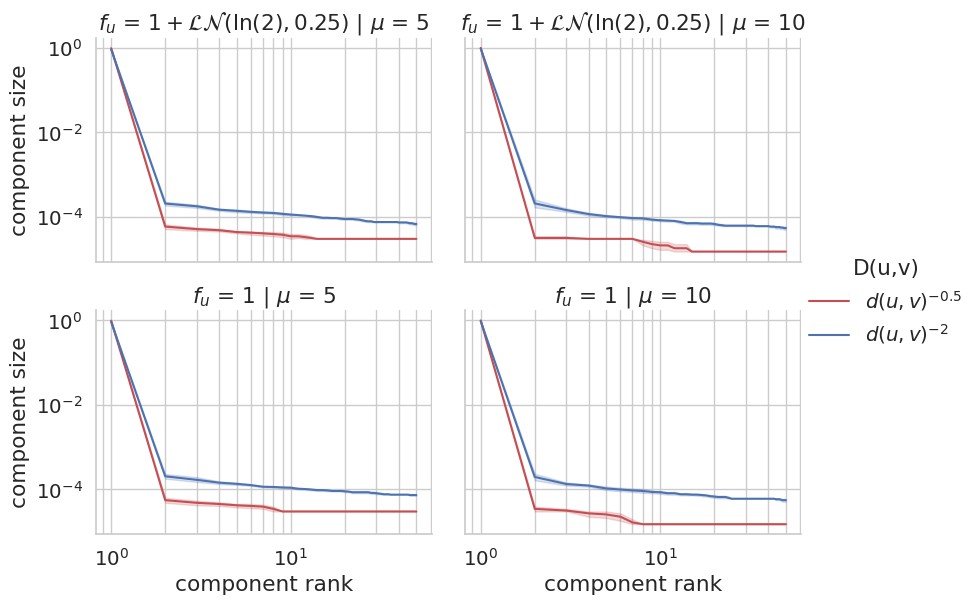}
         \caption{Size of the largest 50 connected components.}
         \label{fig:components_viterbo}
    \end{subfigure}
    \hfill
    \begin{subfigure}[b]{.48\textwidth}
         \centering
         \includegraphics[width=\textwidth]{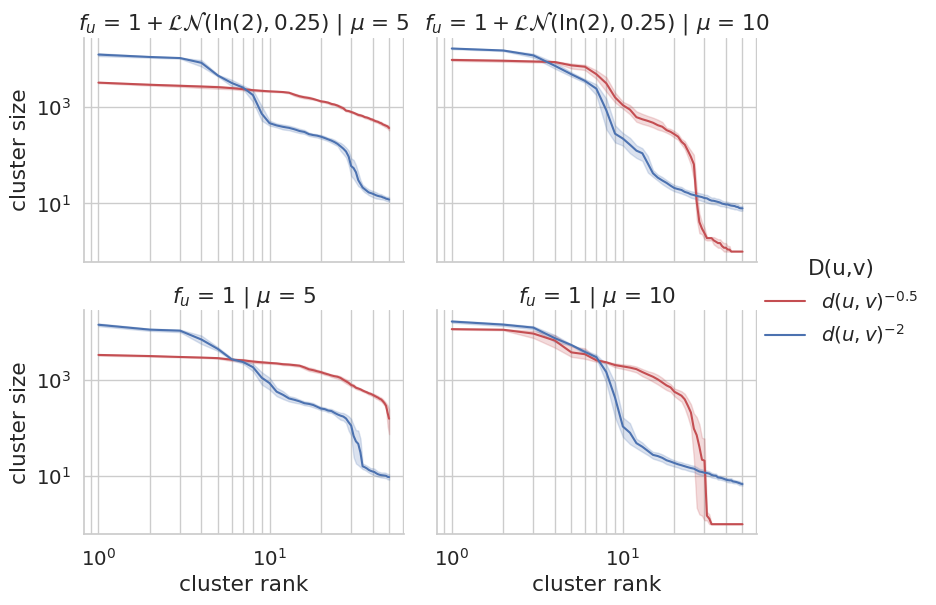}
         \caption{Size of the largest 50 clusters (Louvain).}
         \label{fig:cluster_size_viterbo}
    \end{subfigure}
    
    \begin{subfigure}[b]{.48\textwidth}
         \centering
         \includegraphics[width=\textwidth]{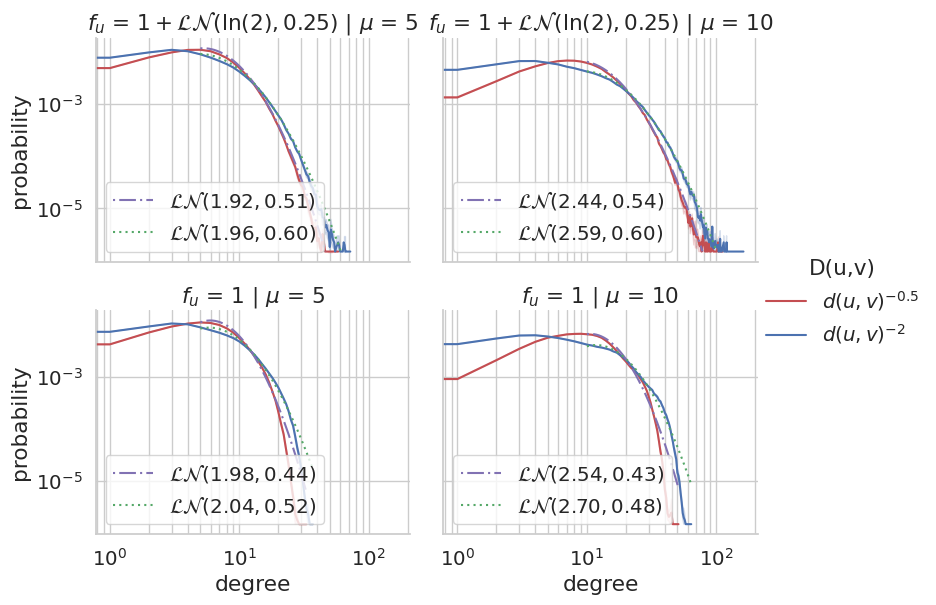}
         \caption{Degree distribution with Lognormal fit.}
         \label{fig:degree_distribution_viterbo}
    \end{subfigure}
    \hfill
    \begin{subfigure}[b]{.48\textwidth}
         \centering
         \includegraphics[width=\textwidth]{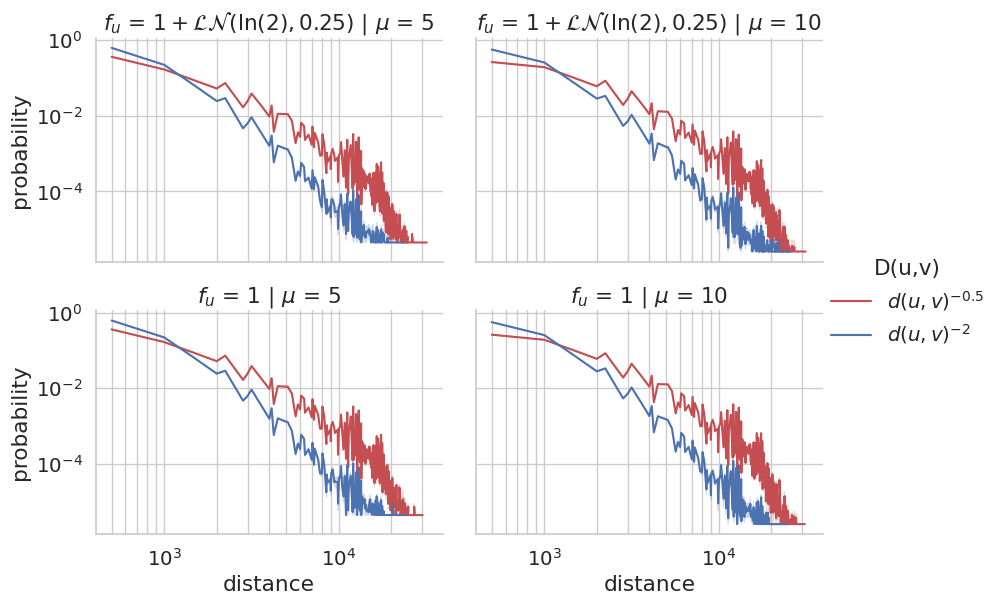}
         \caption{Geographical distance between adjacent vertices.}
         \label{fig:distance_distribution_viterbo}
    \end{subfigure}
    
    \begin{subfigure}[b]{.48\textwidth}
         \centering
         \includegraphics[width=\textwidth]{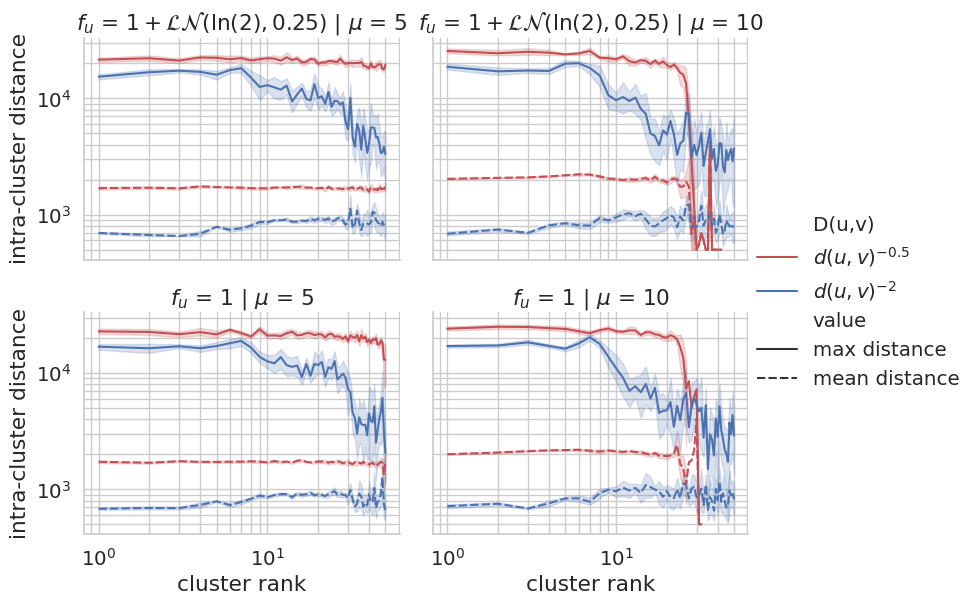}
         \caption{Mean and max intra-cluster geographical distances.}
         \label{fig:cluster_distance_viterbo}
    \end{subfigure}
    \hfill
    \begin{subfigure}[b]{.48\textwidth}
         \centering
         \includegraphics[width=\textwidth]{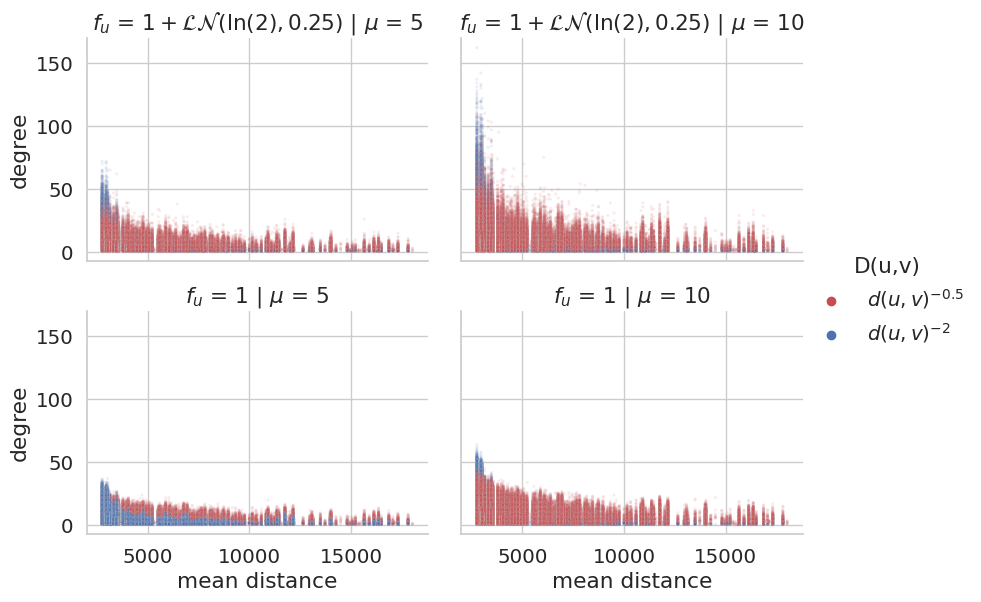}
         \caption{Mean distance to all other vertices \emph{vs.} Degree.}
         \label{fig:degree_distance_viterbo}
    \end{subfigure}
    \caption{[\textbf{Viterbo}] Overview of the urban social graph for the city of Viterbo. Each plot shows the average with confidence interval for 10 independent runs with the same configuration.}
    \label{fig:viterbo}
\end{figure}

\begin{figure}[htbp]
    \centering
    \includegraphics[width=.6\textwidth]{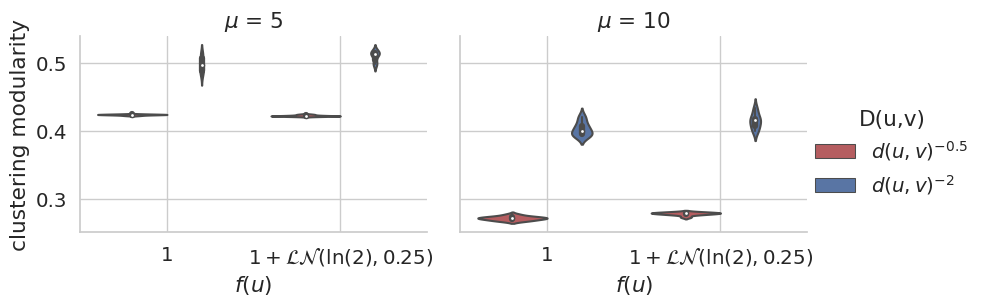}
    \caption{[\textbf{Viterbo}] Modularity of the obtained clustering structure -- distribution for 10 independent runs per configuration.}
    \label{fig:modularity_viterbo}
\end{figure}

\begin{figure}[htbp]
    \centering
    \begin{subfigure}[b]{.35\textwidth}
         \includegraphics[width=\textwidth, trim={100 100 100 100},clip]{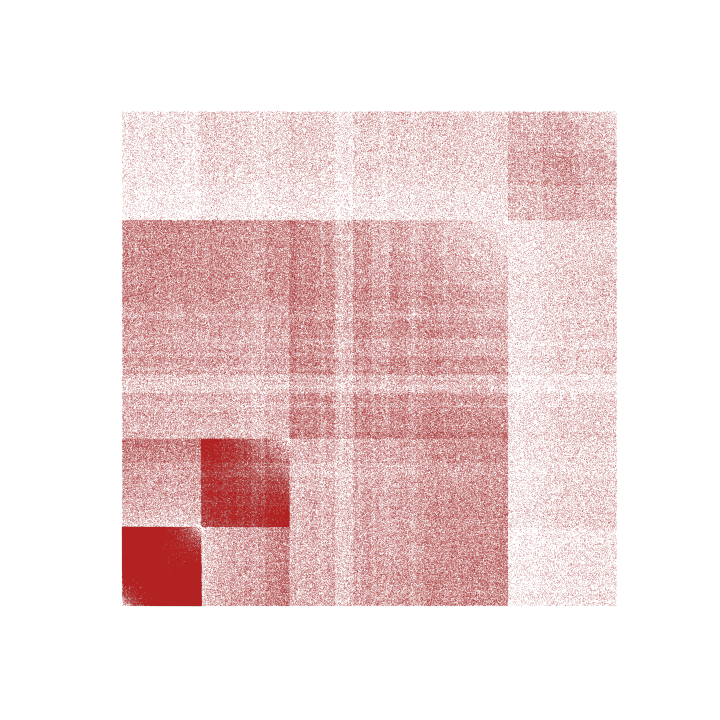}
         \caption{$\beta=0.5$.}
         \label{fig:viterbo-adj-0510}
    \end{subfigure}
    \hspace{20pt}
    \begin{subfigure}[b]{.35\textwidth}
         \includegraphics[width=\textwidth, trim={100 100 100 100},clip]{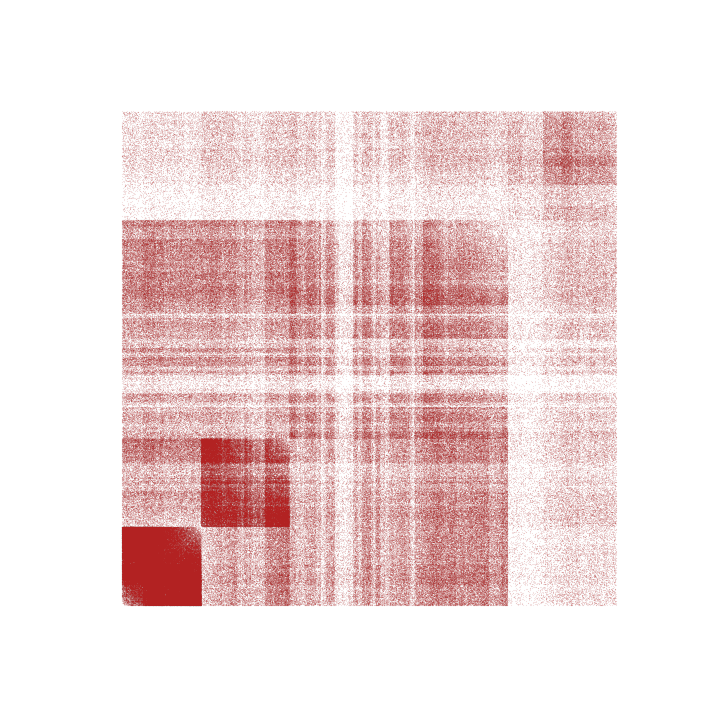}
         \caption{$\beta=2$.}
         \label{fig:viterbo-adj-210}
    \end{subfigure}
    \caption{[\textbf{Viterbo}] Adjacency matrix of the social graph with nodes (people) ordered by age-group.
    In both cases $f_u\sim\fit$, $\mu=10$ and $D(u,v)=d(u,v)^{-\beta}$, but $\beta$ varies between the two figures.}
    \label{fig:adj_matrix_viterbo}
\end{figure}

\begin{figure}[htbp]
    \centering
    \begin{subfigure}[b]{.32\textwidth}
         \includegraphics[width=\textwidth]{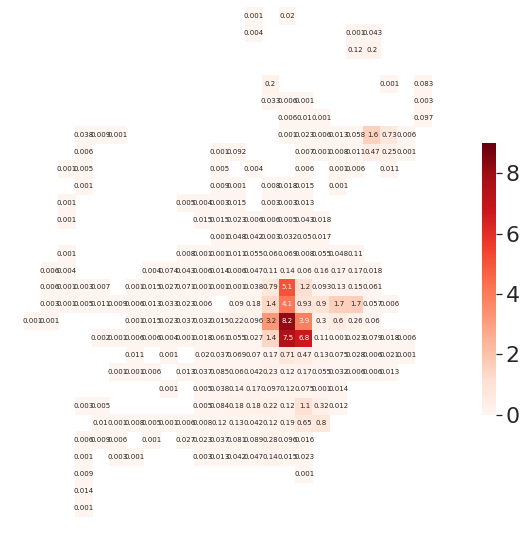}
         \caption{Population in thousands.}
         \label{fig:viterbo-pop-heatmap}
    \end{subfigure}
    \begin{subfigure}[b]{.33\textwidth}
         \includegraphics[width=\textwidth]{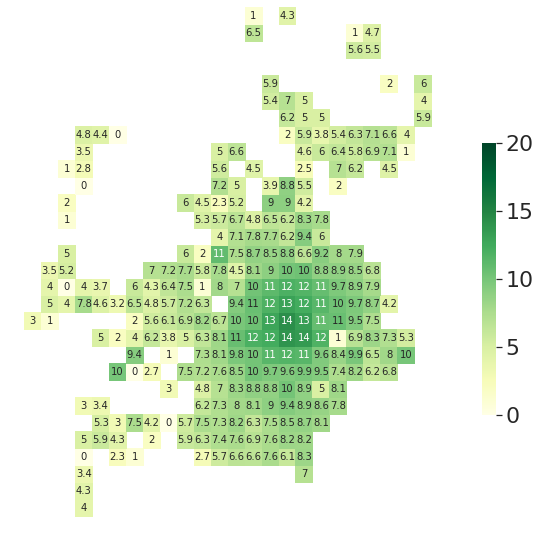}
         \caption{$\beta=0.5$.}
         \label{fig:viterbo-heatmap-0510}
    \end{subfigure}
    \begin{subfigure}[b]{.33\textwidth}
         \includegraphics[width=\textwidth]{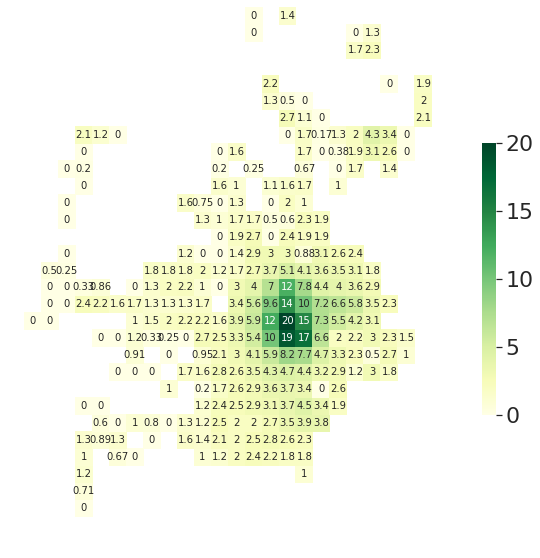}
         \caption{$\beta=2$.}
         \label{fig:viterbo-hetmap-210}
    \end{subfigure}
    \caption{[\textbf{Viterbo}] Heatmaps of the population and of the average degree per tile of the considered territory.
    The average degree is obtained for a graph with $f_u\sim\fit$, $\mu=10$ and $D(u,v)=d(u,v)^{-\beta}$, for both $\beta\in\{0.5,2\}$.
    }
    \label{fig:heatmap_viterbo}
\end{figure}

\begin{figure}[htbp]
    \begin{subfigure}[b]{.48\textwidth}
         \centering
         \includegraphics[width=\textwidth]{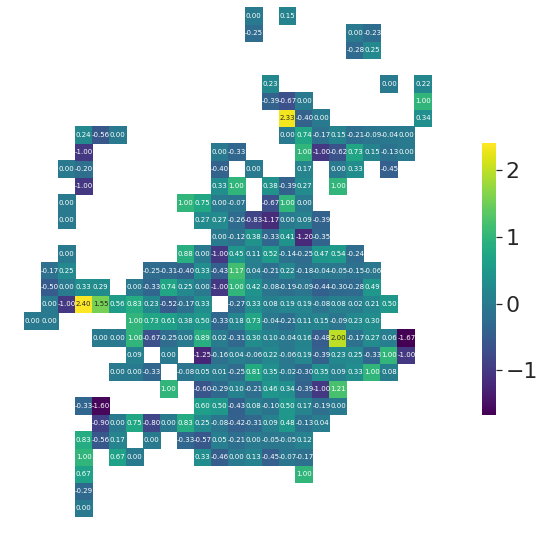}
         \caption{Difference of the \emph{mean} degree of each tile between the configuration with $f_u=\fit$ and the configuration with $f_u\equiv 1$.}
         \label{fig:viterbo_degmean}
    \end{subfigure}
    \hfill
    \begin{subfigure}[b]{.48\textwidth}
         \centering
         \includegraphics[width=\textwidth]{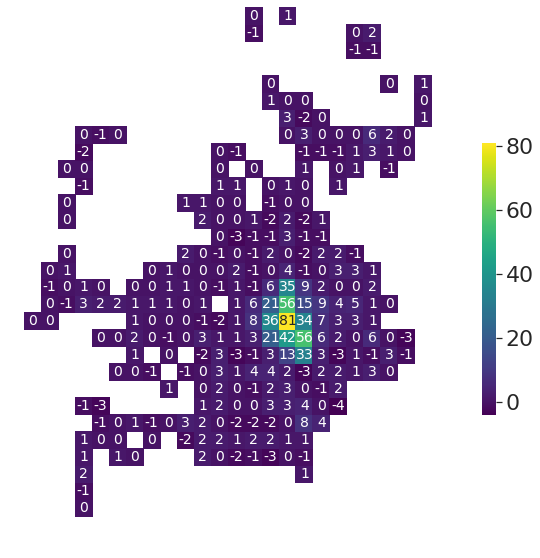}
         \caption{Difference of the \emph{maximum} degree of each tile between the configuration with $f_u=\fit$ and the configuration with $f_u\equiv 1$.}
         \label{fig:viterbo_degmax}
    \end{subfigure}

    \caption{[\textbf{Viterbo}] Impact of switching from $f_u \equiv 1$ to  $f_u\sim\fit$ on the degree distribution of each tile.
    In both cases, $\mu=10$ and $D(u,v)=d(u,v)^{-2}$.}
    \label{fig:heatmaps_degmax_viterbo}
\end{figure}

\section{Results for the City of Sabaudia}\label{app:sabaudia}

In this section we report all plots and tables for the city of Sabaudia, which are by all means analogous to those already presented for Florence and Viterbo.

\begin{table}[htbp]
\caption{[\textbf{Sabaudia}] Percentage of nodes of the graph that belong to the giant component, on average, for the friendship graph $G_F$ and the entire social graph $G$, as $\beta$, $f_u$ and $\mu$ vary.}
\label{tab:sabaudiaGiant}
\centering
\begin{tabular}{llrrrr}
\addlinespace[-\aboverulesep]
\cmidrule[\heavyrulewidth]{3-6}
& & \multicolumn{2}{c}{$\mu=1$} & \multicolumn{2}{c}{$\mu=5$}\\
$\beta$ & $f_u$ & $G_F$ & $G$ & $G_F$ & $G$ \\
\midrule
0.5 & 1                     & 3.8\%  &  75.8\% & 99.3\% & 99.8\% \\
0.5 & $1+\LN(\ln(2), 0.25)$ & 17.4\% &  74.7\% & 98.1\% & 99.4\% \\
2 & 1                       & 6.9\%  &  75.0\% & 99.1\% & 99.7\% \\
2 & $1+\LN(\ln(2), 0.25)$   & 20.3\% &  73.8\% & 97.9\% & 99.3\% \\
\bottomrule
\end{tabular}
\end{table}

\begin{figure}[htbp]
    \centering
    \includegraphics[width=.95\textwidth]{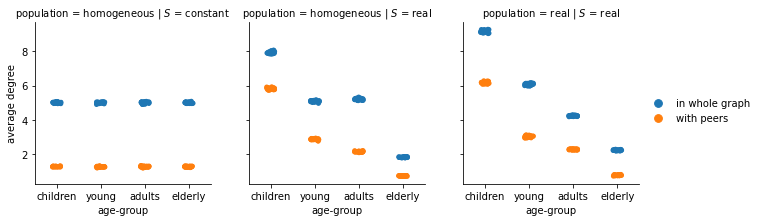}
    \caption{[\textbf{Sabaudia}] Average degree of the individuals of each age group, in the whole friendship graph $G_F$ and with their peers, under different configurations all with $\mu=5$.}
    \label{Figure:deg_sabaudia_avg5}
\end{figure}

\begin{figure}[htbp]
    \centering
    \begin{subfigure}[b]{.35\textwidth}
         \includegraphics[width=\textwidth, trim={50 50 50 50},clip]{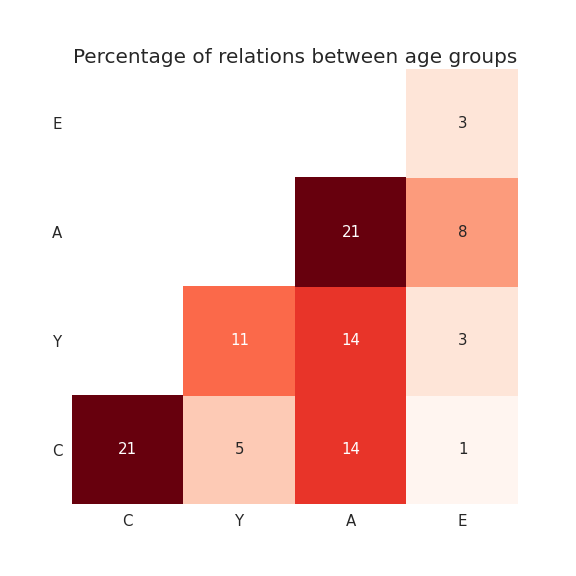}
         \caption{$\beta=0.5$.}
         \label{fig:sabaudia-g2g-hhFalse}
    \end{subfigure}
    \hspace{20pt}
    \begin{subfigure}[b]{.35\textwidth}
         \includegraphics[width=\textwidth, trim={50 50 50 50},clip]{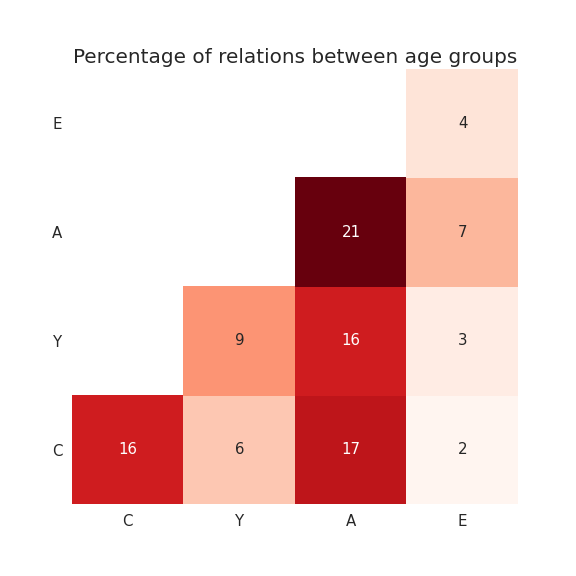}
         \caption{$\beta=2$.}
         \label{fig:sabaudia-g2ghhTrue}
    \end{subfigure}
    \caption{[\textbf{Sabaudia}] Percentage of edges between age groups in a configuration with data-driven population and age-based mixing, for $f_u\sim\fit$, $\mu=5$ and $D(u,v)=d(u,v)^{-2}$.}
    \label{fig:g2g_sabaudia}
\end{figure}

\begin{table}[htbp]
\caption{[\textbf{Sabaudia}] Main features of the friendship graph $G_F$ as the population type, $D(u,v)$ and $\mu$ vary, for constant $f_u$ and $S$. $\dist$ is the average path length, $C$ is the global clustering coefficient, $\rho$ is the degree assortativity, ``\# comp.'' denotes the number of connected components, ``giant \%'' denotes the percentage of nodes in the giant component.}
\label{tab:SabaudiaGeoDist}
\centering 
\begin{subtable}{\textwidth}
\centering
\caption{Friendship network for $\mu=5$. Expected values for an ER graph with $\mu=5$: $\langle dist \rangle \approx 6.2$, $C=0,0002$.}
\begin{tabular}{llrrrrr}
\addlinespace[-\aboverulesep]
\cmidrule[\heavyrulewidth]{3-7}
& & \multicolumn{5}{c}{$\mu=5$} \\
Population  & $D(u,v)$ & $\dist$ & $C$ & $\rho$ & \# comp.  & giant \% \\
\midrule
homogeneous & 1               &   6.37   &   2.55e-04   &  -0.002     &  149.2  & 99.29\%  \\
homogeneous & $d(u,v)^{-0.5}$ &   6.36   &   2.43e-04   &   0.002     &  155.1  & 99.26\%  \\
homogeneous & $d(u,v)^{-2}$   &   6.49   &   0.0017     &   0.04      &  181.4  & 99.13\%  \\
real        & 1               &   6.37   &   2.47e-04   &  -5.67e-04  &  155.6  & 99.26\%  \\
real        & $d(u,v)^{-0.5}$ &   6.25   &   2.95e-04   &   0.06      &  277.1  & 98.67\%  \\
real        & $d(u,v)^{-2}$   &   6.93   &   0.0013     &.  0.4       &  3356.8 & 80.94\%   \\
\bottomrule
\end{tabular}
\end{subtable}

\vspace{10pt}

\begin{subtable}{\textwidth}
\centering
\caption{Friendship network for $\mu=10$. Expected values for an ER graph with $\mu=10$: $\langle dist \rangle \approx 4.32$, $C=0.0005$.}
\begin{tabular}{llrrrrr}
\addlinespace[-\aboverulesep]
\cmidrule[\heavyrulewidth]{3-7}
& & \multicolumn{5}{c}{$\mu=10$}\\
Population & $D(u,v)$ & $\dist$ &
$C$ & $\rho$ & \# comp.  & giant \% \\
\midrule
  homogeneous & 1               &  4.59    &   4.78e-04  &  -0.001       &  1.8    &  100.0\%  \\
  homogeneous & $d(u,v)^{-0.5}$ &  4.59    &   4.71e-04  &   0.004       &  2.2    &  99.99\%  \\
  homogeneous & $d(u,v)^{-2}$   &  4.75    &   0.003     &   0.08        &  3.2    &  99.99\%  \\
  real        & 1               &  4.59    &   4.84e-04  &   5.1e-04     &  2.1    &  99.99\%  \\
  real        & $d(u,v)^{-0.5}$ &  4.56    &   5.72e-04  &   0.12        &  6.3    &  99.98\%  \\
  real        & $d(u,v)^{-2}$   &  5.45    &   0.002     &   0.5         &  1156.6 &  93.96\%  \\
\bottomrule
\end{tabular}
\end{subtable}
\end{table}

\begin{figure}[htbp]
    \centering
    \includegraphics[width=\textwidth]{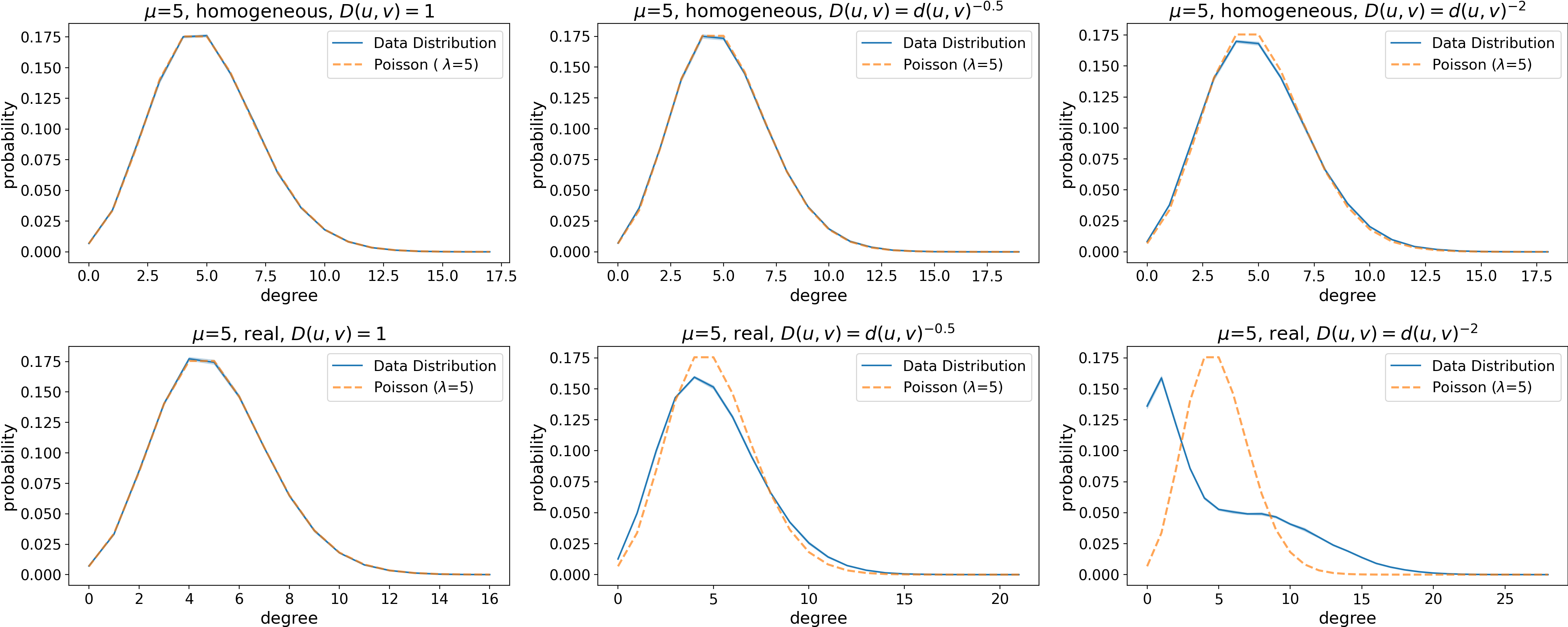}
     \caption{[\textbf{Sabaudia}] Degree distributions of friendship graphs with $\mu=5$ and corresponding Poisson distributions with $\lambda=5$.
    The data distribution lines show the mean, whereas the shaded areas  around the lines represent the 95\% confidence interval.
    The Kullback–Leibler divergence between the Poisson and the empirical distribution are (left to right, top to bottom):
    5.3e-06,
    1.2e-05,
    0.0001, 
    5.8e-06,
    0.001,
    0.05.
    }
    \label{Figure:SabaudiaDegDist5}
\end{figure}

\begin{figure}[htbp]
    \centering
x    \includegraphics[width=\textwidth]{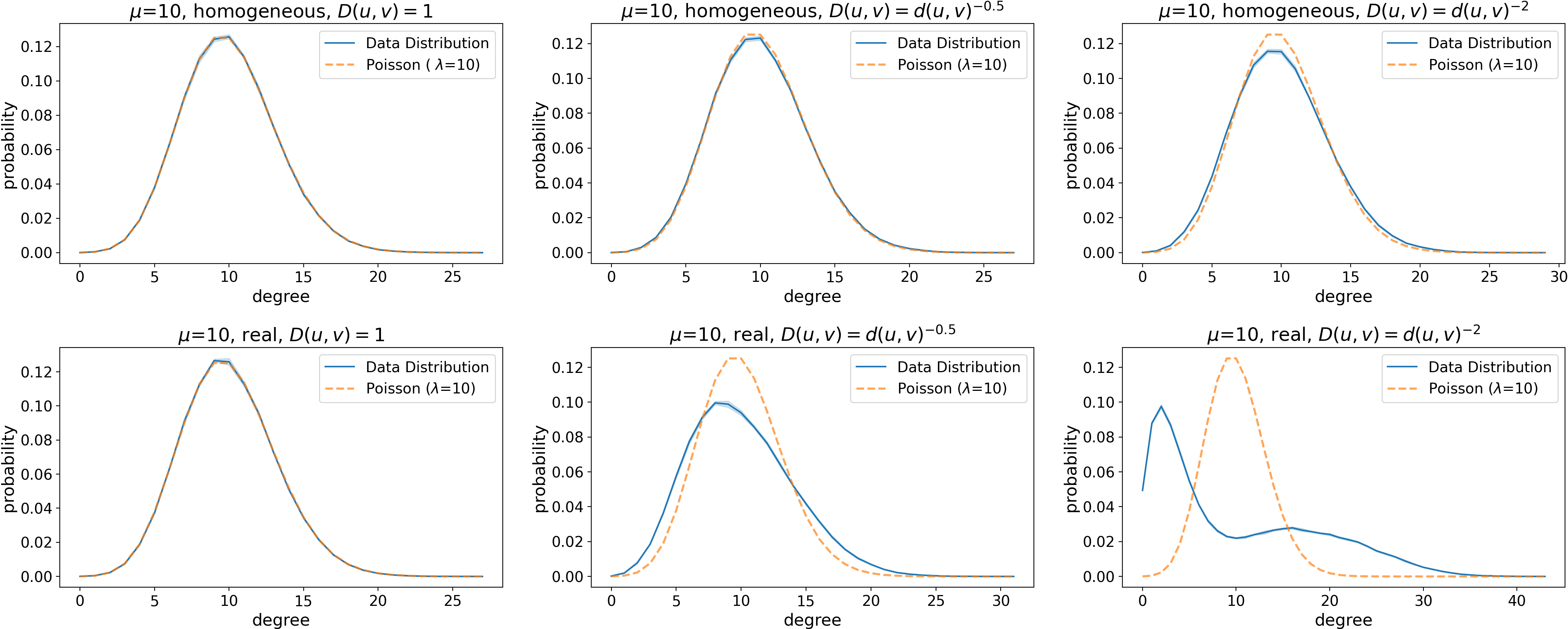}
    \caption{[\textbf{Sabaudia}] Degree distributions of friendship graphs with $\mu=5$ and corresponding Poisson distributions with $\lambda=5$.
    The data distribution lines show the mean, whereas the shaded areas  around the lines represent the 95\% confidence interval.
    The Kullback–Leibler divergence between the Poisson and the empirical distribution are (left to right, top to bottom):
    2.23e-06,
    3.4e-05,
    0.0004,
    3.01e-06,
    0.003,
    0.08.
    }
    \label{Figure:SabaudiaDegDist10}
\end{figure}

\begin{figure}[htbp]
    \centering
    \includegraphics[width=\textwidth]{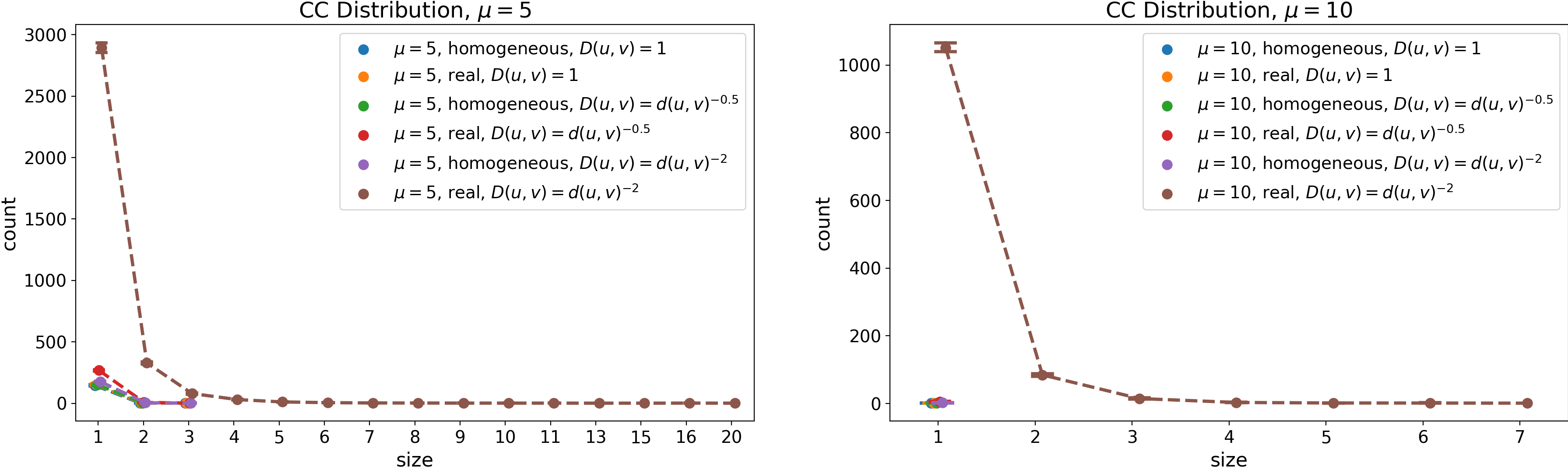}
    \caption{[\textbf{Sabaudia}] Distribution of the size of the connected components other than the giant for friendship graphs with $\mu=5$ and $\mu=10$.
    The lines show the mean, whereas the error bars (with ``caps'') represent the 95\% confidence interval computed over each set of runs.}
    \label{Figure:SabaudiaConnComp}
\end{figure}

\begin{table}[htbp]
\caption{[\textbf{Sabaudia}] Main features of the social graph as $f_u$, $\beta$ and $\mu$ vary, for data-driven population and $S$. $K=\mu+\nu$ is the average degree, $\dist$ is the average path length, $C$ and $C_{\mathrm{loc}}$ are the global and average local clustering coefficients, $\rho$ is the degree assortativity, ``\# comp.'' denotes the number of connected components, ``giant \%'' denotes the percentage of nodes in the giant component.}
\centering
\begin{subtable}{\textwidth}
\centering
\caption{Social graph for $\mu=5$.}
\label{tab:Metrics_Sabaudia_5}
\begin{tabular}{llrrrrrrr}
\addlinespace[-\aboverulesep] 
\cmidrule[\heavyrulewidth]{3-9}
& & \multicolumn{7}{c}{$\mu=5$} \\
$f_u$ & $\beta$ & $K$ & $\dist$ & $C$ & $C_{\mathrm{loc}}$ & $\rho$ & \# comp.  & giant \% \\
\midrule
1      & 0.5   & 6.751 & 5.389 & 0.052 & 0.076 & 0.270  & 238.2  & 98.8 \\
1      & 2     & 6.743 & 6.140 & 0.041 & 0.145 & 0.524  & 1234.3  & 92.0 \\
$\fit$ & 0.5   & 6.733 & 5.297 & 0.048 & 0.087 & 0.206  & 312.3  & 98.4 \\
$\fit$ & 2     & 6.748 & 6.201 & 0.039 & 0.155 & 0.400  & 1309.9  & 91.5 \\
\bottomrule
\end{tabular}
\end{subtable}

\vspace{10pt}

\begin{subtable}{\textwidth}
\centering
\caption{Social graph for $\mu=10$.}
\label{tab:Metrics_Sabaudia_10}
\begin{tabular}{llrrrrrrr}
\addlinespace[-\aboverulesep] 
\cmidrule[\heavyrulewidth]{3-9}
& & \multicolumn{7}{c}{$\mu=10$} \\
$f_u$ & $\beta$ & $K$ & $\dist$ & $C$ & $C_{\mathrm{loc}}$ & $\rho$ & \# comp.  & giant \% \\
\midrule
1      & 0.5 & 11.733  & 4.310 & 0.018 & 0.028 & 0.317  & 28.5   & 99.9 \\
1      & 2   & 11.724  & 5.114 & 0.018 & 0.080 & 0.553  & 543.5  & 96.9 \\
$\fit$ & 0.5 & 11.742  & 4.254 & 0.017 & 0.034 & 0.216  & 54.8   & 99.7 \\
$\fit$ & 2   & 11.759  & 5.033 & 0.018 & 0.089 & 0.393  & 588.7  & 96.6 \\
\bottomrule
\end{tabular}
\end{subtable}
\end{table}

\begin{figure}[htbp]
    \centering
    \begin{subfigure}[b]{.48\textwidth}
         \centering
         \includegraphics[width=\textwidth]{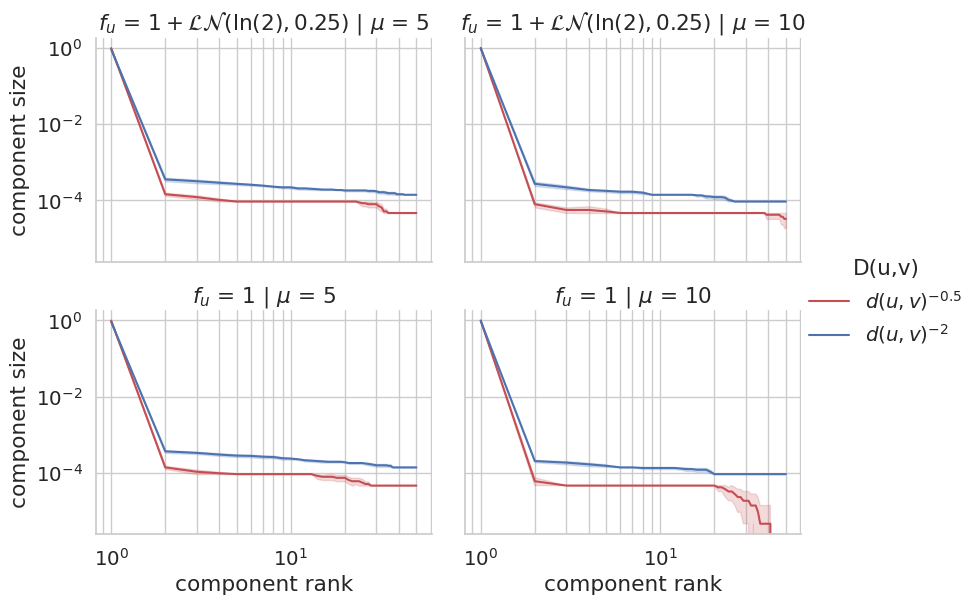}
         \caption{Size of the largest 50 connected components.}
         \label{fig:components_sabaudia}
    \end{subfigure}
    \hfill
    \begin{subfigure}[b]{.48\textwidth}
         \centering
         \includegraphics[width=\textwidth]{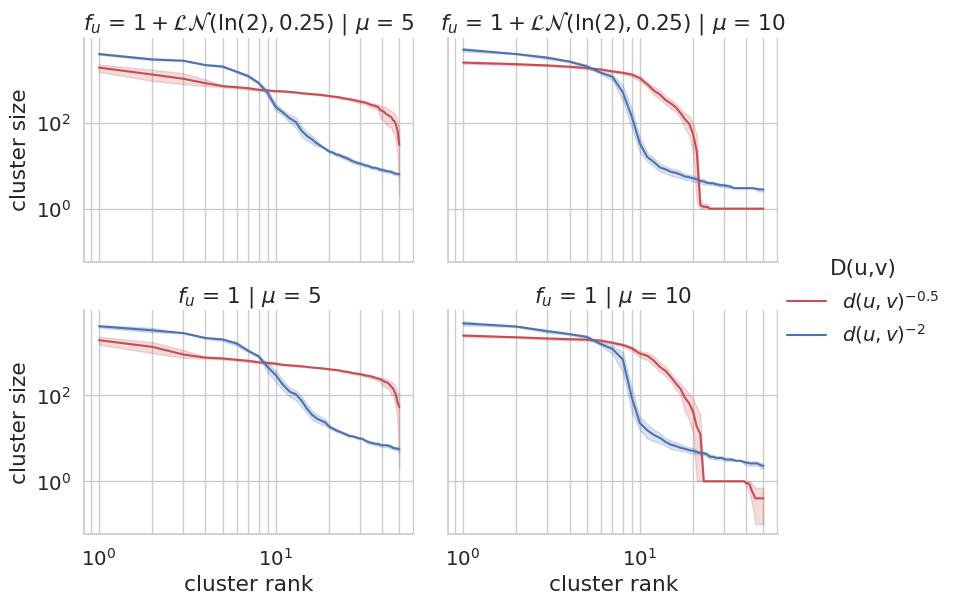}
         \caption{Size of the largest 50 clusters (Louvain).}
         \label{fig:cluster_size_sabaudia}
    \end{subfigure}
    
    \begin{subfigure}[b]{.48\textwidth}
         \centering
         \includegraphics[width=\textwidth]{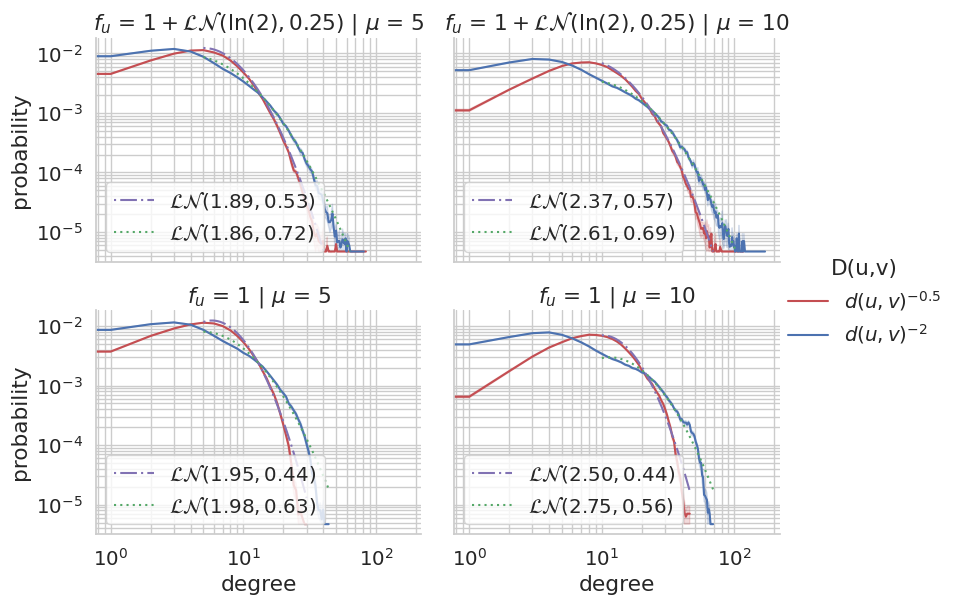}
         \caption{Degree distribution with Lognormal fit.}
         \label{fig:degree_distribution_sabaudia}
    \end{subfigure}
    \hfill
    \begin{subfigure}[b]{.48\textwidth}
         \centering
         \includegraphics[width=\textwidth]{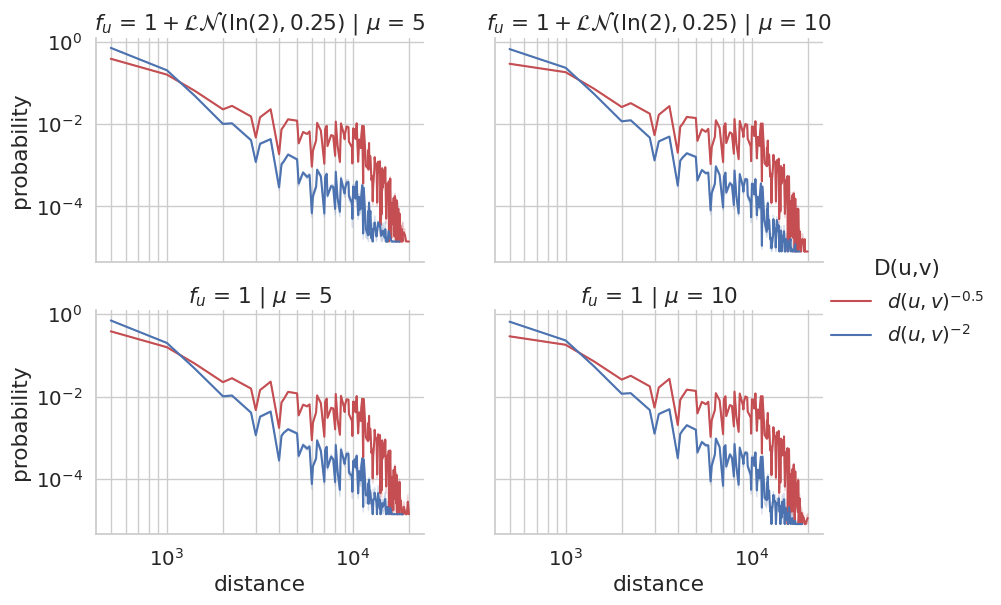}
         \caption{Geographical distance between adjacent vertices.}
         \label{fig:distance_distribution_sabaudia}
    \end{subfigure}
    
    \begin{subfigure}[b]{.48\textwidth}
         \centering
         \includegraphics[width=\textwidth]{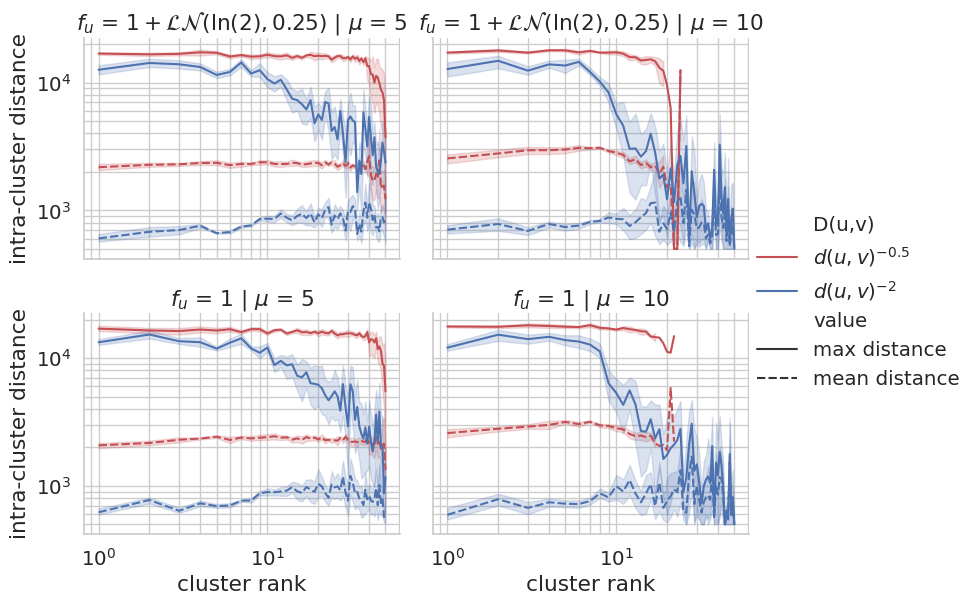}
         \caption{Mean and max intra-cluster geographical distances.}
         \label{fig:cluster_distance_sabaudia}
    \end{subfigure}
    \hfill
    \begin{subfigure}[b]{.48\textwidth}
         \centering
         \includegraphics[width=\textwidth]{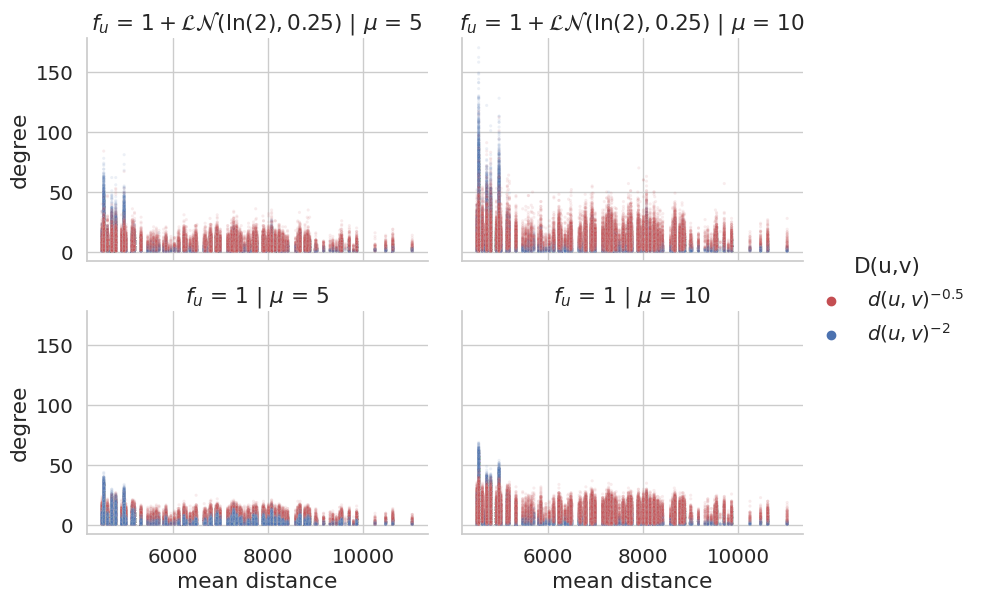}
         \caption{Mean distance to all other vertices \emph{vs.} Degree.}
         \label{fig:degree_distance_sabaudia}
    \end{subfigure}
    \caption{[\textbf{Sabaudia}] Overview of the urban social graph for the city of Sabaudia. Each plot shows the average with confidence interval for 10 independent runs with the same configuration.}
    \label{fig:sabaudia}
\end{figure}

\begin{figure}[htbp]
    \centering
    \includegraphics[width=.6\textwidth]{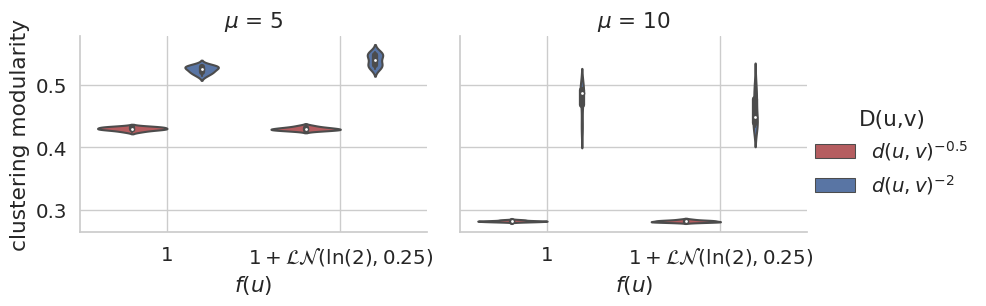}
    \caption{[\textbf{Sabaudia}] Modularity of the obtained clustering structure -- distribution for 10 independent runs per configuration.}
    \label{fig:modularity_sabaudia}
\end{figure}

\begin{figure}[htbp]
    \centering
    \begin{subfigure}[b]{.35\textwidth}
         \includegraphics[width=\textwidth, trim={100 100 100 100},clip]{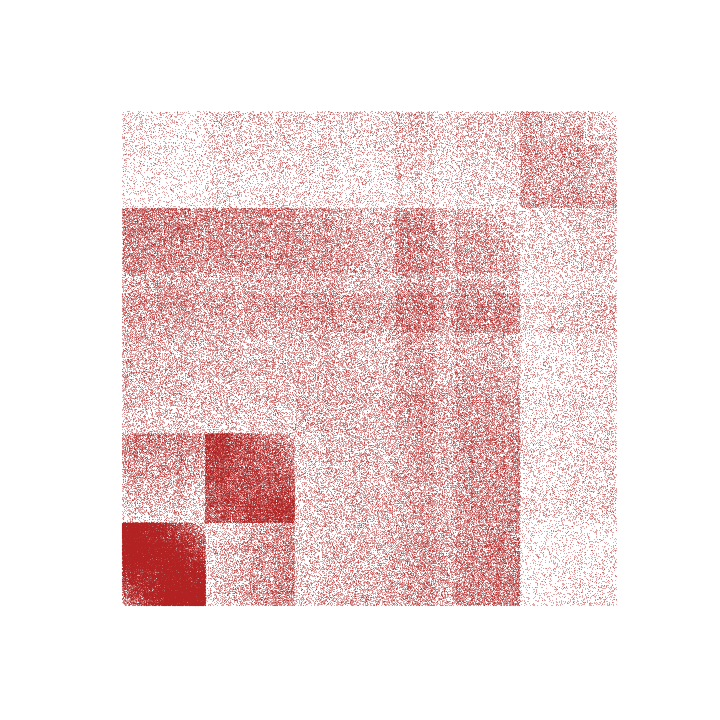}
         \caption{$\beta=0.5$.}
         \label{fig:sabaudia-adj-0510}
    \end{subfigure}
    \hspace{20pt}
    \begin{subfigure}[b]{.35\textwidth}
         \includegraphics[width=\textwidth, trim={100 100 100 100},clip]{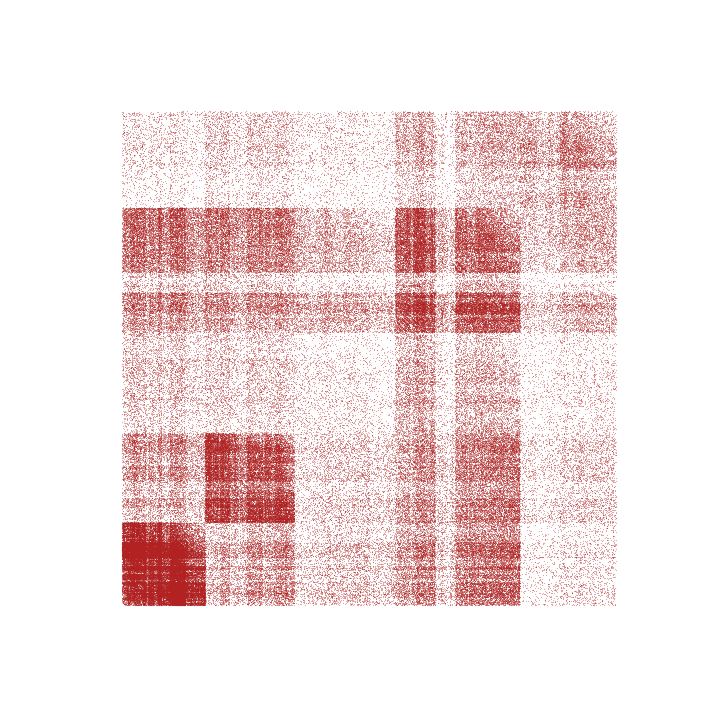}
         \caption{$\beta=2$.}
         \label{fig:sabaudia-adj-210}
    \end{subfigure}
    \caption{[\textbf{Sabaudia}] Adjacency matrix of the social graph with nodes (people) ordered by age-group.
    In both cases $f_u\sim\fit$, $\mu=10$ and $D(u,v)=d(u,v)^{-\beta}$, but $\beta$ varies between the two figures.}
    \label{fig:adj_matrix_sabaudia}
\end{figure}

\begin{figure}[htbp]
    \centering
    \begin{subfigure}{.33\textwidth}
         \includegraphics[width=\textwidth]{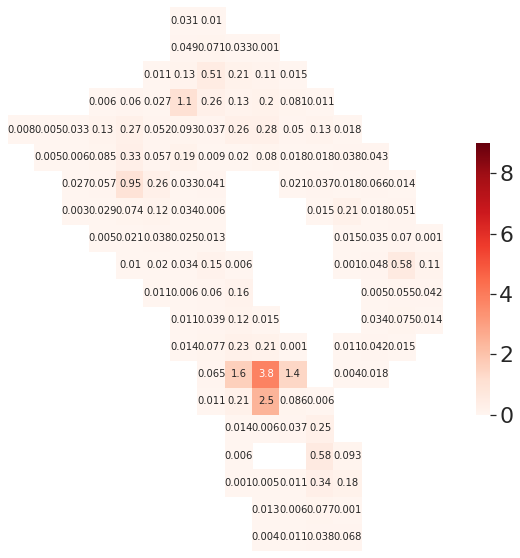}
        \caption{Population in thousands.}
        \label{fig:sabaudia-pop-heatmap}
    \end{subfigure}
    \begin{subfigure}{.33\textwidth}
         \includegraphics[width=\textwidth]{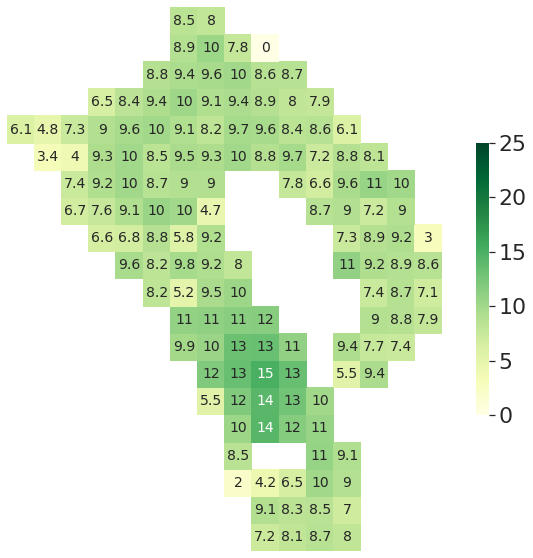}
        \caption{$\beta=0.5$.}
        \label{fig:sabaudia-adj-log10}
    \end{subfigure}
    \begin{subfigure}{.33\textwidth}
         \includegraphics[width=\textwidth]{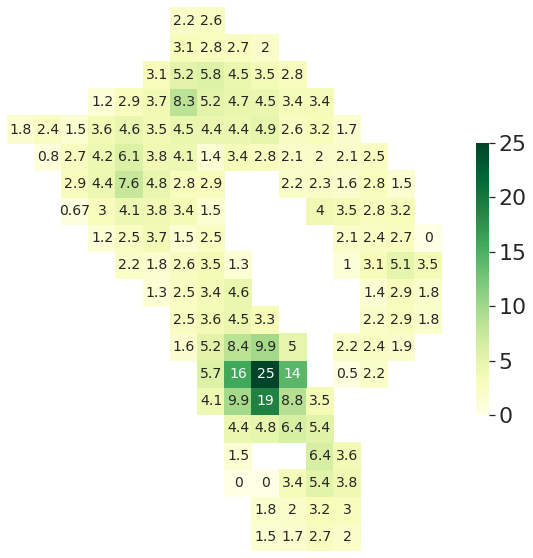}
         \caption{$\beta=2$.}
         \label{fig:sabaudia-adj-one10}
    \end{subfigure}
    \caption{[\textbf{Sabaudia}] Heatmaps of the population and of the average degree per tile of the considered territory.
    The average degree is obtained for a graph with $f_u\sim\fit$, $\mu=10$ and $D(u,v)=d(u,v)^{-\beta}$, for both $\beta\in\{0.5,2\}$.
    }
    \label{fig:heatmap_sabaudia}
\end{figure}

\begin{figure}[htbp]
    \begin{subfigure}[b]{.48\textwidth}
         \centering
         \includegraphics[width=\textwidth]{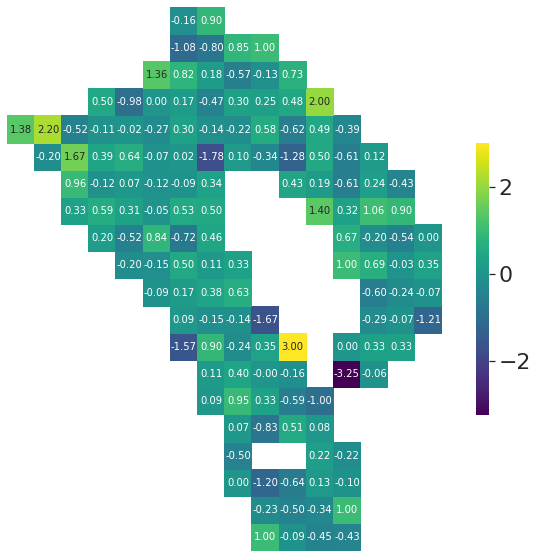}
         \caption{Difference of the \emph{mean} degree of each tile between the configuration with $f_u=\fit$ and the configuration with $f_u\equiv 1$.}
         \label{fig:sabaudia_degmean}
    \end{subfigure}
    \hfill
    \begin{subfigure}[b]{.48\textwidth}
         \centering
         \includegraphics[width=\textwidth]{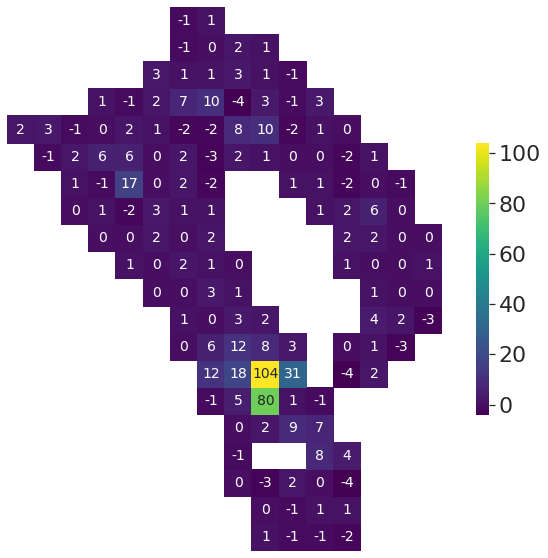}
         \caption{Difference of the \emph{maximum} degree of each tile between the configuration with $f_u=\fit$ and the configuration with $f_u\equiv 1$.}
         \label{fig:sabaudia_degmax}
    \end{subfigure}

    \caption{[\textbf{Sabaudia}] Impact of switching from $f_u \equiv 1$ to  $f_u\sim\fit$ on the degree distribution of each tile.
    In both cases, $\mu=10$ and $D(u,v)=d(u,v)^{-2}$.}
    \label{fig:heatmaps_degmax_sabaudia}
\end{figure}

\end{document}